\documentclass[%
 aip,
 amsmath,amssymb,
preprint,%
]{revtex4-1}

\usepackage{graphicx}% Include figure files
\usepackage{dcolumn}% Align table columns on decimal point
\usepackage{bm}% bold math
\usepackage[utf8]{inputenc}
\usepackage[T1]{fontenc}
\usepackage{mathptmx}
\usepackage{lineno,hyperref}
\usepackage{gensymb} 
\modulolinenumbers[5]
\usepackage{amssymb,amsmath}
\usepackage{multirow}
\usepackage{gensymb}
\usepackage{epstopdf}
\usepackage{float}
\usepackage{subfig}
\usepackage{supertabular}
\usepackage{verbatim}
\usepackage{array}
\usepackage{color}
\usepackage{makecell}
\usepackage{CJKutf8}
\usepackage{natbib}

\begin{document}

\preprint{AIP/123-QED}

\begin{CJK*}{UTF8}{gbsn}
	
\title[Propulsive performance of pitching flexible plates]{A high-fidelity numerical study on the propulsive performance of pitching flexible plates}
% Force line breaks with \\

\author{Guojun Li (李国俊)}
\email{li.guojun@u.nus.edu}
\affiliation{Mechanical Engineering, National University of Singapore, 9 Engineering Drive 1, Singapore 117576}%Lines break automatically or can be forced with \\
 
\author{Ga\"{e}l Kemp}
\affiliation{Engineering, University of Cambridge, Trumpington Street, Cambridge CB2 1PZ, UK}%Lines break automatically or can be forced with \\
 
\author{Rajeev Kumar Jaiman}%
\affiliation{Mechanical Engineering, University of British Columbia, Vancouver, BC Canada V6T 1Z4}%

\author{Boo Cheong Khoo}
\affiliation{Mechanical Engineering, National University of Singapore, 9 Engineering Drive 1, Singapore 117576}

\date{\today}% It is always \today, today,

\begin{abstract}
In this paper, we numerically investigate the propulsive performance of three-dimensional pitching flexible plates with varying flexibility and trailing edge shapes. We employ our recently developed body-conforming fluid-structure interaction solver for our high-fidelity numerical study. To eliminate the effect of other geometric parameters, only the trailing edge angle is varied from $45^\circ$ (concave plate), $90^\circ$ (rectangular plate) to $135^\circ$ (convex plate) while maintaining the constant area of the flexible plate. For a wide range of flexibility, three distinctive flapping motion regimes are classified based on the variation of the flapping dynamics: (i) low bending stiffness $K_B^{low}$, (ii) moderate bending stiffness $K_B^{moderate}$ near resonance, and (iii) high bending stiffness $K_B^{high}$. We examine the impact of the frequency ratio $f^*$ defined as the ratio of the natural frequency of the flexible plate to the actuated pitching frequency. Through our numerical simulations, we find that the global maximum mean thrust occurs near $f^* \approx 1$ corresponding to the resonance condition. However, the optimal propulsive efficiency is achieved around $f^*$=1.54 instead of the resonance condition. While the convex plate with low and high bending stiffness values shows the best performance, the rectangular plate with moderate $K_B^{moderate}$ is the most efficient propulsion configuration. To examine the flow features and the correlated structural motions, we employ the sparsity-promoting dynamic mode decomposition (SP-DMD). We find that the passive deformation induced by the flexibility effect can help in redistributing the pressure gradient thus improving the efficiency and the thrust production. A momentum-based thrust evaluation approach is adopted to link the temporal and spatial evolution of the vortical structures with the time-dependent thrust. When the vortices detach from the trailing edge, the instantaneous thrust shows the largest values due to the strong momentum change and convection process. Moderate flexibility and convex shape help transfer momentum to the fluid, thereby improving the thrust generation and promoting the transition from drag to thrust. The increase of the trailing edge angle can broaden the range of flexibility that produces positive mean thrust. The role of added mass effect on the thrust generation is quantified for different pitching plates and the bending stiffness. These findings are of great significance to the optimal design of propulsion systems with flexible wings.
\end{abstract}

\maketitle
\end{CJK*}

\section{Introduction}
% Background
Biological species in nature have evolved over millions of years to possess superior propulsive performance and high maneuverability for locomotion. These traits are achieved by various flapping-wing-like surfaces with a wide range of shapes and flexibility in different fliers and swimmers \cite{katz1978hydrodynamic,lauder2000function}. These natural fliers and swimmers can inspire the design of highly-efficient self-propelled propulsors and human-made vehicles by searching for the optimal combination of wing geometry and fluid-structure parameters. Towards this goal, a vast body of work has been carried out during the past decades \cite{fish2006passive,muijres2008leading,wang2020optimal,manjunathan2020thrust,cheng2021wing}. Particularly, flexibility and trailing edge (TE) shape were found to play important roles in improving propulsive performance by affecting the surrounding flow features \cite{shyy2008aerodynamics,shyy2010recent,shahzad2018effects,shi2020effects,han2020hydrodynamics}. However, the large physical parameter space poses a serious challenge to characterize the impact of each parameter on the propulsive performance. The thrust-generating mechanism and the efficiency gain of flapping wings with varying trailing edge shapes and flexibility by correlating with the flow features are not fully understood, which motivates the present computational study.

% brief introduction of self-propelled foils
During the past decades, a plethora of early research on the flapping rigid wings has been performed to characterize the effects of various geometric and physical parameters on the thrust generation and the propulsive efficiency \cite{triantafyllou1991wake,anderson1998oscillating,dong2006wake,zhang2018effects,zhang2019effects}. To simplify the flapping dynamics, rigid wing models were used to understand the thrust-generating mechanism of actual biological wings \cite{triantafyllou1991wake,anderson1998oscillating,manjunathan2020thrust}. In reality, biological wings have a variety of flexibility and wing shapes that can meet the desired performance requirements. Inspired by intelligent and efficient biological flight, a series of studies considering the shape and flexibility of the wings were carried out to optimize the performance of the rigid counterparts with simplified shapes \cite{quinn2014scaling,chao2018propulsive,manjunathan2020thrust}.
However, when examining the impact of wing shape or flexibility, in most cases one of the parameters was fixed. This limitation results in a poor understanding of the optimal combination of the wing shape and flexibility that can maximize the propulsive efficiency. In flapping flight, the vortical structure properties are closely connected with the propulsive performance and thrust generation \cite{dabiri2009optimal,lyu2019aerodynamic,zhang2019role}. The flapping wing accelerates the unsteady flow to form vortices containing high velocities, thereby generating thrust by transferring momentum to the fluid \cite{koochesfahani1989vortical,green2011unsteady,park2016vortical}. The creation and the transport process of vortices are strongly influenced by the wing shape and flexibility. While most studies focused on the variation of the vortical structures induced by the flapping wing, very few of them have explored the thrust-generating mechanism by directly correlating the temporal and spatial evolution of the vortical structures and the time-dependent thrust forces, and even fewer examined the effects of wing shape and flexibility from this perspective.

\subsection{Effect of trailing edge shape and flexibility}
\begin{figure*}
	\centering
	\subfloat[][$\Phi = 45 \degree$]{\includegraphics[scale = 1,trim = 0cm 0cm 0cm 0cm, clip]{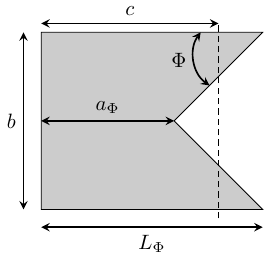} \label{Geometry45}} 
	\subfloat[][$\Phi = 90\degree$]{\includegraphics[scale = 1,trim = -1cm -0.8cm 0cm 0cm, clip]{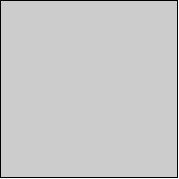} \label{Geometry90}} 
	\subfloat[][$\Phi = 135 \degree$]{\includegraphics[scale = 1,trim = -1cm -0.8cm 0cm 0cm, clip]{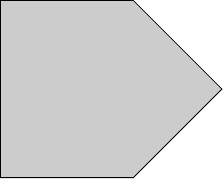} \label{Geometry135}} 
	\caption{\label{fig:3Geometries}Schematic of the plate geometry with the varying trailing edge angles of $\Phi =$ (a) $45 \degree$, (b) $90 \degree$ and (c) $135 \degree$.}
\end{figure*}

% effect of shape and trailing edge shape
To gain further insight into the role of trailing edge shape played in the thrust generation and the optimal propulsive efficiency achievement, a number of studies have been carried out for flapping rigid plates with non-flat trailing edge. One of the pioneers in this research area focused on the lunate tails with varying trailing edge shapes to determine the optimal combination with maximum efficiency \cite{chopra1974hydromechanics,chopra1977hydromechanics}. Liu et al. \cite{liu2016effects} investigated the hydrodynamic performance and wake patterns for various caudal fin shapes over a wide range of Strouhal numbers. Krishnadas et al. \cite{krishnadas2018analysis} numerically studied the propulsive efficiency of biomimetic trapezoidal wings with different trailing edge angles undergoing pitching and heaving motion. The effect of the trailing edge shape on the wake behaviors for a trapezoidal pitching panel was explored through a series of experiments \cite{king2017experimental,king2019experimental}. Although the studies mentioned above aimed to investigate the impact of trailing edge shape on the propulsive performance, other geometric parameters (e.g., wing length, span and leading edge shape) were modified simultaneously. Such simultaneous modifications hinder to isolate the only effect of the trailing edge shape on the propulsive performance and the wake topology of flapping plates. 
Recently, Van Buren et al. \cite{VanBuren2017} reported their experimental work on the effect of trailing edge shape on the wake evolution and the propulsive performance of pitching rigid plates. To eliminate the impact of other geometric parameters on propulsive performance, a constant area, $S=0.1$ m$^2$, and the same mean aspect ratio, $AR=1$ were maintained, but only the trailing edge angle $\Phi$ was modified from $45^\circ$ to $135^\circ$ in intervals of $15^\circ$ for different plates. A schematic of three representative plate geometries with varying chevron-shaped trailing edges is shown in Fig.~\ref{fig:3Geometries}. It can be concluded from this experiment that pitching plates with convex shape ($\Phi>90^\circ$) exhibited larger thrust production and superior propulsive efficiency than the plates with rectangular ($\Phi=90^\circ$) and concave ($\Phi<90^\circ$) trailing edge angles. However, the wake topology was only measured at a Reynolds number of $Re=6000$ and the propulsive performance was available at $Re=10000$ in the experiment. Hence, it is hard to establish a connection between the wake structures and the propulsive performance. Further mechanism study on the thrust transition was restricted to the existing experimental results. To avoid this limitation, some recent numerical studies \cite{hemmati2017performance,hemmati2019effects} were performed to relate the thrust generation, the wake topology and the instantaneous flow features around a moving rigid plate.

% effect of flexibility
With respect to the self-propelled plates with different trailing edge shapes, the experimental work done by Van Buren et al. \cite{VanBuren2017} was restricted to a rigid plate. However, the caudal fin of aquatic animals and the flapping wing of flying species exhibit various flexible properties \cite{fish2006passive}. Through numerous studies on the role of flexibility, it was found that passive flexibility can help in redistributing the pressure gradient on the plate surface and regulating the vortical structures \cite{shyy2008aerodynamics,marais2012stabilizing,paraz2016thrust,hoover2018swimming}. As a result, the thrust coefficient and the propulsive efficiency of a flexible plate were enhanced to different degrees compared to its rigid counterpart \cite{michelin2009resonance,floryan2019large,floryan2020distributed}. The resonance between the actuated frequency and the natural frequency of the coupled flapping system governed by flexibility was found to be beneficial to the thrust generation \cite{quinn2014scaling,paraz2016thrust,goza2020connections}. However, the optimal propulsive efficiency did not strongly depend on the occurrence of resonance \cite{dewey2013scaling,moored2014linear,floryan2017scaling,floryan2018clarifying}. Considering the benefits of flexibility and trailing edge shape for enhanced propulsion performance, the participation between these two factors should be considered when designing an effective propulsion system. Only a handful of publications on the joint effects of trailing edge shape and flexibility on propulsive performance can be found in the literature. Zhang et al. \cite{zhang2020effect} numerically investigated the self-propulsive performance for flexible plates undergoing heaving motion as a function of trailing edge angle $\Phi$. As the flexibility increased, the optimal performance was achieved for the concave, convex and square panels, respectively. In the present study, we investigate the propulsive performance for pitching flexible plates with varying trailing edge shapes over a wide range of flexibility based on the experimental plate models \cite{VanBuren2017}. Three representative plates with a concave shape ($\Phi=45^\circ$), a square shape ($\Phi=90^\circ$) and a convex shape ($\Phi=135^\circ$) shown in Fig.~\ref{fig:3Geometries} are considered for simplicity in the current paper. 

\subsection{Drag-thrust transition }
% drag-thrust transition
The transition between drag and thrust is widely observed in the self-propelled plates with different propulsive parameters. Biological species and bio-inspired vehicles may choose proper parameter combinations to achieve the goal of moving forward, backward and rapid turn (i.e., maneuverability). Godoy-Diana et al. \cite{godoy2008transitions} experimentally investigated the vortex streets formed behind a flapping rigid foil. The results revealed that the drag-thrust transition mechanism was governed by the transition from a Bénard–von Kármán (BvK) to a reverse BvK. Andersen et al. \cite{andersen2017wake} reported a new drag-thrust transition mechanism caused by the formation of two vortex pairs per oscillation period for rigid foils undergoing heaving motion with high amplitude and low frequency. Most of the studies on the drag-thrust transition were limited to rigid foils. Recently, Marais et al. \cite{marais2012stabilizing} discovered that flexibility suppressed the symmetry-breaking process of the reverse BvK to improve thrust production. Tzezana et al. \cite{tzezana2019thrust} investigated the drag-thrust transition phenomenon for flapping compliant membrane models with different values of wing compliance, flapping kinematics and inertia. Flapping wings with larger flexibility were prone to trigger the thrust-drag transition earlier at a higher flapping frequency, compared to the wings with smaller wing compliance. A handful of studies can be found to investigate the effect of the trailing edge shape on the drag-thrust transition. The wake structures behind flexible wings with varying wing shapes may become complex and cannot be simply regarded as the typical von Kármán wakes or other common regular wake structures. The discovered drag-thrust transition mechanism based on the identification of the typical wake patterns is not suitable for the coupled fluid-flexible plate system with complex wake structures \cite{floryan2020swimmers}. Some advanced approaches are desirable to establish a direct correlation between the flexible plate deformation, the temporal and spatial evolution of the vortical structures and the time-dependent fluid loads to reveal the thrust-generating mechanism. 

% connection between displacement, vortical structures and thrust (control volume analysis and SP-DMD)
To explore the variation of the generated thrust related to the structural displacement, the correlated dominant modes of the wake structures and the structural motions need to be identified from the coupled system. Goza et al. \cite{goza2018modal} proposed combined fluid-structure formulations based on proper orthogonal decomposition (POD) and dynamic mode decomposition (DMD) to extract the correlated fluid and structural modes from physical fields. In this study, we extend the combined formulation to an improved mode decomposition method called sparsity-promoting dynamic mode decomposition (SP-DMD) \cite{jovanovic2014sparsity}. The SP-DMD method can overcome the limitations of the traditional DMD method to efficiently select the most influential DMD modes from all decomposed modes \cite{liu2016interaction}. The drag-thrust transition process is usually accompanied by complex changes in vortical structures, which are potentially related to the transition mechanism. The velocity/pressure method \cite{noca1997evaluation} and the vorticity/added-mass approach \cite{wu1981theory} as well as their variants \cite{noca1999comparison} derived from the momentum balance equation offer a practical way to reveal the thrust-generating mechanism by connecting with the instantaneous flow features. These non-intrusive methods were successfully adopted to evaluate the time-dependent forces exerted on a body by integrating the fluid variables within the control volume obtained from experimental measurements \cite{van2007evaluation,kim2011flexibility} or numerical simulations \cite{wu2007integral,andersen2017wake}. For the first time, we apply the velocity/pressure method to the fluid-pitching flexible plate coupled system with varying flexibility and trailing edge shapes. Herein, the main purpose is to explore the relationship between the instantaneous produced thrust and the induced vortical structures. Through the decomposition of the resulting total thrust into four terms with obvious physical significance, the role of flexibility and trailing edge shape in the thrust generation is quantitatively examined. From the perspective of the unsteady force dynamics, two types of the thrust-generating mechanisms have been summarized based on a vast body of works \cite{wu1981theory,bottom2016hydrodynamics,smits2019undulatory,dynnikova2019added,guvernyuk2020contribution}: (i) the lift-based mechanism and (ii) the added mass mechanism. The thrust generated by the lift-based mechanism primarily relies on the circulatory forces due to vorticity generation. However, for a pitching flexible plate immersed in the unsteady flow, the reactive force associated with the acceleration due to the pitching motion and the passive deformation due to the added mass effect can play an important role in the thrust generation. An approximate analytical formulation of the added mass coefficient evaluation \cite{yadykin2003added,jaiman2014added} is employed to investigate the added mass effect on the thrust generation for pitching flexible plates with varying flexibility and the physical conditions.

% research gap and research question
\subsection{Current work and contributions}
In this study, we numerically investigate the propulsive performance of flexible plates undergoing prescribed pitching motion with different trailing edge shapes and flexibilities. A recently developed three-dimensional partitioned aeroelastic framework is adopted to simulate the pitching flexible plates \cite{li2018novel}. With the aid of the combined mode decomposition technique, the momentum-based thrust evaluation approach and the analytical added mass model, the following key questions concerning the thrust generation, the propulsive efficiency and the drag-thrust transition are addressed: (i) How do flexibility and trailing edge shape affect the propulsive performance and the wake topology of pitching flexible plates? (ii) What is the optimal combination of the trailing edge shape and flexibility to maximize the efficiency of flexible propulsors? (iii) What is the drag-thrust transition mechanism for pitching flexible plates with varying flexibility and trailing edge shapes? To address (i), we perform a series of numerical simulations for the pitching flexible plates with three representative trailing edge shapes shown in Fig.~\ref{fig:3Geometries} and varying bending stiffnesses at a moderate Reynolds number of $Re=1000$ and a fixed Strouhal number of $St=0.3$ that $St$ approaches the optimal efficiency range. The comparisons of the propulsive performance, the flapping dynamics and the flow features are performed to understand the role of trailing edge shape and flexibility. The optimal combination of these two parameters for the pitching plates with maximum thrust and efficiency is determined from the propulsive performance map. The natural frequency of the flexible plate immersed in the unsteady flow is evaluated to explore the effect of flexibility on the structural resonance and the propulsive performance. The combined SP-DMD method is employed to correlate the coherent vortical structures and the structural motions. The relationship between the unsteady momentum transfer and the thrust generation is established via the momentum-based thrust evaluation approach. The added mass force contributions are quantified to assess the effects of flexibility and the trailing edge shape on the generated thrust force. To understand the feedback connection between the pitching motion and the flow features, we finally examine the mechanism of the drag-thrust transition for various pitching plates by the decomposition of the time-averaged thrust terms.

% organization
The reminder of this paper is organized as follows. The governing equations for the fluid-flexible plate coupled system are described in Section~\ref{sec:section2}. The problem set-up for the pitching plate is described in Section~\ref{sec:section3}. In Section~\ref{sec:section4}, we study the effect of trailing edge shape and flexibility on the propulsive performance and the drag-thrust transition mechanism is explored with the aid of the SP-DMD method, the momentum-based equation and the analytical added mass model. Main conclusions are summarized in Section~\ref{sec:section5}.

\section{Governing equation} \label{sec:section2}
To simulate the coupled fluid-flexible structure system, the incompressible Navier-Stokes equations are coupled with the nonlinear structure equations via a partitioned iterative scheme. The Navier-Stokes equations are discretized via a stabilized Petrov-Galerkin finite element method in an arbitrary Lagrangian-Eulerian (ALE) reference frame and the structure motion equations are solved via nonlinear co-rotational finite element method in a Lagrangian coordinate \cite{jaiman2016stable,li2018novel}. The body-fitted moving boundary is applied between the interface of the fluid domain and the structure domain. The turbulent flow is modeling by the delayed detached eddy simulation (DDES) model using the positivity preserving variational (PPV) scheme. For the sake of completeness, we present the variational formulations for the coupled fluid-structure system. The generalized-$\alpha$ method which can ensure unconditionally stable is utilized to update the fluid variables in the time domain
\begin{eqnarray}
\bm{u}^{f,n+1} = \bm{u}^{f,n} + \Delta t \partial_t \bm{u}^{f,n} + \gamma^f \Delta t (\partial_t \bm{u}^{f,n+1}- \partial_t \bm{u}^{f,n})
\\
\partial_t \bm{u}^{f,n+\alpha_m^f} = \partial_t \bm{u}^{f,n} + \alpha_m^f (\partial_t \bm{u}^{f,n+1}- \partial_t \bm{u}^{f,n})
\\
\bm{u}^{f,n+\alpha^f} = \bm{u}^{f,n} + \alpha^f (\bm{u}^{f,n+1}- \bm{u}^{f,n})
\\
\bm{u}^{m,n+\alpha^f} = \bm{u}^{m,n} + \alpha^f (\bm{u}^{m,n+1}- \bm{u}^{m,n})
\end{eqnarray}
where $\partial_t$ and $\Delta t$ are the partial derivative of a physical variable in time and the time step size. $\bm{u}^{f,n}$ and $\bm{u}^{m,n}$ denote the fluid and mesh velocities at the time step $n$ at every spatial node $\bm{x}^f$ in the fluid domain $\Omega^f(t)$. The three coefficients $\alpha^f$, $\alpha_m^f$ and $\gamma^f$ related to the spectral radius $\rho_{\infty}$ are the generalized-$\alpha$ parameters defined by
\begin{equation}
\alpha^f = \frac{1}{1+\rho_{\infty}}, \quad  \alpha_m^f = \frac{1}{2} \left( \frac{3-\rho_{\infty}}{1+\rho_{\infty}} \right), \quad  \gamma^f = \frac{1}{2} + \alpha_m^f - \alpha^f
\end{equation}

Suppose ${\mathcal{S}_{\bm{u}^f}^f}$ and ${\mathcal{S}_{p}^f}$ are the test function spaces for fluid velocity and pressure, which are defined as
\begin{eqnarray}
{\mathcal{S}_{\bm{u}^f}^f} = \{\bm{u}^f | \bm{u}^f \in H^1(\Omega^f(t)), \bm{u}^f = \bm{u}^f_d \ \text{on} \ \Gamma^f_d(t)\}
\\
{\mathcal{S}_{p}^f} = \{ p | p \in L^2(\Omega^f(t)) \}
\end{eqnarray}
where $H^1(\Omega^f(t))$ and $L^2(\Omega^f(t))$ are the square-integrable $\mathbb{R}^d$-valued function space and the scalar-valued function space with square-integrable derivatives in the fluid domain $\Omega^f(t)$, respectively. $\bm{u}^f_d $ denotes the velocity on the Dirichlet boundary $\Gamma^f_d(t)$ of the fluid domain. The corresponding test function spaces for fluid velocity $\mathcal{V}^{f}_{\boldsymbol{\phi}^f}$ and pressure $\mathcal{V}^{f}_q$ are defined as
\begin{eqnarray}
\mathcal{V}^{f}_{\boldsymbol{\phi}^f} = \{ \boldsymbol{\phi}^f | \boldsymbol{\phi}^f \in  H^1(\Omega^f(t)), \boldsymbol{\phi}^f = \bm{0} \ \text{on} \ \Gamma^f_d(t) \}
\\
\mathcal{V}^{f}_q = \{ q | q \in L^2(\Omega^f(t)) \}
\end{eqnarray}
where $\boldsymbol{\phi}^f$ and $q$ are the weighting-function counterparts of fluid velocity $\bm{u}^f$ and pressure $p$. 

The fluid equations in the variational statement can be written as: find the velocity and pressure fields $[\overline{\bm{u}}^{f}(t^{n+\alpha^f}),\overline{p}(t^{n+1})] \in {\mathcal{S}_{\bm{u}^f}^f} \times {\mathcal{S}_{p}^f}$ such that $\forall[\boldsymbol{\phi}^f,q] \in \mathcal{V}^{f}_{\boldsymbol{\phi}^f} \times \mathcal{V}^{f}_q  $
\begin{eqnarray}
 & &\int_{\Omega^f(t)} \rho^f (\partial_t \overline{\bm{u}}^{f}+(\overline{\bm{u}}^{f}-\bm{u}^{m}) \cdot \nabla \overline{\bm{u}}^{f}) \cdot \boldsymbol{\phi}^f {\rm{d}\Omega} \nonumber\\
 & &+\int_{\Omega^f(t)} \overline{\boldsymbol{\sigma}}^{f}:\nabla \boldsymbol{\phi}^f {\rm{d}\Omega} + \int_{\Omega^f(t)} {\boldsymbol{\sigma}^{\text{ddes}}}^{f}:\nabla \boldsymbol{\phi}^f {\rm{d}\Omega} \nonumber\\
 & & + \sum_{e=1}^{n^f_{el}} \int_{\Omega^e} \tau_m (\rho^f (\overline{\bm{u}}^{f}-\bm{u}^{m}) \cdot \nabla \boldsymbol{\phi}^f+\nabla q) \cdot \boldsymbol{\mathcal{R}}_m {\rm{d}\Omega^e} \nonumber\\
 & &-\int_{\Omega^f(t)} \nabla \cdot \overline{\bm{u}}^{f} q  {\rm{d}\Omega}  +\sum_{e=1}^{n^f_{el}} \int_{\Omega^e} \nabla \cdot \boldsymbol{\phi}^f \tau_c \nabla \cdot \overline{\bm{u}}^{f} {\rm{d}\Omega^e} \nonumber\\
 & &-\sum_{e=1}^{n^f_{el}} \int_{\Omega^e} \tau_m \boldsymbol{\phi}^f \cdot (\boldsymbol{\mathcal{R}}_m \cdot \nabla \overline{\boldsymbol{u}}^{f}) {\rm{d}\Omega^e}  \nonumber\\
& &- \sum_{e=1}^{n^f_{el}} \int_{\Omega^e} \nabla \boldsymbol{\phi}^f : (\tau_m \boldsymbol{\mathcal{R}}_m\otimes \tau_m \boldsymbol{\mathcal{R}}_m) {\rm{d}\Omega^e} \nonumber\\
& & = \int_{\Omega^f(t)} \bm{b}^f(t^{n+\alpha^f}) \cdot \boldsymbol{\phi}^f {\rm{d}\Omega} + \int_{\Gamma} \bm{h}^f \cdot \boldsymbol{\phi}^f {\rm{d} \Gamma} 
\label{eq:eqGA6} 
\end{eqnarray}
where $\rho^f$ denotes the fluid density. $\overline{\boldsymbol{\sigma}}^{f}$ and ${\boldsymbol{\sigma}^{\text{ddes}}}^{f}$ are the Cauchy stress tensor for a Newtonian fluid and the turbulent stress term, respectively. The Galerkin terms for the momentum equation and the viscous and turbulent stress terms are presented in the first and second lines. The third line represents the integral of the Petrov-Galerkin stabilization terms for the momentum equation on the total number of $n^f_{el}$ element domains $\Omega^e$. The Galerkin and the Galerkin/least-squares stabilization terms for the continuity equation are shown in the fourth line. In the fifth and sixth lines, the approximation of the fine scale velocity on element interiors based on the multiscale argument forms two residual terms. On the right-hand side of Eq.~(\ref{eq:eqGA6}) in the seventh line, the two terms represent the body forces $\bm{b}^f(t^{n+\alpha^f})$ and the Neumann boundary conditions $\bm{h}^f$. $\boldsymbol{\mathcal{R}}_c$ and $\boldsymbol{\mathcal{R}}_m$ denote the element-wise residuals of the continuity and the momentum equations, given by
\begin{eqnarray}
\boldsymbol{\mathcal{R}}_c = \nabla \cdot \overline{\bm{u}}^{f}
\\
\boldsymbol{\mathcal{R}}_m = \rho^f \partial_t \overline{\bm{u}}^{f}+\rho^f (\overline{\bm{u}}^{f}-\bm{u}^{m}) \cdot \nabla \overline{\bm{u}}^{f} - \nabla \cdot \overline{\boldsymbol{\sigma}}^{f} - \nabla \cdot {\boldsymbol{\sigma}^{\text{ddes}}}^{f} - \bm{b}^f(t^{n+\alpha^f}) 
\end{eqnarray}
$\tau_c$ and $\tau_m$ are the stabilization parameters for the continuity and the momentum equations, which adds the least-squares metrics to the element level integrals in the stabilized formulation. The stabilization parameters are expressed as
\begin{eqnarray}
\tau_m = \left[ \left( \frac{2 \rho^f}{\Delta t} \right)^2 + (\rho^f)^2 (\overline{\bm{u}}^{f}-\bm{u}^{m}) \cdot \bm{G} (\overline{\bm{u}}^{f}-\bm{u}^{m}) + C_I (\mu^f + \mu_T)^2 \bm{G}:\bm{G}  \right]^{-1/2}
\\
\tau_c = \frac{1}{\text{tr}(\bm{G}) \tau_m}
\end{eqnarray}
where $\Delta t$ represents the time increment. $C_I$ is a constant value based on the element-wise inverse estimates. $\bm{G}$ denotes the element contravariant metric tensor defined as
\begin{equation}
\bm{G} = \frac{\partial \boldsymbol{\xi}^T}{\partial \bm{x}^f} \frac{\partial \boldsymbol{\xi}}{\partial \bm{x}^f} 
\end{equation} 
where $\bm{x}^f$ is the physical coordinate system and $\boldsymbol{\xi}$ denotes the local element coordinate system. $\text{tr}(\bm{G})$ represents the trace of the contravariant metric tensor.

Suppose ${\mathcal{S}_{\bm{u}^s}^s}$ and $\mathcal{V}^{s}_{\boldsymbol{\phi}^s}$ are the trial solution and the test function spaces, which are defined as
\begin{eqnarray}
{\mathcal{S}_{\bm{u}^s}^s} = \{\bm{u}^s | \bm{u}^s \in H^1(\Omega^s(t)), \bm{u}^s = \bm{u}^s_d \ \text{on} \ \Gamma^s_d(t)\}
\\
\mathcal{V}^{s}_{\boldsymbol{\phi}^s} = \{ \boldsymbol{\phi}^s | \boldsymbol{\phi}^s \in  H^1(\Omega^s(t)), \boldsymbol{\phi}^s = \bm{0} \ \text{on} \ \Gamma^s_d(t) \}
\end{eqnarray}
where $\bm{u}^s = \frac{\partial \bm{d}^s}{\partial t}$ is the structural velocity and $\bm{d}^s$ denotes the structural displacement. $H^1(\Omega^s(t))$ and $L^2(\Omega^s(t))$ are the square-integrable $\mathbb{R}^d$-valued function space and the scalar-valued function space with square-integrable derivatives in the structural domain $\Omega^s(t)$, respectively. $\bm{u}^s_d $ denotes the velocity on the Dirichlet boundary $\Gamma^s_d(t)$ of the structural domain. $\boldsymbol{\phi}^s$ represents the weighting-function of the structural velocity $\bm{u}^s$.
The variational formulation of the motion equations for a flexible structure is given as: find $\bm{u}^s \in  {\mathcal{S}_{\bm{u}^s}^s}$ such that $\forall \boldsymbol{\phi}^s \in \mathcal{V}^{s}_{\boldsymbol{\phi}^s} $
\begin{eqnarray}
&&\int^{t^{n+1}}_{t^n} 
\left( \int_{\Omega^s_i} \rho^s \frac{\partial \bm{u}^s}{\partial t} \cdot \boldsymbol{\phi}^s {\rm{d}\Omega} + \int_{\Omega^s_i} \boldsymbol{\sigma}^s : \nabla \boldsymbol{\phi}^s  {\rm{d}\Omega}
\right) {\rm{d}}t \nonumber\\
&&= \int^{t^{n+1}}_{t^n} 
\left(
\int_{\Omega^s_i} \bm{b}^s \cdot \boldsymbol{\phi}^s  {\rm{d}\Omega} + \int_{\Gamma^s_i} \bm{h}^s \cdot \boldsymbol{\phi}^s {\rm{d} \Gamma}
\right) {\rm{d}}t   
\label{eq:eqMB1} 
\end{eqnarray}
where $\rho^s$ denotes the structural density. $\boldsymbol{\sigma}^s$ and $\bm{h}^s=\boldsymbol{\sigma}^s \cdot \bm{n}^s$ are the stress tensor and the Neumann condition at the boundary $\Gamma^s_{i}$, respectively. On the right-hand side of Eq.~(\ref{eq:eqMB1}), $\bm{b}^s$ is the body force acting on the flexible structures $\Omega^s_i$.

The velocity and traction continuity along the fluid-structure interface $\Gamma^{fs}_{i}$ is satisfied for the coupled fluid and structural motion equations
\begin{eqnarray}
\overline{\bm{u}}^f(\boldsymbol{\varphi}^s(\bm{x}^s,t),t)=\bm{u}^s(\bm{x}^s,t) \quad \forall \bm{x}^s \in \Gamma^{fs}_i  
\label{eq:eqFSI1}
\\
\int_{\boldsymbol{\varphi}^s(\gamma^{fs},t)}\overline{\boldsymbol{\sigma}}^f(\bm{x}^f,t) \cdot \bm{n}^f {\rm{d}}\Gamma+\int_{\gamma^{fs}}\boldsymbol{\sigma}^s(\bm{x}^s,t) \cdot \bm{n}^s {\rm{d}}\Gamma=0  \quad \forall \gamma^{fs} \in \Gamma^{fs}_i
\label{eq:eqFSI2}
\end{eqnarray}
where $\bm{u}^s$ is the structural velocity for the initial Lagrangian point $\bm{x}^s \in \Omega^s_i$ at time instant $t$. $\boldsymbol{\varphi}^s$ denotes the mapping function between the structural point $\bm{x}^s$ and its deformed position. $\bm{n}^f$ and $\bm{n}^s$ are the outer normals to the interface boundaries in the fluid and structural domains, respectively. $\boldsymbol{\varphi}^s(\gamma^{fs},t)$ represents the fluid domain at time $t$ associated with any part $\gamma^{fs}$ of the interface $\Gamma^{fs}_{i}$.

A partitioned iterative coupling algorithm is adopted to couple the fluid and structural motion equations. A predictor-corrector approach is utilized to solve the coupled framework advanced in time. The compactly-supported radius basis function (RBF) is employed to transfer the fluid loads and the structural displacements along the non-matching fluid-solid interface, which naturally ensures the energy conservation. The body-fitted spatial fluid meshes are updated based on the efficient RBF remeshing method to preserve the high mesh quality. The recently developed nonlinear interface force correction (NIFC) scheme \cite{jaiman2016stable} is implemented in the coupled framework to correct the fluid forces at each iterative step to avoid the numerical instability caused by the significant added mass effect. This high-fidelity fluid-structural interaction solver has been applied to the studies on flexible flapping wings \cite{li2018novel} and fluid-membrane interaction \cite{li2020flow}.

\section{Problem description} \label{sec:section3}
\begin{figure*}
	\centering
	\subfloat[][]{\includegraphics[width=0.49\textwidth]{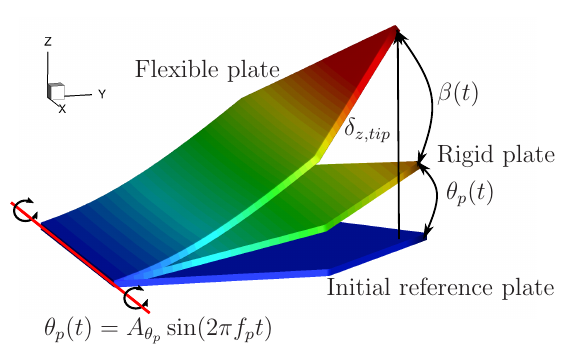}\label{fig:computational domaina}}
	\subfloat[][]{\includegraphics[width=0.49\textwidth]{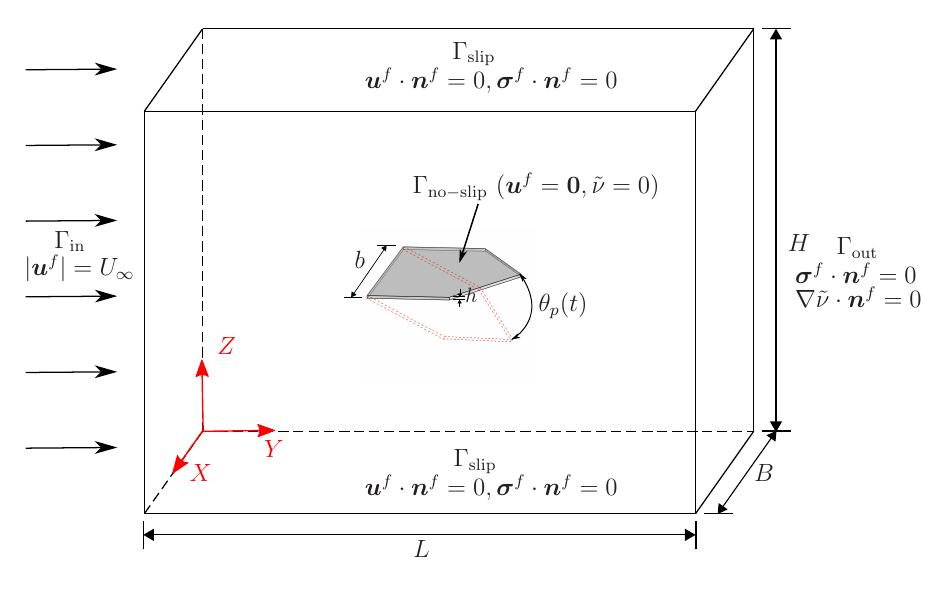}\label{fig:computational domainb}}
	\caption{\label{fig:computational domain} Problem set-up for a pitching plate: (a) illustrations of prescribed pitching motion at the leading edge $\theta_p(t)$, the deflected bending angle $\beta(t)$ and the wing tip transverse displacement $\delta_{z,tip}$ at the middle point of the trailing edge and (b) three-dimensional computational setup and boundary condition for the uniform flow past a pitching plate. The plates in (a) are colored by the transverse displacement $\delta_{z}$.}
\end{figure*}

In the current study, we consider a series of flexible plates with varying trailing edge shapes and flexibility to investigate their effects on the propulsive performance. The identical plate sizes are adopted as those in the water tunnel experiment done by Van Buren \cite{VanBuren2017}. The mean chord of the plate is $c = 0.1$ m and the width is set to $b = 0.1$ m, resulting in a plate area of $S=bc=0.01$ m$^2$. The thickness of this thin plate is $h = 2.54 \times 10^{-3}$ m. The thin plate is placed in the unsteady fluid medium with a uniform oncoming flow. As illustrated in Fig.~\ref{fig:computational domain} \subref{fig:computational domaina}, the plate is clamped at the LE to restrict the displacement of the LE at any direction, but allow the relative rotation around the $X$-axis with a prescribed pitching angle of $\theta_p(t)$, which is defined as follows
\begin{equation}
\theta_p(t) = A_{\theta_p} \sin(2 \pi f_p t)
\end{equation}
where $A_{\theta_p}$ represents the amplitude of the pitching angle and $f_p$ is the pitching frequency. Due to flexibility effect, the flexible plate deforms during the prescribed pitching motion, which leads to a deflected bending angle $\beta(t)$ with respect to its rigid counterpart
\begin{equation}
\beta(t) = A_{\beta} \sin(2 \pi f_p t - \gamma)
\end{equation}
where $A_{\beta}$ is the amplitude of the deflected bending angle and $\gamma$ represents the phase lag between the pitching angle and the deflected bending angle. Thus, the effective pitching angle is defined as $\beta_{eff}=\theta_p + \beta$, which is measured as the angle between the initial reference plate and the trailing edge of the flexible plate shown in Fig.~\ref{fig:computational domain} \subref{fig:computational domaina}.

The complex dynamics of the pitching plate is mainly governed by four key non-dimensional fluid-structure interaction parameters, namely Reynolds number $Re$, mass ratio $m^*$, bending stiffness $K_B$ and Strouhal number $St$ \cite{connell2007flapping}, which are defined as
\begin{eqnarray}
&Re=\frac{\rho^f U_{\infty } c}{\mu^f},\quad \quad m^*=\frac{\rho^s h}{\rho^f c} \nonumber\\ 
&K_B=\frac{B}{\rho^f U_{\infty}^2 c^3}, \quad \quad St = \frac{2 f_p c \sin(A_{\theta_p})}{U_{\infty}}
\end{eqnarray}
where $\rho^f$ represents the fluid density and $U_{\infty}$ denotes the oncoming flow velocity. $\mu^f$ is the dynamic viscosity of the fluid and $\rho^s$ is the plate density. The flexural rigidity $B=\frac{E h^3}{12 (1-(\nu^s)^2)}$ characterizes the flexibility of the plate, where $E$ and $\nu^s$ represent the Young's modulus and the Poisson's ratio, respectively. 

To quantitatively evaluate the propulsive performance and the dynamics of the pitching plate, we calculate the thrust coefficient $C_T$, the input power coefficient $C_{power}$ and the propulsive efficiency $\eta$ from the numerical simulations, which are given as
\begin{eqnarray}
& C_T = \frac{T}{\frac{1}{2} \rho^f U_{\infty}^2 S}=-\frac{1}{\frac{1}{2} \rho^f U_{\infty}^2 S} \int_{\Gamma} (\boldsymbol{\bar{\sigma}}^f \cdot \bm{n}) \cdot \bm{n}_y {\rm{d \Gamma}}, \quad  C_{power}=\frac{P_{input}}{\frac{1}{2} \rho^f U_{\infty}^3 S} \nonumber \\
&C_L = \frac{1}{\frac{1}{2} \rho^f U_{\infty}^2 S} \int_{\Gamma} (\boldsymbol{\bar{\sigma}}^f \cdot \bm{n}) \cdot \bm{n}_z {\rm{d \Gamma}}, \quad  \eta = \frac{\overline{C}_T}{\overline{C}_{power}}
\label{eq:force_coefficient}
\end{eqnarray}
where $T$ is the thrust force of the pitching plate. $\bm{n}_y$ and $\bm{n}_z$ represent the Cartesian components of the unit outward normal $\bm{n}$ to the plate surface $\Gamma$. $P_{input}=\int_{\Gamma^{fs}} \bm{F}^f \cdot \frac{ \partial \bm{d}^s}{\partial t} {\rm{d}} \Gamma$ represents the instantaneous input power of the pitching plate. $\bm{F}^f$ is the driving force acting on the surrounding fluid by the pitching plate along the interface $\Gamma^{fs}$.

A schematic of the three-dimensional computational domain constructed for the flexible pitching plate is shown in Fig.~\ref{fig:computational domain} \subref{fig:computational domainb}. The midpoint of the leading edge (LE) of the plate is located at the origin of the computational domain. The length $L$, the width $B$ and the height $H$ of the computational domain are all set to $40c$. The unsteady fluid flows into the computational domain through the inlet boundary $\Gamma_{\rm{in}}$ with a uniform velocity $|\bm{u}^f|=U_{\infty}$. A traction-free boundary condition is considered at the outlet boundary $\Gamma_{\rm{out}}$. We apply the slip-wall boundary conditions on the four sides of the computational domain and set the no-slip boundary condition for all plate surfaces. The plate filled with gray color and surrounded by the black color edge in Fig.~\ref{fig:computational domain} \subref{fig:computational domainb} represents the neutral position of the pitching plate. The rigid plate indicated by the red dash line denotes the instantaneous position with a pitching angle of $\theta_p(t)$. Before we proceed to examine the effect of trailing shape and flexibility on the propulsive performance of pitching plates, a mesh convergence study is conducted in Appendix A for a rigid plate with a trailing edge angle of $\Phi=45^\circ$ to ensure sufficient mesh resolutions for the numerical simulations. We further validate the coupled fluid-structure solver for the flexible plates with varying Strouhal numbers, which are compared against the available thrust and efficiency statistics at $Re=10000$ obtained from the experiments \cite{VanBuren2017}.

To study the impact of trailing edge shape and flexibility, we select 13 groups of bending stiffness over a wide range of parameter space of $K_B \in [9.86, 1.53 \times 10^5]$ for a concave plate with $\Phi$=$45^\circ$, a rectangular plate with $\Phi$=$90^\circ$ and a convex plate with $\Phi$=$135^\circ$. To isolate the effect of other physical parameters, we examine the propulsive performance for a moderate Reynolds number of $Re$=1000 at a fixed Strouhal number of $St$=0.3, which falls into the parameter range of fish swimming with high propulsive efficiency \cite{triantafyllou1995efficient}.

\section{Results and discussion} \label{sec:section4}
In this study, we explore the underlying mechanism of how flexibility affects the thrust generation and the propulsive performance of pitching flexible plates with varying trailing edge shapes. The flapping dynamics and the flow features associated with the thrust generation are investigated in detail. With the aid of the SP-DMD method, the correlated vortical structures and structural motions are identified together from the complex spatial-temporal coupled physical fields. The relationship between the unsteady momentum transfer and the thrust generation is examined by the momentum-based thrust evaluation approach. An analytical added mass model is employed to evaluate the generated thrust due to the added mass effect. The effects of trailing edge shape and flexibility on the thrust generation and the drag-thrust transition are studied in detail.

\subsection{Thrust production and propulsive efficiency} \label{sec:section4.1}
\begin{figure*}
	\subfloat[][]{\includegraphics[width=0.49\textwidth]{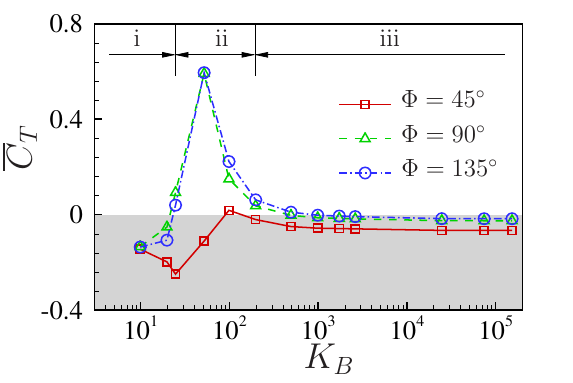}\label{fig:thrust_efficiencya}}
   \subfloat[][]{\includegraphics[width=0.49\textwidth]{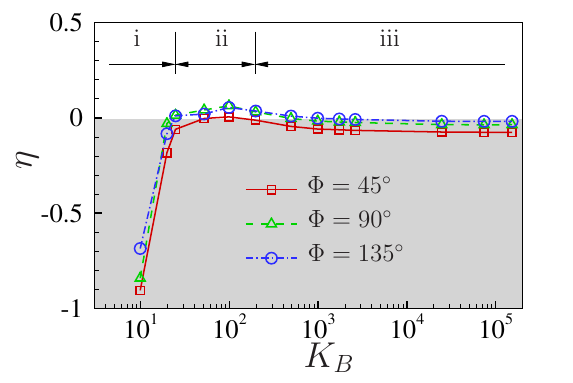}\label{fig:thrust_efficiencyb}}
	\caption{\label{fig:thrust_efficiency} (a) Mean net thrust coefficient $\overline{C}_T$ and (b) propulsive efficiency $\eta$ as a function of bending stiffness $K_B$ for pitching plates with varying trailing edge angles $\Phi$=$45^\circ$, $90^\circ$ and $135^\circ$ at $Re$ = 1000 and $St$ = 0.3.}
\end{figure*}

The mean net thrust coefficient $\overline{C}_T$ and the propulsive efficiency $\eta$ produced by the pitching flexible plates with three representative trailing edge angles $\Phi$ as a function of bending stiffness $K_B$ are shown in Fig.~\ref{fig:thrust_efficiency}. Three classified flapping motion regimes, namely (i) low bending stiffness $K_B^{low}$, (ii) moderate bending stiffness $K_B^{moderate}$ near resonance and (iii) high bending stiffness $K_B^{high}$, are added in Fig.~\ref{fig:thrust_efficiency} to study the characteristics of the produced thrust and propulsive efficiency. The classification of the flapping motion regimes and their relationship with varying bending stiffnesses are discussed in Section~\ref{sec:section4.2} in detail. It can be seen from Fig.~\ref{fig:thrust_efficiency} \subref{fig:thrust_efficiencya} that the mean net thrust coefficient grows up rapidly to achieve its peak value at moderate $K_B$ and then decreases to a plain when the plates become stiffer. In Fig.~\ref{fig:thrust_efficiency} \subref{fig:thrust_efficiencyb}, the propulsive efficiency changes from negative values to the optimal values and then reduces gradually to almost constant values as $K_B$ increases. The rectangular plate can produce the largest thrust within the low bending stiffness regime. The thrust generated by the convex plate is the largest among the three types of plates at moderate and high $K_B$ values. Regarding propulsive efficiency, the convex plate is the most efficient propulsion system within the low and high bending stiffness regimes. The rectangular plate achieves the optimal efficiency gain at moderate $K_B$ values. It is worth noting that the concave plate has the poorest ability in thrust generation and propulsive efficiency gain within the studied $K_B$ range. The bending stiffness value corresponding to the overall largest thrust is smaller than that related to the optimal efficiency. The transition between thrust and drag is observed in Fig.~\ref{fig:thrust_efficiency} \subref{fig:thrust_efficiencya} when the flexible plate with a moderate $K_B$ value becomes more flexible or more rigid. The convex plate shows the largest transition region with positive thrust within the studied $K_B$ range. The mechanism of the thrust generation and the drag-thrust transition will be discussed in the next sections.

\subsection{Flapping dynamics} \label{sec:section4.2}
The generated thrust and the propulsive efficiency of the flexible plates with varying trailing edge angles are closely associated with flexibility. In this section, three distinctive flapping motion regimes are firstly classified based on the flapping dynamics. We further discuss how flexibility affects the propulsive performance by examining the interplay between the natural frequency of the flexible plate and the pitching frequency. 
\subsubsection{Classification of flapping motion regime} \label{sec:section4.2.1}
\begin{figure*}
	\subfloat[][]{\includegraphics[width=0.3\textwidth]{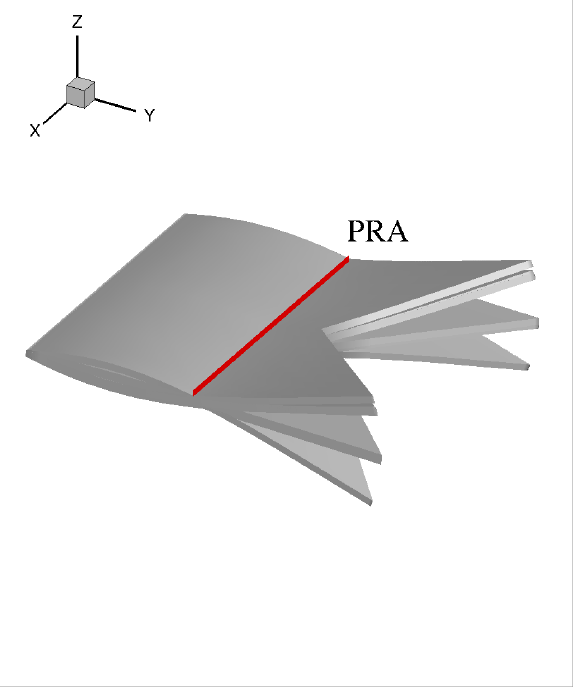}\label{fig:plate_modea}}
	\subfloat[][]{\includegraphics[width=0.3\textwidth]{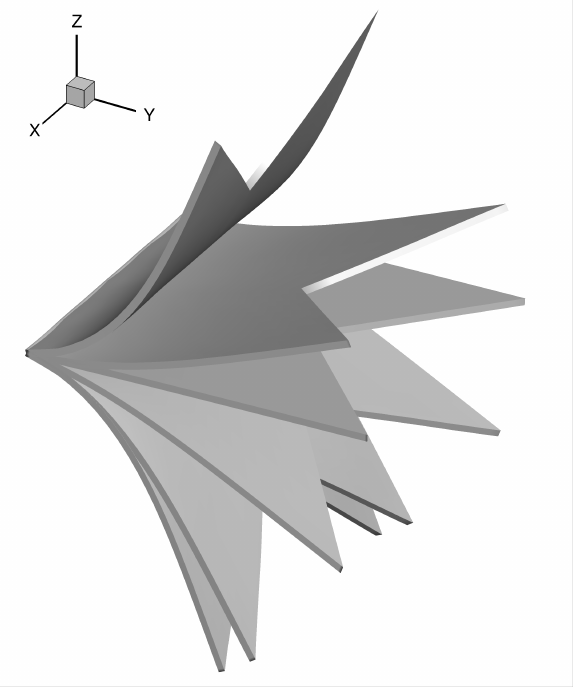}\label{fig:plate_modeb}}
	\subfloat[][]{\includegraphics[width=0.3\textwidth]{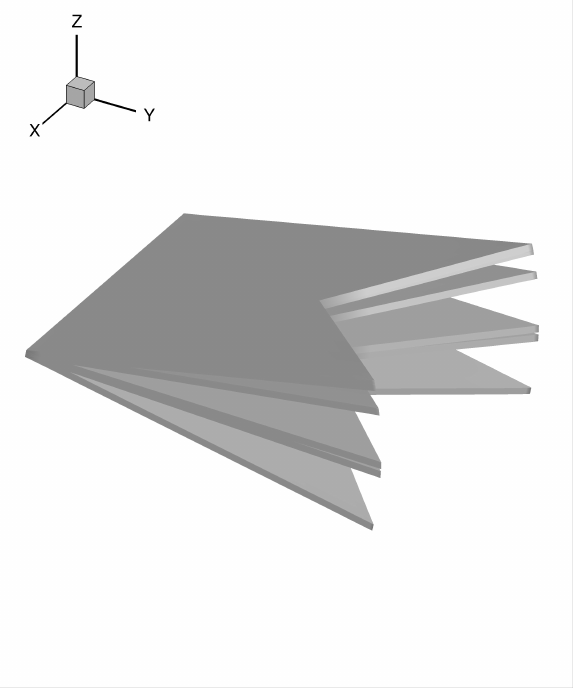}\label{fig:plate_modec}}
	\caption{\label{fig:plate_mode} Instantaneous deformation of pitching plate with trailing edge angle of $\Phi=45^\circ$ at $K_B$= (a) 9.86, (b) 51.8 and (c) 73998.}
\end{figure*}

The pitching flexible plates with varying trailing edge shapes exhibit different types of flapping motions as a function of flexibility. Based on the variation of the propulsive performance and the related dynamic responses of the flexible plates, three distinctive regimes are classified from the coupled system: (i) low bending stiffness $K_B^{low}$, (ii) moderate bending stiffness $K_B^{moderate}$ near resonance and (iii) high bending stiffness $K_B^{high}$. From the analysis of the propulsive performance as a function of flexibility in Section~\ref{sec:section4.1}, it can be seen that the propulsive performance is strongly affected by flexibility. The generated thrust and the propulsive efficiency can be enhanced at moderate bending stiffness values. Regardless of the trailing edge shape, the flexible plate exhibits similar flapping motion modes within the same regime. Here, we plot the instantaneous deformations of the pitching plate with a trailing edge angle of $\Phi=45^\circ$ in Fig.~\ref{fig:plate_mode} briefly for the regime classification purpose. It can be seen from Fig.~\ref{fig:plate_mode} \subref{fig:plate_modea} that the flexible plate exhibits a chord-wise second flexural mode during pitching motion for low $K_B$ values. The deformation of the trailing edge shows an opposite direction to the pitching motion at the leading edge. As a result, the whole motion of the pitching plate can be divided into two parts by an almost fixed passive rotation axis (PRA) along the span-wise direction. The front portion follows the pitching motion applied along the leading edge, and the rear portion deforms passively under the action of inertial, elastic and aerodynamic forces. As $K_B$ further increases to moderate values, the dominant structural mode changes from the chord-wise second flexural mode to the chord-wise first flexural mode as shown in Fig.~\ref{fig:plate_mode} \subref{fig:plate_modeb}. Moreover, the deformation amplitude increases dramatically. In Fig.~\ref{fig:plate_mode} \subref{fig:plate_modec}, the whole plate follows the prescribed pitching motion within the high bending stiffness regime. The passive deformation is significantly suppressed by the elastic forces at higher $K_B$ values.

\subsubsection{Structural resonance and role of flexibility} \label{sec:section4.2.2}
Based on the classification of the flapping motion regime, the passive deformation of the pitching plate is dramatically enhanced within a certain range of bending stiffness values. In this range, the natural frequency of the flexible plate immersed in the unsteady fluid approaches the fixed pitching frequency. A natural question to ask is whether the ratio between these two frequencies plays an important role in the dynamic characteristics. Here, we first calculate the frequency ratio, and then examine the connection between the flapping dynamics and the frequency ratio for flexible plates with different bending stiffnesses. As suggested in Van Eysden et al. \cite{van2006resonant}, the relationship between the natural frequencies of the chord-wise first flexural mode in the fluid $f_1^{f}$ and that in vacuum $f_1^{vac}$ can be considered
%calculation of resonance frequency
\begin{equation}
f_1^{f}= f_1^{vac} \left[ 1 + \frac{\pi \rho^f b}{4 \rho^s h} \Omega(\kappa) \right]^{-0.5}
\label{eq:resonance}
\end{equation}

The approximate hydrodynamic function $\Omega(\kappa)$ as a function of the coefficient $\kappa=1.8751 \frac{b}{c}$ is given as
\begin{equation}
\Omega(\kappa) = \frac{1+0.74273 \kappa + 0.14862 \kappa^2}{1+0.74273 \kappa +0.35004 \kappa^2+0.058364 \kappa^4}
\end{equation}

In Eq.~(\ref{eq:resonance}), the natural frequency of the plate in vacuum $f_1^{vac}$ is directly calculated from the structural motion equations shown in Eq.~(\ref{eq:eqMB1}) via the modal analysis. The frequency ratio $f^*$ between the first natural frequency of the plate in the fluid and the fixed pitching frequency is defined as 
\begin{equation}
f^*=\frac{f_1^{f}}{f_p}=\frac{f_1^{vac}}{f_p} \left[ 1 + \frac{\pi \rho^f b}{4 \rho^s h} \Omega(\kappa) \right]^{-0.5}
\label{eq:frequency_ratio}
\end{equation}

\begin{figure*}
	\subfloat[][]{\includegraphics[width=0.49\textwidth]{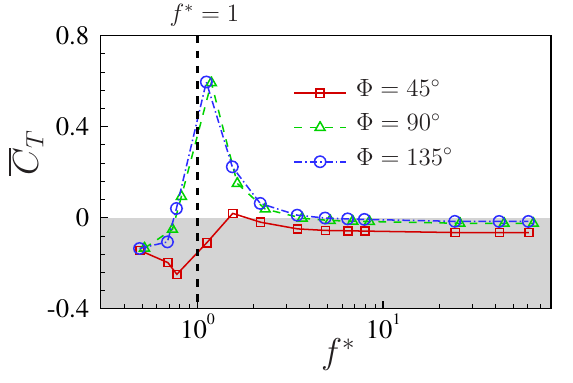}\label{fig:thrust}}
	\subfloat[][]{\includegraphics[width=0.49\textwidth]{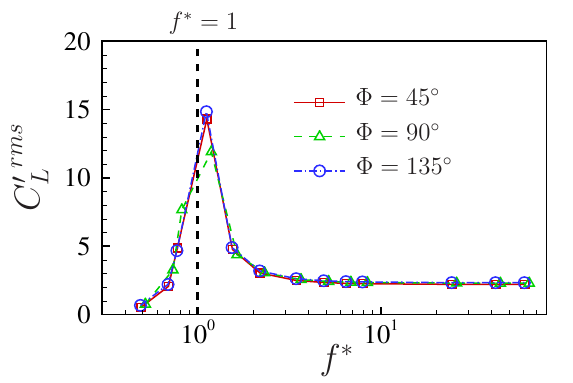}\label{fig:CLrms}}
	\\
	\subfloat[][]{\includegraphics[width=0.49\textwidth]{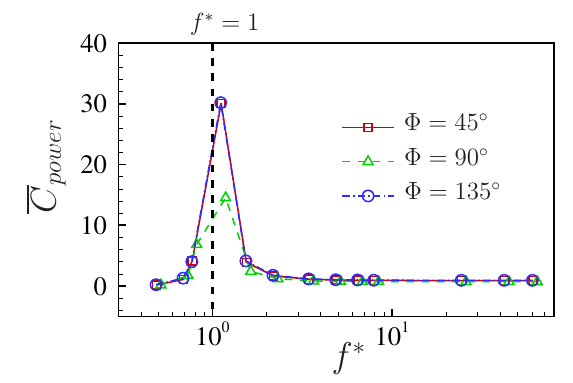}\label{fig:cpower}}
	\subfloat[][]{\includegraphics[width=0.49\textwidth]{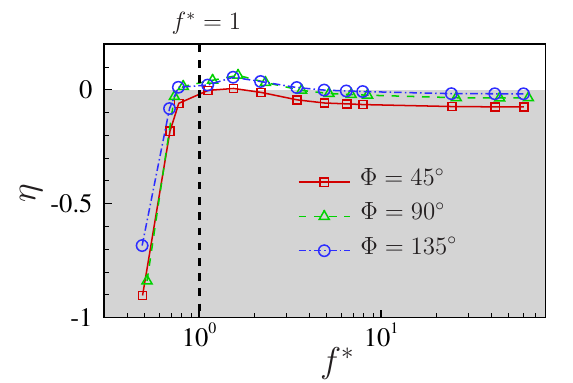}\label{fig:efficiency}}
	\caption{\label{fig:thrust_efficiency_norm} (a) Mean thrust coefficient $\overline{C}_T$, (b) r.m.s of the lift coefficient fluctuation ${C_L^{\prime}}^{rms}$, (c) mean input power coefficient $\overline{C}_{power}$ and (d) propulsive efficiency $\eta$ as a function of the frequency ratio $f^*$ for pitching plate with varying trailing edge angles $\Phi$=$45^\circ$, $90^\circ$ and $135^\circ$ at $Re$ = 1000 and $St$ = 0.3.}
\end{figure*}

As a function of the nondimensional frequency ratio $f^*$ , the mean net thrust coefficient, the root-mean-squared value of the lift coefficient fluctuation, the mean input power coefficient and the propulsive efficiency are summarized in Fig.~\ref{fig:thrust_efficiency_norm} for the concave, rectangular and convex plates. The vertical black dash line is located at $f^*=1$, which indicates the structural resonance between the first natural frequency of the plate and the actuated frequency. It can be seen from Fig.~\ref{fig:thrust_efficiency_norm} \subref{fig:thrust} that all three types of flexible plates with different $\Phi$ exhibit the global maximum mean net thrust forces when the frequency ratio gets close to $f^*=1$. The amplitude of the lift force is greatly enhanced within the near resonance regime. As observed in Fig.~\ref{fig:plate_mode}, a large amplitude of passive deformation relative to the active pitching motion is excited by the resonance effect. According to the formulation of the input power, the flexible plate requires more input power to maintain the prescribed pitching motion due to the increased fluid loads acting on the plate surface and the improved velocity of the plate motion under the resonance condition. Although the flexible plate can produce the largest thrust forces at $K_B=51.8$ (close to $f^*=1$), the optimal propulsive efficiency is achieved at a higher $K_B$ of 98.66. This is mainly because the pitching flexible plate with $K_B=51.8$ can only improve the thrust forces at most four times, but it needs at least six times the input power than the plate with $K_B=98.66$. As a result, the propulsive efficiency for the flexible plate with the largest vibration amplitude under the resonance condition is reduced and its optimal value is achieved for the flexible plate with moderate passive deformations.

\begin{figure*}
	\subfloat[][]{\includegraphics[width=0.49\textwidth]{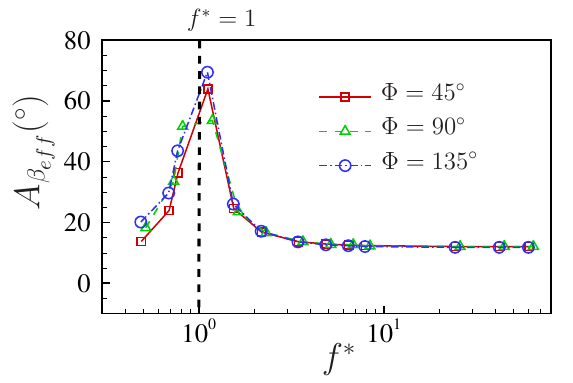}\label{fig:structure_dynamica}}
	\subfloat[][]{\includegraphics[width=0.49\textwidth]{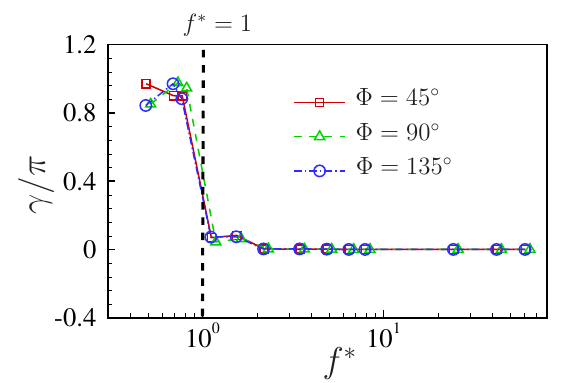}\label{fig:structure_dynamicb}}
	\\
	\subfloat[][]{\includegraphics[width=0.49\textwidth]{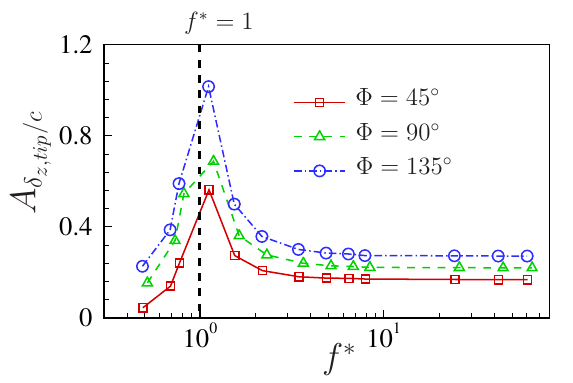}\label{fig:structure_dynamicc}}
	\caption{\label{fig:structure_dynamic} (a) Amplitude of effective pitching angle $A_{\beta_{eff}}$,  (b) phase lag $\gamma$ and (c) amplitude of non-dimensional transverse displacement $A_{\delta_{z,tip}/c}$ as a function of the frequency ratio $f^*$ for pitching plate with varying trailing edge angles $\Phi$=$45^\circ$, $90^\circ$ and $135^\circ$ at the middle point of the TE at $Re$ = 1000 and $St$ = 0.3.}
\end{figure*}

We further examine the role of flexibility in the dynamic responses of the pitching flexible plates with varying trailing edge angles. In Fig.~\ref{fig:structure_dynamic} \subref{fig:structure_dynamica}, the amplitude of the effective pitching angle grows up rapidly and reaches its peak when the frequency ratio $f^*$ approaches 1. As $f^*$ further increases, the amplitude decreases sharply and finally maintains a value close to the pitching angle amplitude of $12^\circ$ applied at the LE for high $K_B$ cases. The variation of the phase lag $\gamma$ between the deflected bending angle and the prescribed pitching angle as a function of the frequency ratio is presented in Fig.~\ref{fig:structure_dynamic} \subref{fig:structure_dynamicb}. The phase lag is related to the relative direction of the motion at the TE and the required input power. The flexible plate shows a phase lag close to $\pi$ within the low bending stiffness regime. This is caused by the excited chord-wise second flexural mode at lower $K_B$ values. When the resonance between the first natural frequency of the flexible plate and the pitching frequency is exited at moderate $K_B$ values, the phase lag reduces rapidly. Synchronization between the pitching angles at the LE and TE is maintained by reducing the phase lag to zero at larger $K_B$ values. The amplitude of the transverse displacement at the TE exhibits a similar trend to the effective pitching angle as a function of $f^*$. A large displacement amplitude is excited near $f^*=1$ by the resonance effect. It can be concluded that the enhanced thrust forces and the increased input power are strongly associated with the passive deformation of the flexible plate and its relative phase lag to the prescribed pitching motion at the LE. Since the passive deformation can mutually affect the flow field, the generated fluid forces are directly related to the flow features and wake structures around the plate. In the next section, we further analyze the relationship between the flow characteristics and the propulsive performance of the flexible plates with varying bending stiffness and trailing edge shapes.

\subsection{Flow field and wake structures}
In this section, the time-averaged and the instantaneous flow features are analyzed to understand the thrust generation mechanism. To link the flow features with the motion of the flexible plate, the correlated wake and structural modes are extracted from the simulation data with the aid of the SP-DMD method.

\subsubsection{Time-averaged flow features}
\begin{figure*}
	\subfloat[][]{\includegraphics[width=0.25\textwidth]{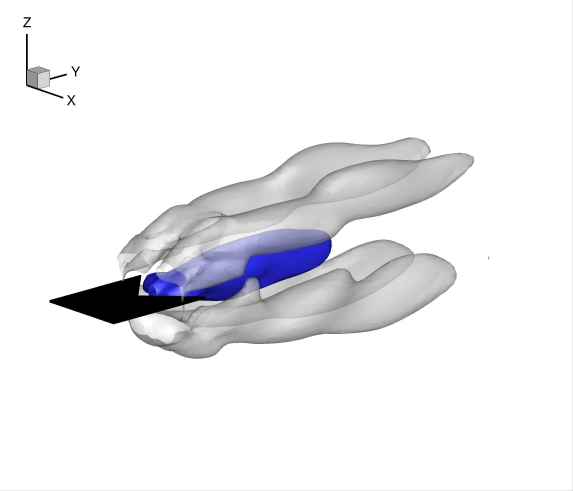}}
	\subfloat[][]{\includegraphics[width=0.25\textwidth]{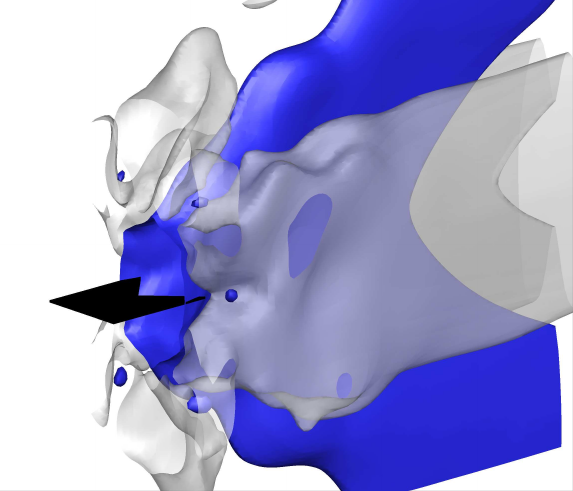}}
	\subfloat[][]{\includegraphics[width=0.25\textwidth]{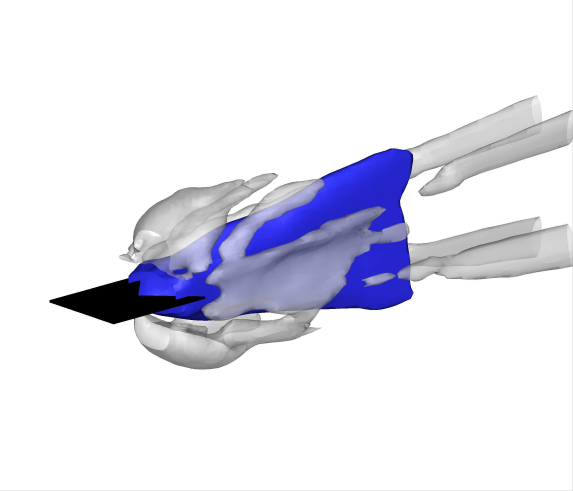}}
	\subfloat[][]{\includegraphics[width=0.25\textwidth]{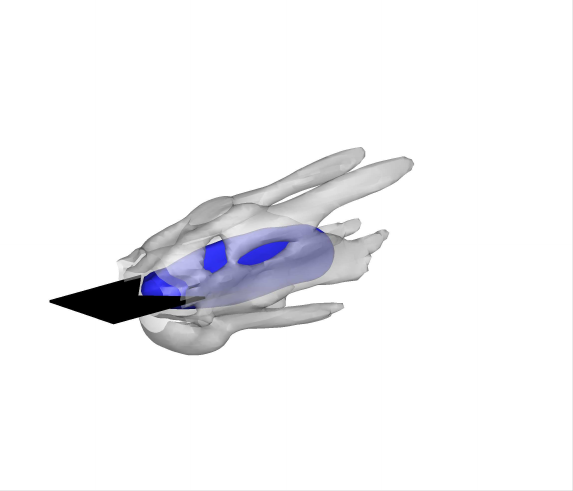}}
	\\
	\subfloat[][]{\includegraphics[width=0.25\textwidth]{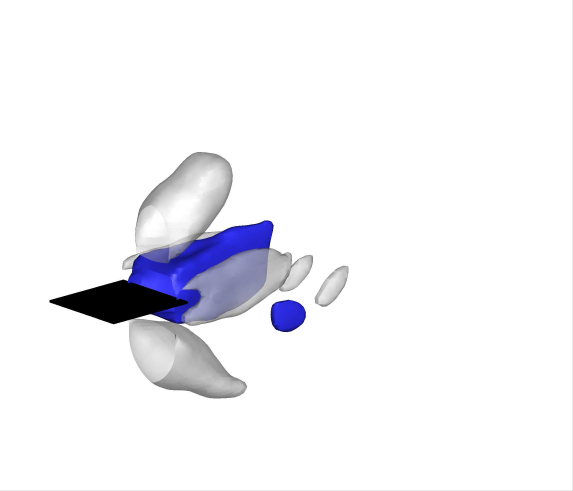}}
	\subfloat[][]{\includegraphics[width=0.25\textwidth]{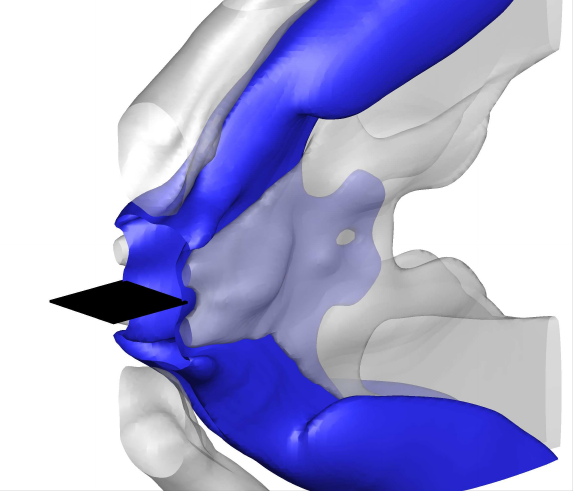}}
	\subfloat[][]{\includegraphics[width=0.25\textwidth]{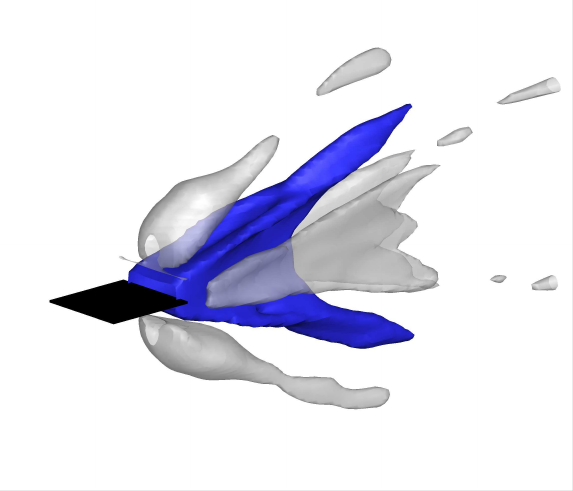}}
	\subfloat[][]{\includegraphics[width=0.25\textwidth]{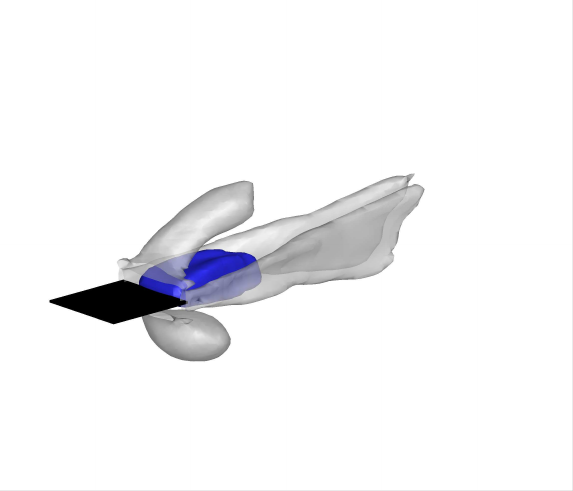}}
	\\
	\subfloat[][]{\includegraphics[width=0.25\textwidth]{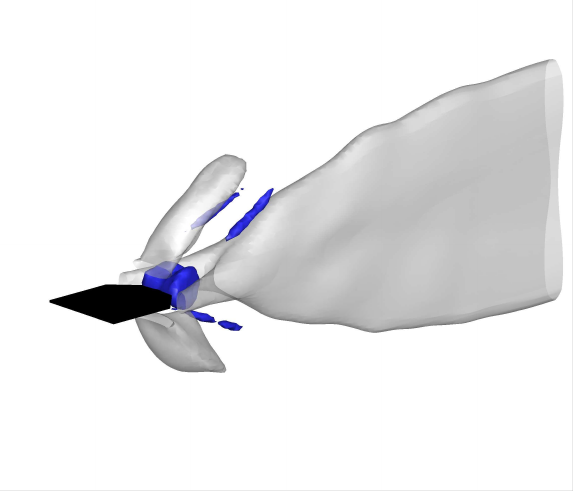}}
	\subfloat[][]{\includegraphics[width=0.25\textwidth]{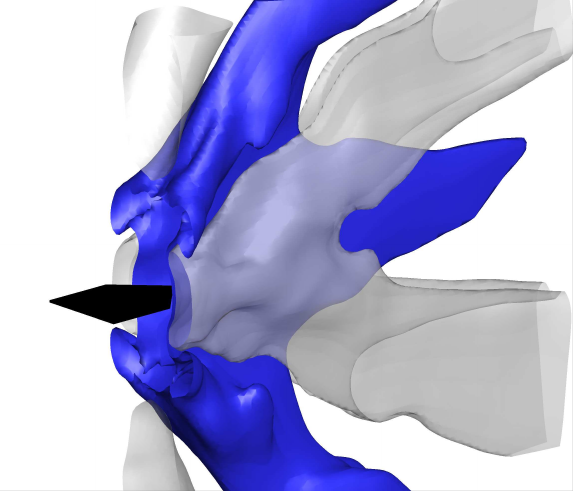}}
	\subfloat[][]{\includegraphics[width=0.25\textwidth]{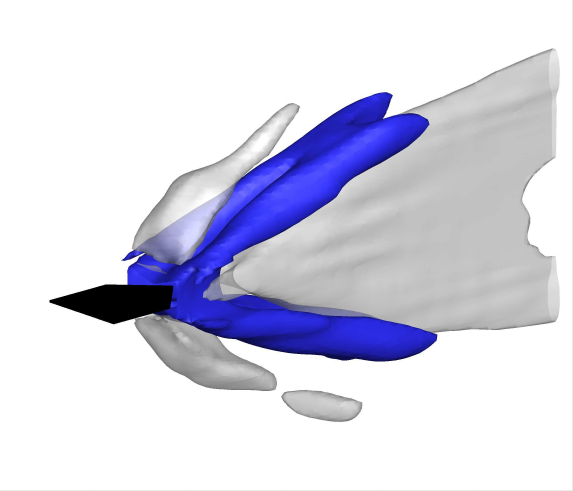}}
	\subfloat[][]{\includegraphics[width=0.25\textwidth]{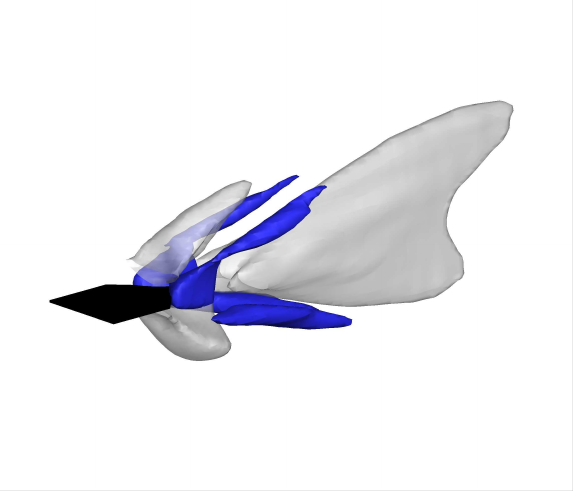}}
	\caption{\label{fig:iso_vvel_rotate} Isosurfaces of time-averaged streamwise velocity $\bar{v} / U_{\infty}$ thresholded at 0.95 (gray) and 1.15 (blue) of a pitching plate at $K_B$= (a,e,i) 19.73, (b,f,j) 51.8, (c,g,k) 197.3 and (d,h,l) 73998 with trailing edge angle of $\Phi$= (a,b,c,d) $45^\circ$, (e,f,g,h) $90^\circ$ and (i,j,k,l) $135^\circ$.}
\end{figure*}

To explore the effect of trailing edge shape and flexibility on the unsteady momentum imparted by the plate into the wake, the iso-surfaces of the time-averaged streamwise velocity $\bar{v} / U_{\infty}$ are plotted for three types of flexible plates with four representative bending stiffness values in Fig.~\ref{fig:iso_vvel_rotate}. For the extraction of time-averaged streamwise velocity, the numerical results on the body-fitted moving mesh are projected to a reference stationary mesh, and then averaged over five pitching cycles. In Fig.~\ref{fig:iso_vvel_rotate}, the iso-surfaces in the gray color with a threshold of $\bar{v} / U_{\infty}=0.95$ indicate the deceleration flow region. The iso-surfaces in the blue color thresholded at 1.15 represent the high-velocity jet produced by the pitching plate. The plate in the black color is plotted at the neutral position. The iso-surface is cut within $y/c \in [1,5]$ to concentrate on the velocity distribution in the near wake behind the plate.

In Fig.~\ref{fig:iso_vvel_rotate}(a,e,i), the flexible plate produces small regions of high-velocity jet flows in the wake within the low bending stiffness regime. It can be inferred from the dynamic response shown in Fig.~\ref{fig:plate_mode} \subref{fig:plate_modea} that the very flexible plate with the chord-wise second mode is unable to produce large thrust. As $K_B$ increases to 51.8 to meet the resonance condition, the near-wake flow which contains high streamwise velocities accelerated by the pitching flexible plate expands wider in the transverse direction and farther in the streamwise direction. Consequently, the generated thrust is significantly enhanced when more energies are transferred to the fluid. The inclination and the length of the jet behind the flexible plate become smaller gradually when the coupled system enters the off-resonance states. Meanwhile, the passive deflection of the plate reduces significantly, compared to the case under the resonance condition. As a result, the net thrust decreases sharply. It can be observed from Fig.~\ref{fig:iso_vvel_rotate}(d,h,l) that the high-velocity region behind the plate is suppressed dramatically as $K_B$ further increases to 73998. The passive deformation of the plate is almost negligible and less momentum is transferred to the fluid, resulting in lower thrust generation.

The trailing edge shape also affects the topology of the high-velocity region to further govern the thrust generation. The deceleration flow changes from a quadfurcated shape to a compressed shape when the trailing edge angle increases from $45^\circ$ to $135^\circ$ at $K_B$=19.73, 197.3 and 73998, respectively. Meanwhile, the shape of the high-velocity jet shows an opposite change. The inclination of the low-velocity and high-velocity flows becomes larger when the plate changes from the concave shape to the convex shape. When the resonance is excited, the high-velocity jet presents a compressed shape without obvious bifurcation for all three types of plates but shows a larger inclination for the convex plate. As a result, the convex plate can produce the overall largest thrust within the studied $K_B$ range and the concave plate is the least efficient propulsion system.

\begin{figure*}
%	\subfloat[][]{\includegraphics[width=0.3\textwidth]{./Figures/SchematicsSlicesUavg3D.pdf}}
    \includegraphics[width=0.8\textwidth]{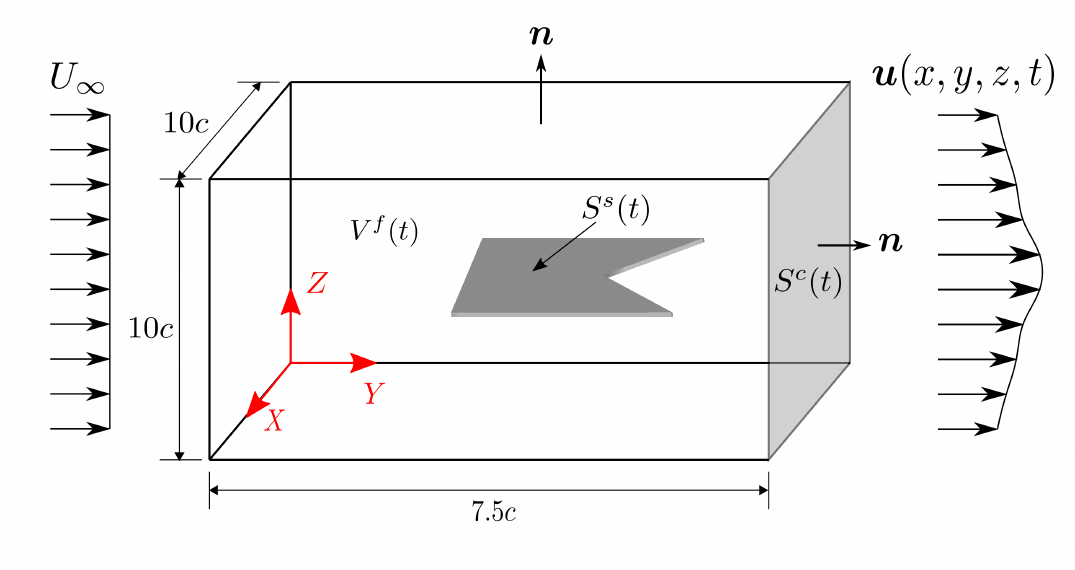}
	\caption{\label{fig:velocity_in_wake} Schematic of monitoring surface for plotting time-averaged streamwise velocity and control volume enclosed the flexible plate. The control surface in gray color is used to extract the streamwise velocity profiles.}
\end{figure*}

\begin{figure*}
	\subfloat[][]{\includegraphics[width=0.25\textwidth]{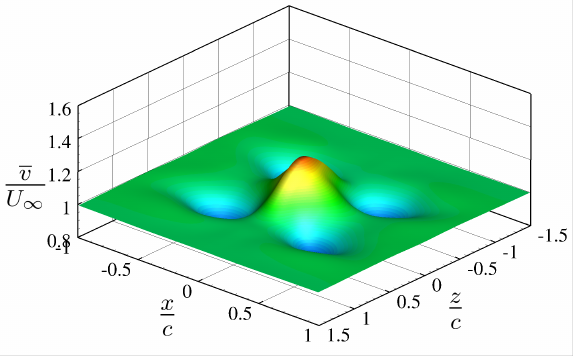}}
	\subfloat[][]{\includegraphics[width=0.25\textwidth]{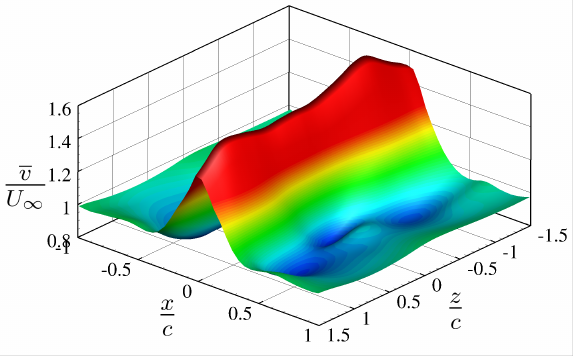}}
	\subfloat[][]{\includegraphics[width=0.25\textwidth]{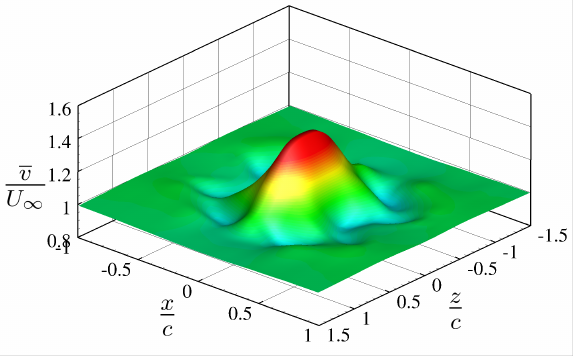}}
	\subfloat[][]{\includegraphics[width=0.25\textwidth]{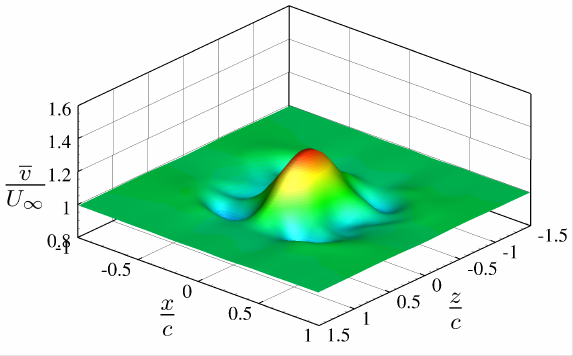}}
	\\
	\subfloat[][]{\includegraphics[width=0.25\textwidth]{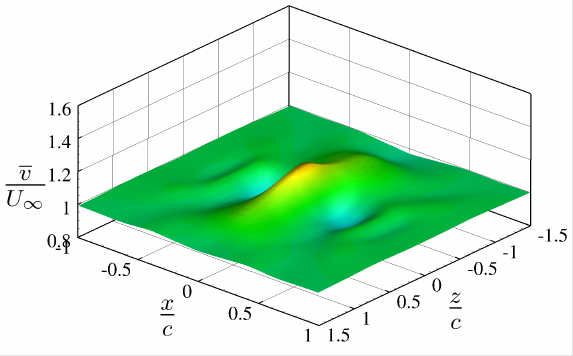}}
	\subfloat[][]{\includegraphics[width=0.25\textwidth]{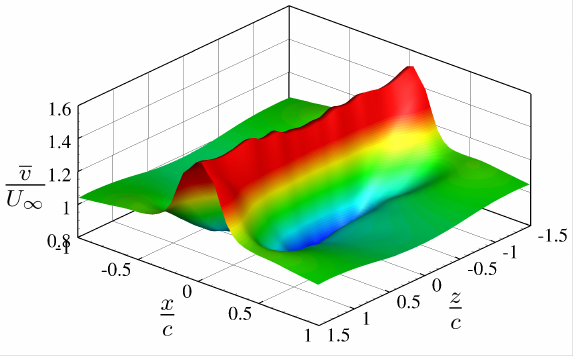}}
	\subfloat[][]{\includegraphics[width=0.25\textwidth]{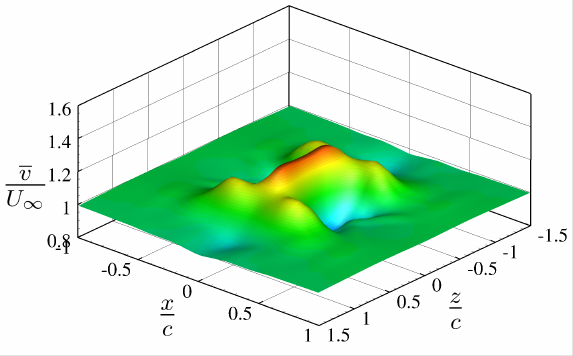}}
	\subfloat[][]{\includegraphics[width=0.25\textwidth]{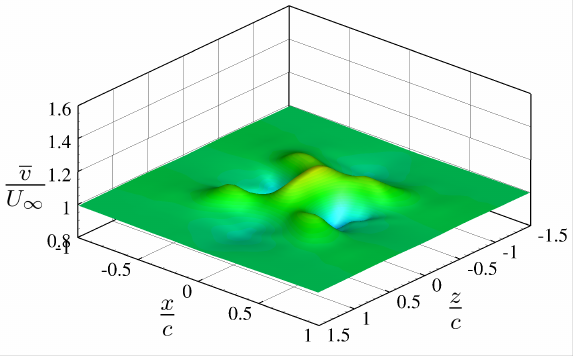}}
	\\
	\subfloat[][]{\includegraphics[width=0.25\textwidth]{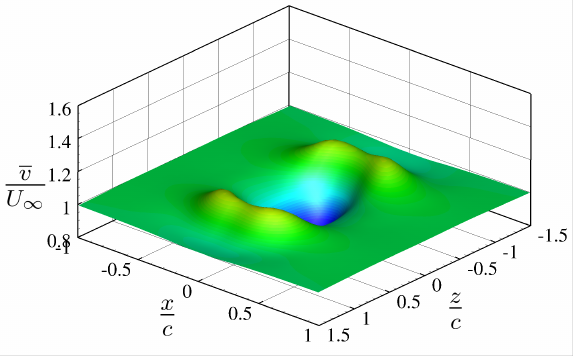}}
	\subfloat[][]{\includegraphics[width=0.25\textwidth]{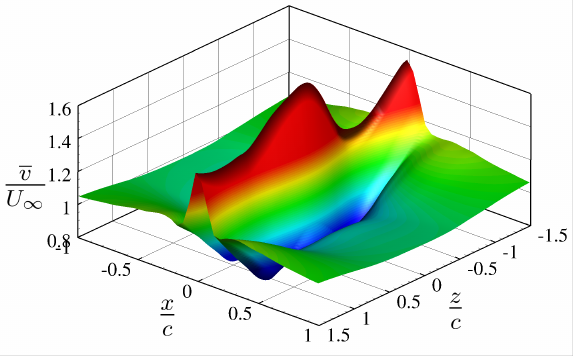}}
	\subfloat[][]{\includegraphics[width=0.25\textwidth]{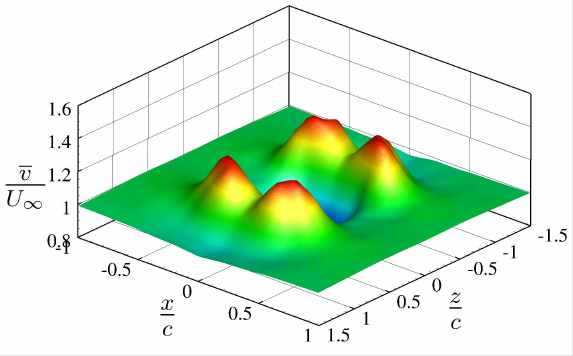}}
	\subfloat[][]{\includegraphics[width=0.25\textwidth]{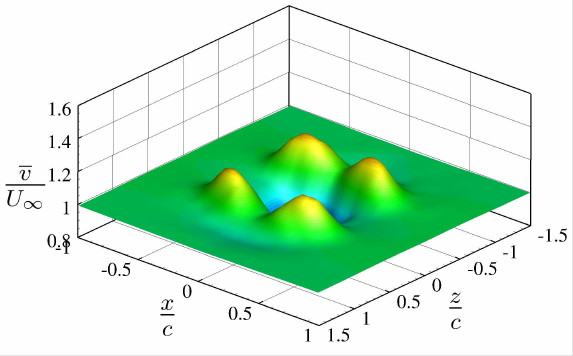}}
	\\
	\includegraphics[width=0.5\textwidth]{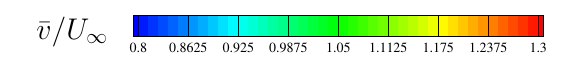}
	\caption{\label{fig:wake_profile} Time-averaged streamwise velocity $\bar{v} / U_{\infty}$ on a plane at $y/c=2.5$ normal to streamwise of a pitching plate at $K_B$= (a,e,i) 19.73, (b,f,j) 51.8, (c,g,k) 197.3 and (d,h,l) 73998 with trailing edge angle of $\Phi$= (a,b,c,d) $45^\circ$, (e,f,g,h) $90^\circ$ and (i,j,k,l) $135^\circ$.}
\end{figure*}

A schematic of a monitoring surface in the gray color for plotting the energetic jet flows is presented in Fig.~\ref{fig:velocity_in_wake}. This monitoring surface normal to the streamwise direction is located at $y/c$=2.5 behind the plate. The uniform oncoming flow $U_{\infty}$ is accelerated or decelerated by the pitching flexible plate to modulate the flow with a redistributed velocity profile $\bm{u}(x,y,z,t)$ in the wake, which is directly related to the thrust generation. We extract the time-averaged streamwise velocity distribution on the monitoring plane to quantitatively compare the thrust-generating momentum. In Fig.~\ref{fig:wake_profile}, the high-velocity region ($\overline{v}/U_{\infty}$>1) expands both in the transverse and span-wise directions and the magnitude increases significantly when the resonance condition is achieved at $K_B$=51.8. With the further increase of $K_B$, the jet region reduces and the magnitude decreases continuously. According to Newton's third law, the maximum thrust is produced under the resonance condition and the propulsive performance deteriorates when the plate becomes more rigid. Except for the propulsive system near resonance, the jet changes from a compressed shape with one peak in the center to a bifurcated shape with four peaks as the trailing edge shape alters from concave to convex. The bifurcated jet contains more imparted momentum, which can improve the thrust generation for a convex plate. 

\subsubsection{Instantaneous flow features associated with fluid loads}
\begin{figure*}
	\subfloat[][]{\includegraphics[width=0.49\textwidth]{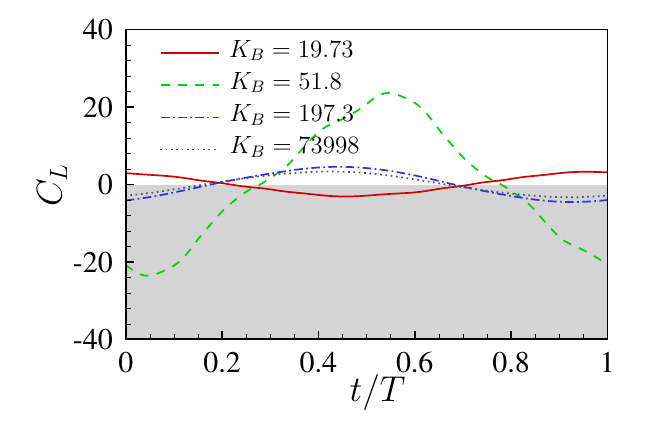}}
	\subfloat[][]{\includegraphics[width=0.49\textwidth]{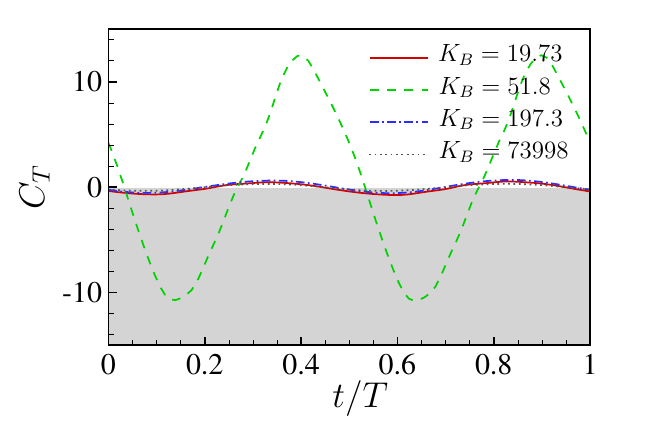}}
	\caption{\label{fig:CT} Comparison of (a) instantaneous lift coefficient and (b) instantaneous thrust coefficient of pitching plate with trailing edge angle of $\Phi=135^\circ$ within one completed pitching cycle. $t/T=0$ corresponds to the pitching upward from the neutral position.}
\end{figure*}

The time-averaged flow features reflect the connection with the time-averaged net thrust, but it lacks information about the instantaneous generated fluid loads. To fully understand the effect of trailing edge shape and flexibility on the propulsive performance, we further examine the instantaneous flow features associated with the instantaneous resulting fluid loads. It can be seen from Fig~.\ref{fig:thrust_efficiency_norm} that the force statistics show similar trends as a function of $K_B$ for flexible plates with different trailing edge shapes. Thus, the convex plates with four representative flexibility values are selected for comparison purpose to examine the role of flexibility in the temporal and spatial evolution of the flow features. Figure~\ref{fig:CT} presents the comparison of the instantaneous lift and thrust coefficients within one completed pitching cycle. The lift coefficient of the too-flexible plate at $K_B$=19.73 shows an opposite phase to the plate with higher $K_B$ values, which is affected by the chord-wise second mode. As $K_B$ increases to 51.8, the amplitudes of the lift and thrust coefficients are significantly increased due to the excited large passive deformation under the resonance condition. This type of burst aerodynamic force is helpful for the bio-inspired locomotion with high maneuverability. The amplitude of the fluid loads keeps decreasing when the plate becomes more rigid, resulting in lower propulsive performance. 

\begin{figure*}
	\subfloat[][]{\includegraphics[width=0.23\textwidth]{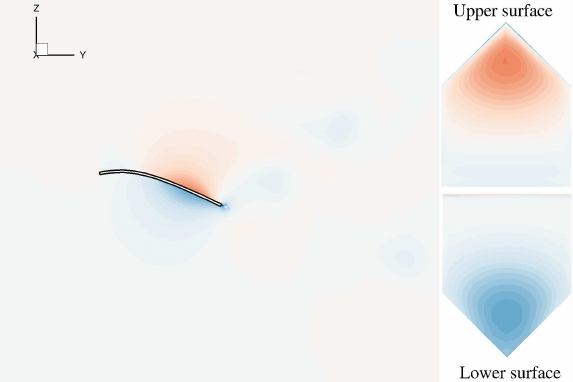}}
	\quad
	\subfloat[][]{\includegraphics[width=0.23\textwidth]{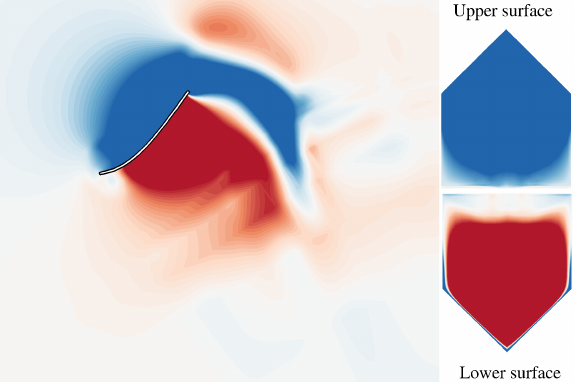}}
	\quad
	\subfloat[][]{\includegraphics[width=0.23\textwidth]{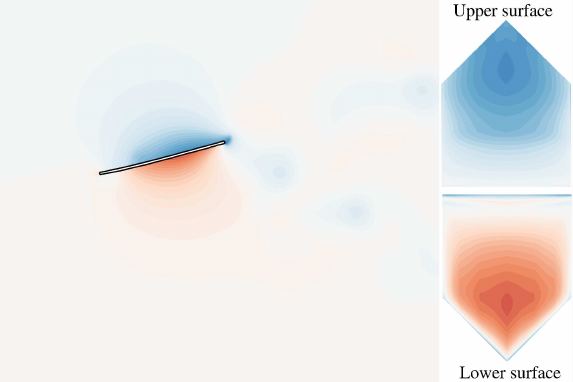}}
	\quad
	\subfloat[][]{\includegraphics[width=0.23\textwidth]{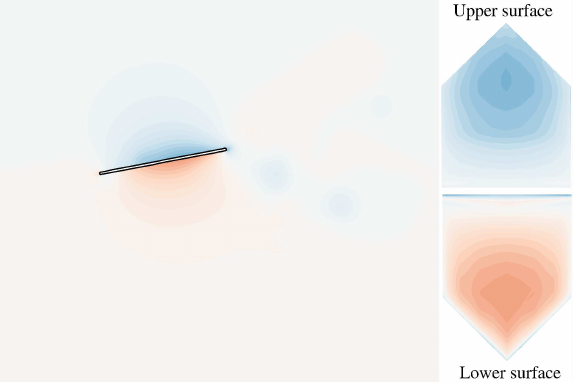}}
	\\
	\subfloat[][]{\includegraphics[width=0.23\textwidth]{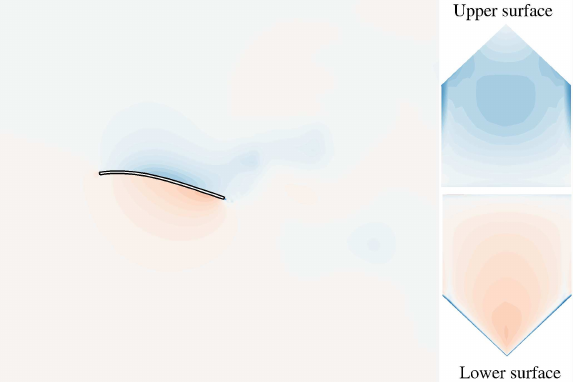}}
	\quad
	\subfloat[][]{\includegraphics[width=0.23\textwidth]{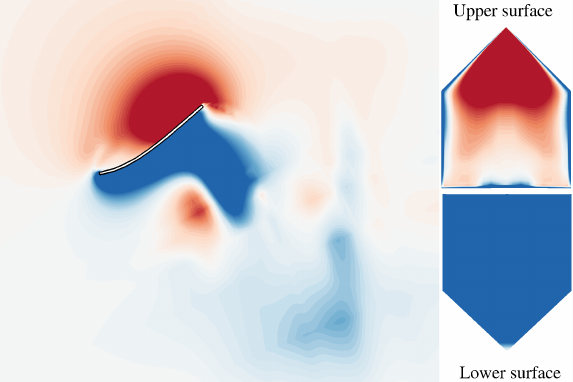}}
	\quad
	\subfloat[][]{\includegraphics[width=0.23\textwidth]{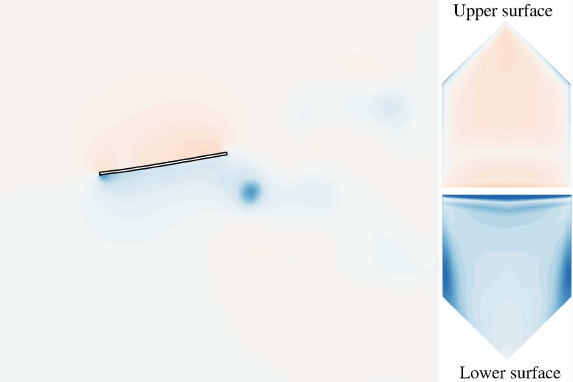}}
	\quad
	\subfloat[][]{\includegraphics[width=0.23\textwidth]{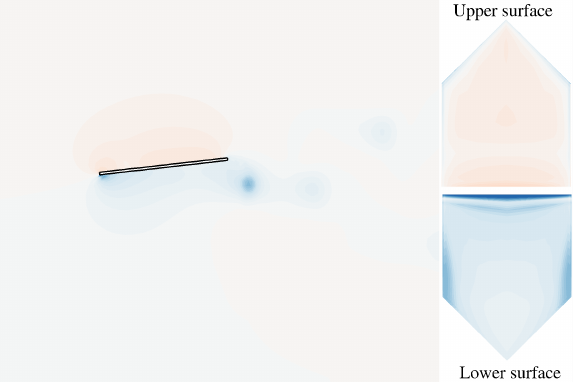}}
	\\
	\subfloat[][]{\includegraphics[width=0.23\textwidth]{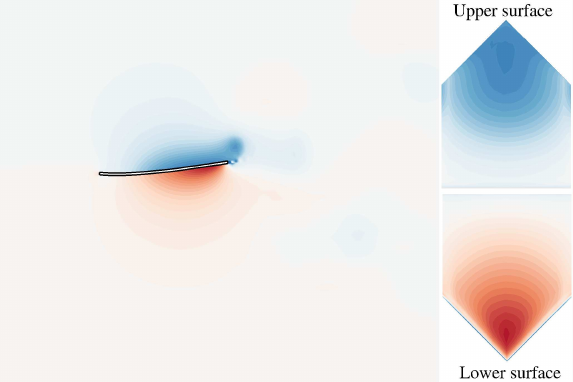}}
	\quad
	\subfloat[][]{\includegraphics[width=0.23\textwidth]{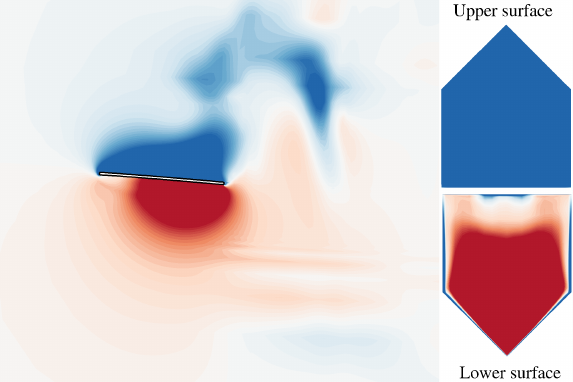}}
	\quad
	\subfloat[][]{\includegraphics[width=0.23\textwidth]{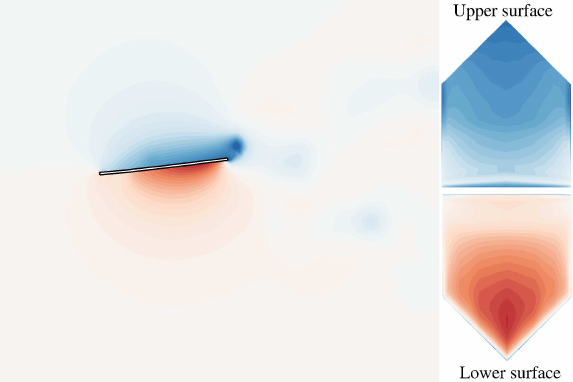}}
	\quad
	\subfloat[][]{\includegraphics[width=0.23\textwidth]{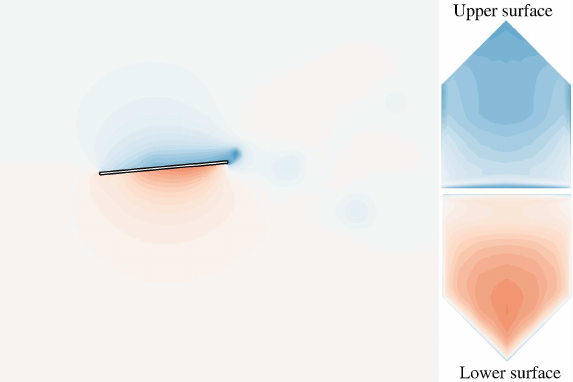}}
	\\
	\includegraphics[width=0.45\textwidth]{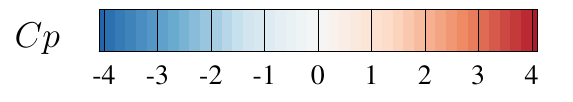}
	\caption{\label{fig:pressure} Instantaneous pressure coefficient contour on the span-wise symmetry plane and pressure coefficient distribution on the upper and lower surfaces of pitching plate with trailing edge angle of $\Phi=135^\circ$ at $K_B$= (a,e,i) 19.73, (b,f,j) 51.8, (c,g,j) 197.3 and (d,h,l) 73998 at the moment with (a,b,c,d) the largest thrust, (e,f,g,h) the lowest thrust and (i,j,k,l) the largest lift.}
\end{figure*}

To gain further insight into the fluid load generation, the instantaneous pressure coefficient distribution around the plate and on the plate surfaces at the moment with the largest thrust, the lowest thrust and the largest lift is presented in Fig.~\ref{fig:pressure}. It can be seen from Fig.~\ref{fig:pressure} (a-d) that the too-flexible plate shows opposite pressure distributions on the upper and lower surfaces compared to the plates with high $K_B$ values. This is mainly caused by the inversed phase difference between the applied pitching motion at the LE and the deflected motion at the TE when the chord-wise second mode is excited. When the plate flaps under the resonance condition at $K_B$=51.8, the plate produces much larger positive and negative pressures on the surfaces due to the higher acceleration to the fluid by the plate. The large pressure difference between the upper and lower surfaces leads to the maximum thrust value near resonance. Furthermore, the large passive deformation of the flexible plate helps in orienting the decomposed component of the pressure gradient in the chord-wise direction, which further enhances the generated thrust. The pressure difference becomes weaker and the pressure gradient component that contributes to the thrust generation decreases when the passive deformation of the plate is suppressed at higher $K_B$ values. As a result, the largest thrust value reduces for a stiffer plate. Similar conclusions can be drawn for the variation of the lowest thrust and the largest lift as a function of $K_B$. 

\begin{figure*}
	\subfloat[][]{\includegraphics[width=0.25\textwidth]{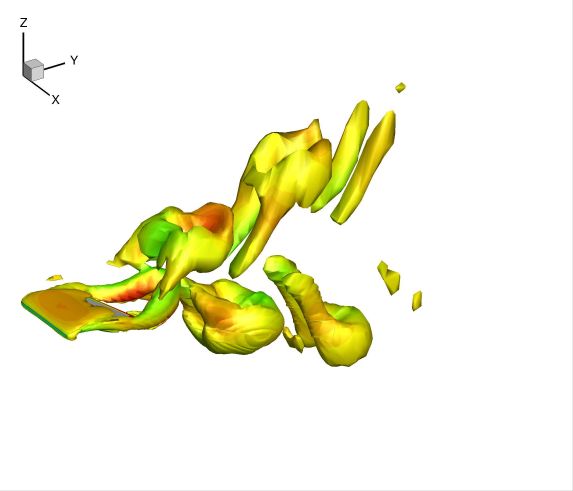}}
	\subfloat[][]{\includegraphics[width=0.25\textwidth]{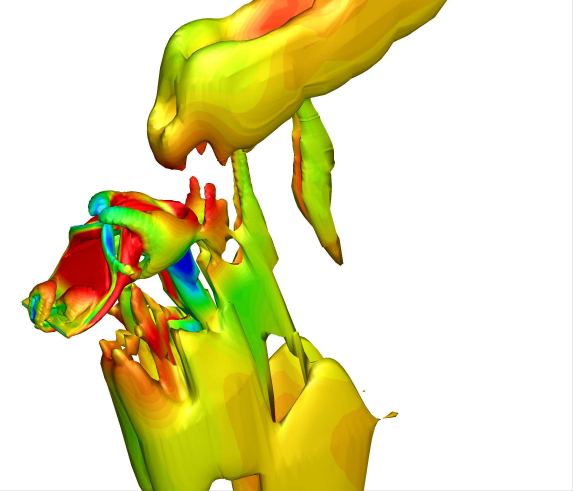}}
	\subfloat[][]{\includegraphics[width=0.25\textwidth]{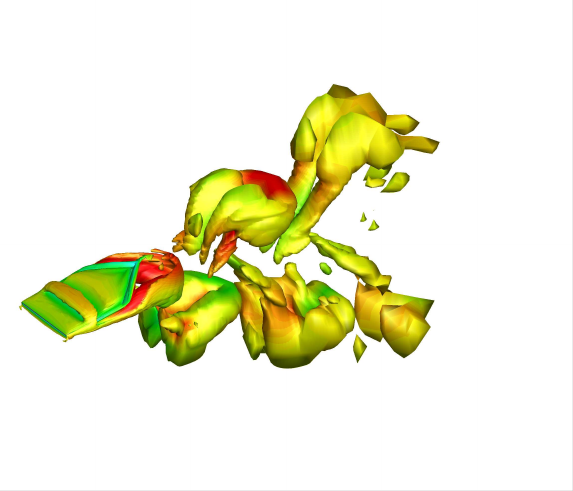}}
	\subfloat[][]{\includegraphics[width=0.25\textwidth]{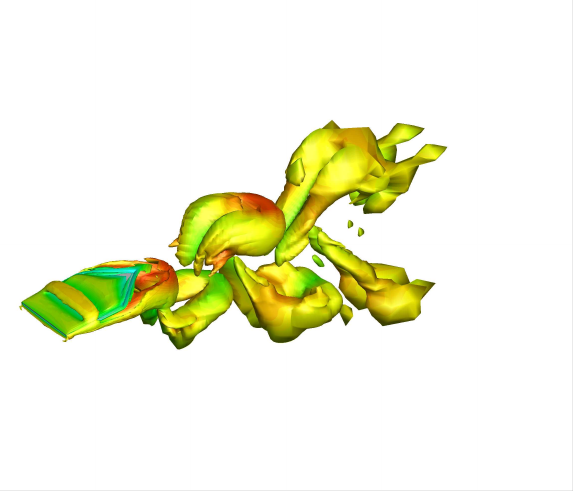}}
	\\
	\includegraphics[width=0.5\textwidth]{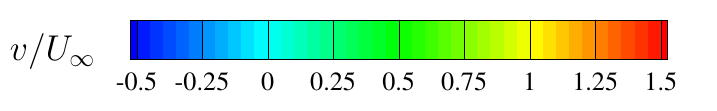}
	\caption{\label{fig:Qcriterion_KB} Instantaneous vortical structures based on the iso-surfaces of $Q \left( = -\frac{1}{2} \frac{\partial u^f_i}{\partial x_j} \frac{\partial u^f_j}{\partial x_i} \right)$ value of a pitching plate with trailing edge angle of $\Phi=135^\circ$ at $K_B$= (a) 19.73, (b) 51.8, (c) 197.3 and (d) 73998 at the moment with the largest thrust. Iso-surfaces of non-dimensional $Q^+ = Q(c/U_{\infty})^2 = 0.25$ are colored by the normalized streamwise velocity $v/U_{\infty}$.}
\end{figure*}

The instantaneous vortical structures indicated by the iso-surfaces of the Q criterion behind convex plates at different $K_B$ values at the moment with the largest thrust are plotted in Fig~.\ref{fig:Qcriterion_KB}. The iso-surfaces are colored by the normalized streamwise velocity. The wake behind the convex pitching plate shows horseshoe-like structures, but the size and the direction of the vortical structures are varied with flexibility. The horseshoe-like structures are elongated in the transverse and streamwise directions when $K_B$ increases from 19.73 to 51.8, which is caused by the strong acceleration near resonance. The inclination between the formed vortical structures and the centerline becomes smaller for stiffer plates due to the suppressed passive deformation. It is worth noting that the largest thrust is generated when the trailing edge vortex (TEV) detaches from the TE and convects downstream by transferring momentum to the wake.

\begin{figure*}
	\subfloat[][]{\includegraphics[width=0.49\textwidth]{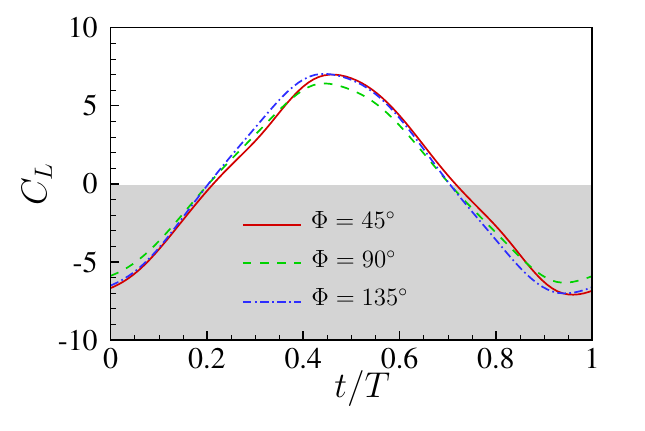}\label{fig:CT_phia}}
	\subfloat[][]{\includegraphics[width=0.49\textwidth]{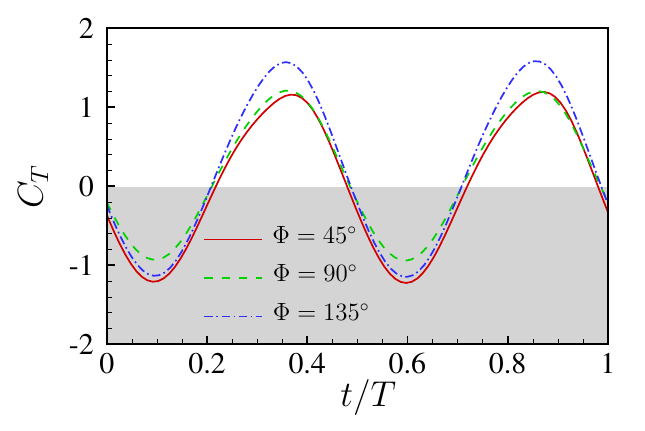}\label{fig:CT_phib}}
	\caption{\label{fig:CT_phi} Comparison of (a) instantaneous lift coefficient and (b) instantaneous thrust coefficient of pitching plate with $K_B=98.66$ within one complete pitching cycle. $t/T=0$ corresponds to the pitching upward from the neutral position.}
\end{figure*}

To understand the effect of the trailing edge shape on the propulsive performance, the comparison of the instantaneous lift and thrust coefficients within one completed pitching cycle for a concave plate, a rectangular plate and a convex plate is shown in Fig.~\ref{fig:CT_phi}. It can be seen from Fig.~\ref{fig:thrust_efficiency} that the convex plate produces the overall largest thrust and the concave plate is the least efficient in thrust generation at most $K_B$ values. Thus, we choose a representative bending stiffness of $K_B$=98.66 which is slightly far away from the resonance condition for simplicity. In Fig.~\ref{fig:CT_phi} \subref{fig:CT_phia}, the convex plate and the concave plate produce similar amplitudes of the lift coefficient. The rectangular plate is less efficient in generating lift. The convex plate produces the largest instantaneous positive thrust at $t/T$=0.36. The largest instantaneous positive thrust values of the rectangular plate and the concave plate are similar. On the contrary, the rectangular plate has the smallest instantaneous drag and the other two plates show more drag penalties. As a result, the convex plate generates slightly larger net thrust, compared to the rectangular plate. The net thrust produced by the concave plate is the smallest one among the three plates.

\begin{figure*}
	\subfloat[][]{\includegraphics[width=0.31\textwidth]{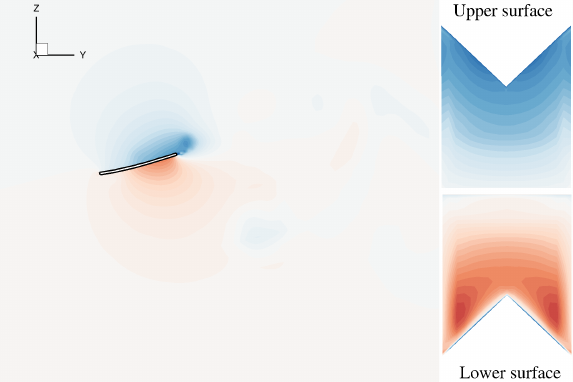}}
	\quad
	\subfloat[][]{\includegraphics[width=0.31\textwidth]{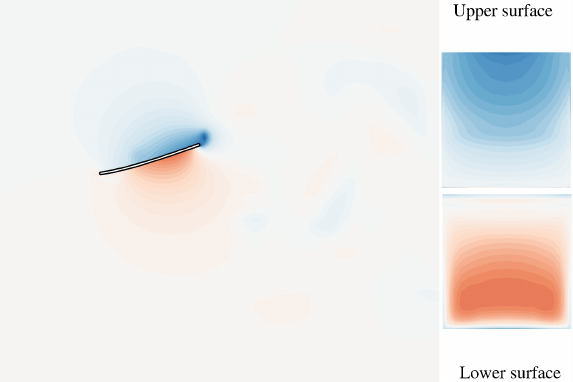}}
	\quad
	\subfloat[][]{\includegraphics[width=0.31\textwidth]{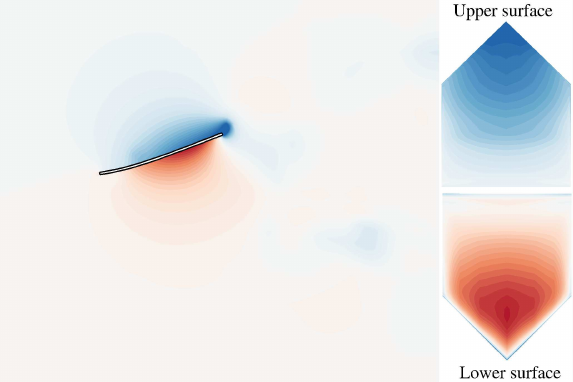}}
	\\
	\subfloat[][]{\includegraphics[width=0.31\textwidth]{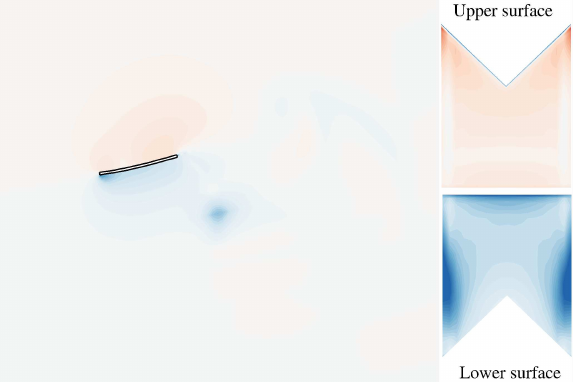}}
	\quad
	\subfloat[][]{\includegraphics[width=0.31\textwidth]{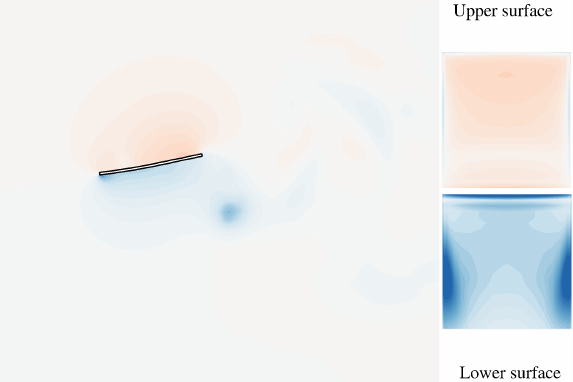}}
	\quad
	\subfloat[][]{\includegraphics[width=0.31\textwidth]{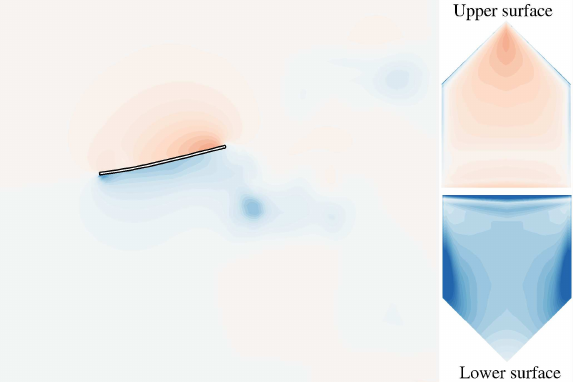}}
	\\
	\subfloat[][]{\includegraphics[width=0.31\textwidth]{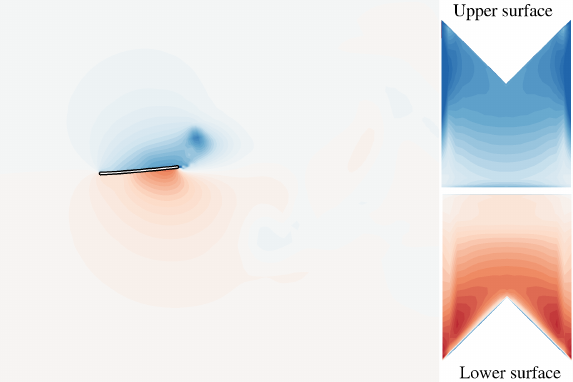}}
	\quad
	\subfloat[][]{\includegraphics[width=0.31\textwidth]{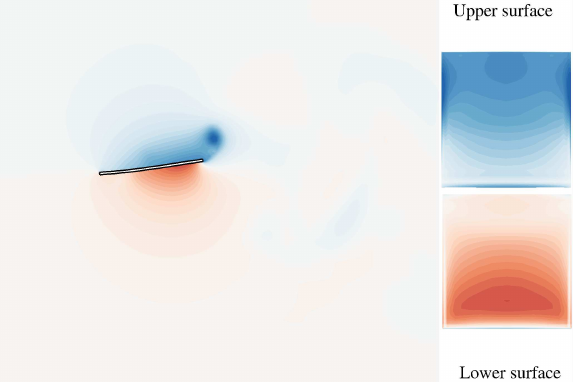}}
	\quad
	\subfloat[][]{\includegraphics[width=0.31\textwidth]{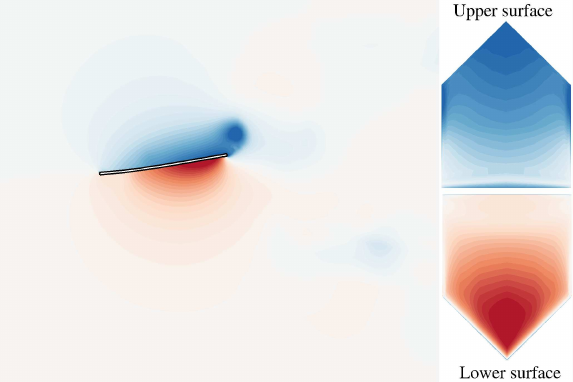}}
	\\
	\includegraphics[width=0.45\textwidth]{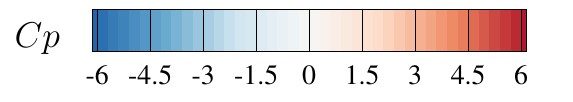}
	\caption{\label{fig:pressure_phi} Instantaneous pressure coefficient contour on the span-wise symmetry plane and pressure coefficient distribution on the upper and lower surfaces of pitching plate with $K_B=98.66$ for $\Phi$= (a,d,g) $45^\circ$, (b,e,h) $90^\circ$ and (c,f,i) $135^\circ$ at the moment with (a,b,c) the largest thrust, (d,e,f) the lowest thrust and (g,h,i) the largest lift.}
\end{figure*}

The comparison of the instantaneous pressure coefficient distributions around the pitching plate with different trailing edge shapes is shown in Fig.~\ref{fig:pressure_phi}. It can be seen from Fig.~\ref{fig:pressure_phi} (a-c) that the pressure difference between the upper and lower surfaces of a convex plate is the largest. Because the convex plate and the concave plate have the longest local chord at the mid-span location and the sides, the flows are strongly accelerated at these locations to induce the largest pressure gradient. Compared to the pressure distribution of the concave plate, the flow with bifurcated four jets on both sides of the convex plate can produce the highest pressure difference. Consequently, the convex plate can generate the largest instantaneous thrust. It can be inferred from Fig.~\ref{fig:pressure_phi} (d-f) that the rectangular plate produces the smallest instantaneous drag due to the less pressure difference. In Fig.~\ref{fig:pressure_phi} (g-i), the rectangular plate shows smaller pressure gradients in the transverse direction. Since the effective projection area of these three plates is similar, the amplitude of the lift coefficient of the rectangular plate is smaller than that of the concave and convex plates.

\begin{figure*}
	\subfloat[][]{\includegraphics[width=0.32\textwidth]{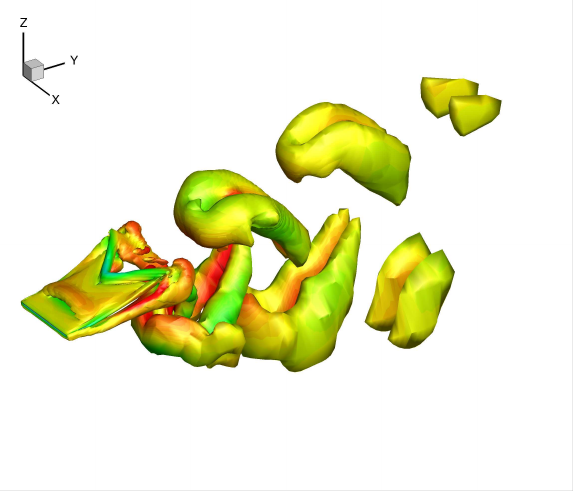}}
	\subfloat[][]{\includegraphics[width=0.32\textwidth]{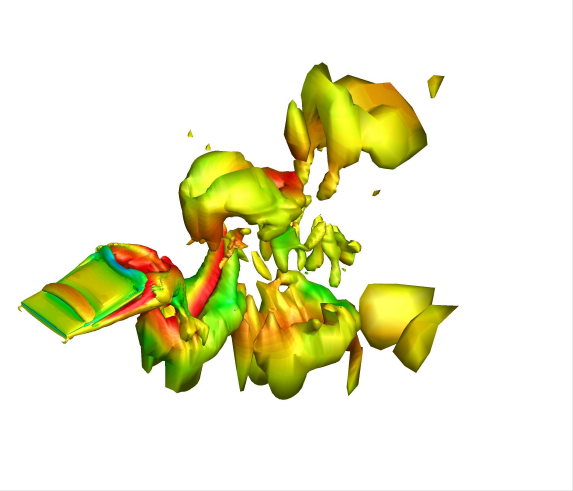}}
	\subfloat[][]{\includegraphics[width=0.32\textwidth]{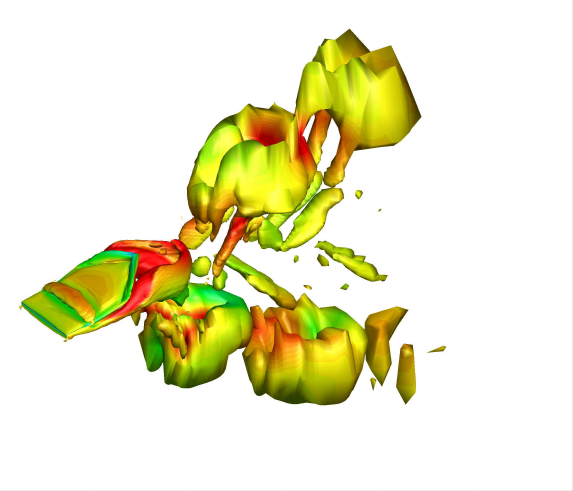}}
	\\
	\includegraphics[width=0.5\textwidth]{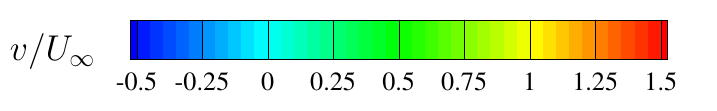}
	\caption{\label{fig:Qcriterion_phi} Instantaneous vortical structures based on the iso-surfaces of $Q \left( = -\frac{1}{2} \frac{\partial u^f_i}{\partial x_j} \frac{\partial u^f_j}{\partial x_i} \right)$ value of a pitching plate with $K_B=98.66$ for $\Phi$= (a) $45^\circ$, (b) $90^\circ$ and (c) $135^\circ$ at the moment with the largest thrust. Iso-surfaces of non-dimensional $Q^+ = Q(c/U_{\infty})^2 = 0.25$ are colored by the normalized streamwise velocity $v/U_{\infty}$.}
\end{figure*}

The wake structures of these three plates are compared in Fig.~\ref{fig:Qcriterion_phi}. At the moment with the largest thrust, the convection of the detached vortices from the TE is observed for all plates. The wake structures behind the concave plate behave as a reversed horseshoe-like structure. Conversely, the horseshoe-like structure is observed behind the rectangular plate and the convex plate. The vortical structures behind the convex plate are stretched to form the widest vortex ring in the transverse direction, which is caused by the largest acceleration at the longest local chord location. Furthermore, the inclination between the vortical structure and the centerline is largest for the convex plate, which helps in orienting the pressure gradient to generate more thrust.

\subsubsection{Wake mode and structural mode}
\begin{figure*}
	\subfloat[][]{\includegraphics[width=0.24\textwidth,height=0.13\textwidth]{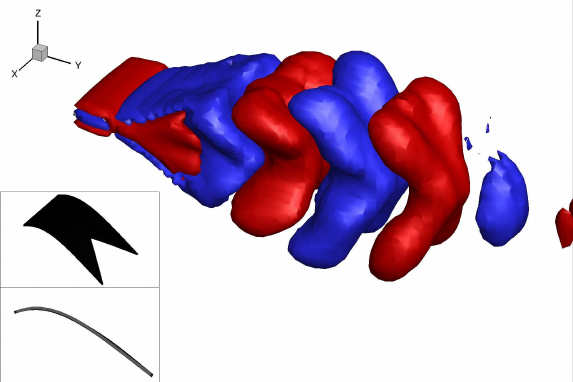}}
	\
	\subfloat[][]{\includegraphics[width=0.24\textwidth,height=0.13\textwidth]{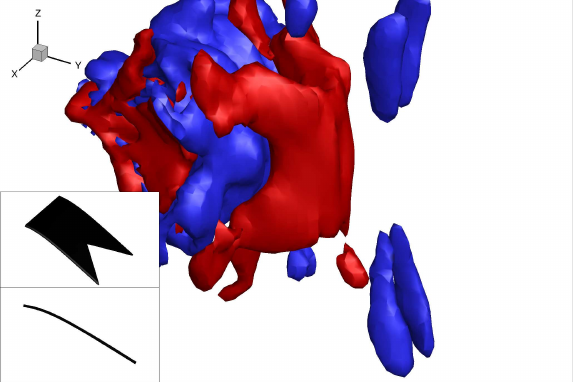}}
	\
	\subfloat[][]{\includegraphics[width=0.24\textwidth,height=0.13\textwidth]{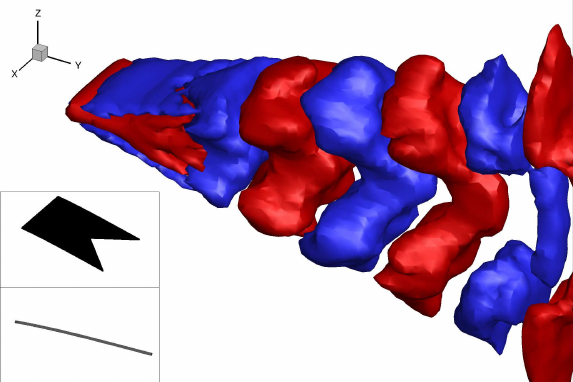}}
	\
	\subfloat[][]{\includegraphics[width=0.24\textwidth,height=0.13\textwidth]{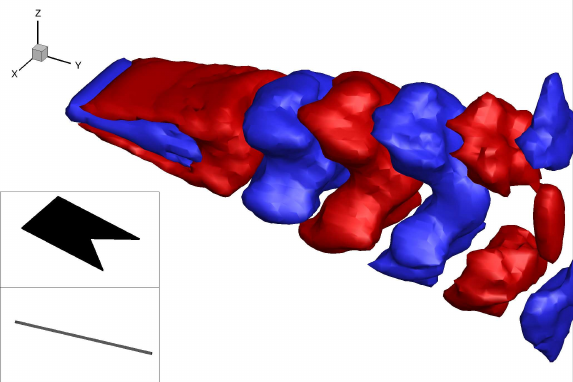}}
	\\
	\subfloat[][]{\includegraphics[width=0.24\textwidth,height=0.13\textwidth]{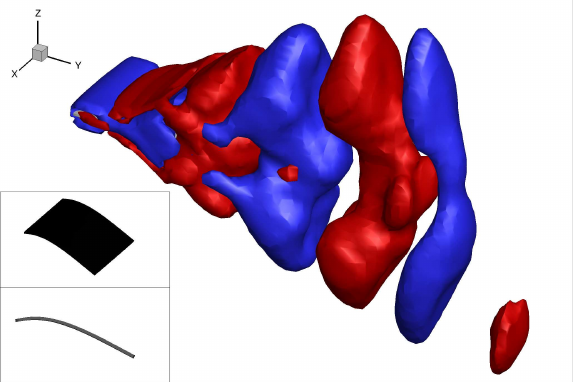}}
	\
	\subfloat[][]{\includegraphics[width=0.24\textwidth,height=0.13\textwidth]{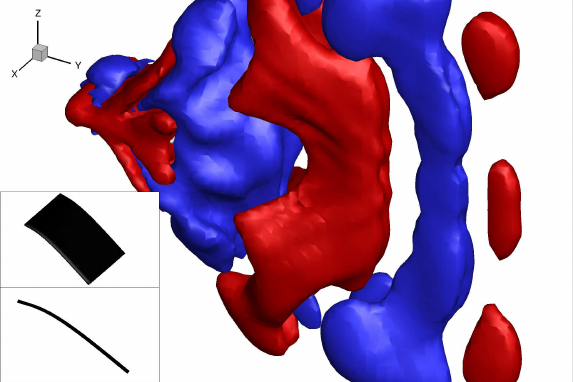}}
	\
	\subfloat[][]{\includegraphics[width=0.24\textwidth,height=0.13\textwidth]{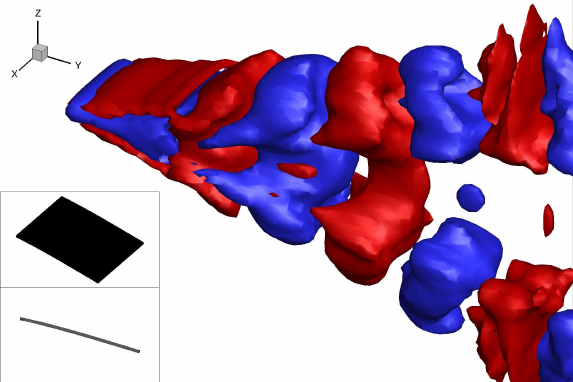}}
	\
	\subfloat[][]{\includegraphics[width=0.24\textwidth,height=0.13\textwidth]{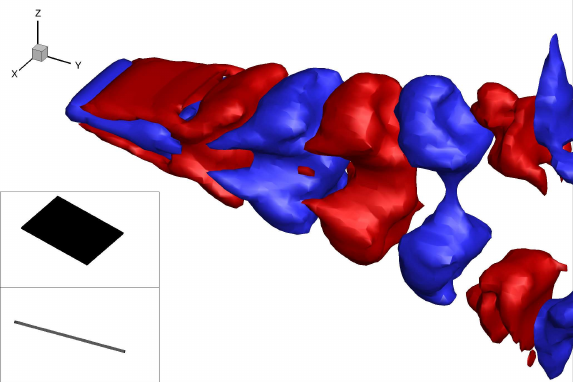}}
	\\
	\subfloat[][]{\includegraphics[width=0.24\textwidth,height=0.13\textwidth]{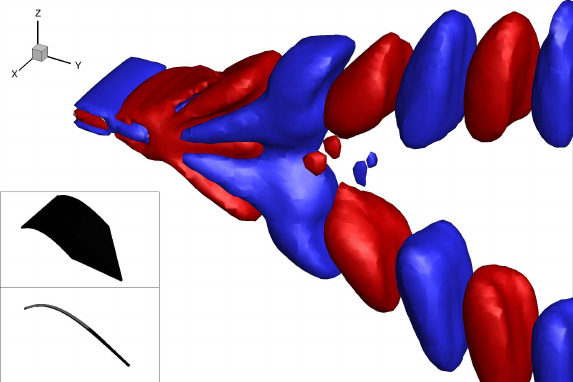}}
	\
	\subfloat[][]{\includegraphics[width=0.24\textwidth,height=0.13\textwidth]{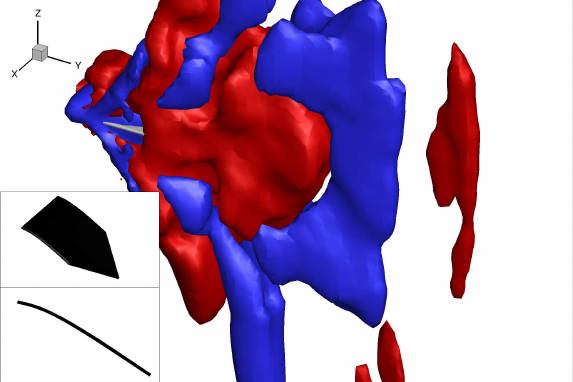}}
	\
	\subfloat[][]{\includegraphics[width=0.24\textwidth,height=0.13\textwidth]{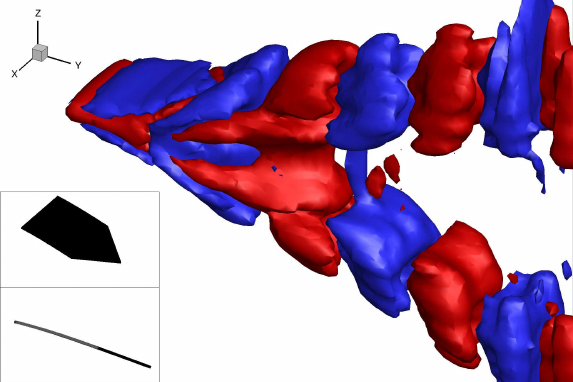}}
	\
	\subfloat[][]{\includegraphics[width=0.24\textwidth,height=0.13\textwidth]{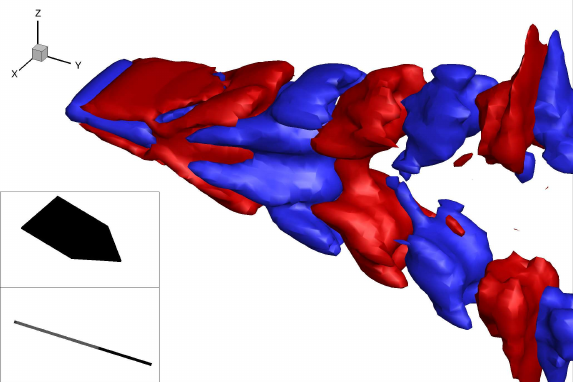}}
	\caption{\label{fig:DMD_mode3} DMD modes of a pitching plate with a mode frequency of $f_p$=1.1 Hz at $K_B$= (a,e,i) 19.73, (b,f,j) 51.8, (c,g,k) 197.3 and (d,h,l) 73998 with trailing edge angle of $\Phi$= (a,b,c,d) $45^\circ$, (e,f,g,h) $90^\circ$ and (i,j,k,l) $135^\circ$. The iso-surfaces in red color is thresholded at $\omega_x=0.0015$ and the iso-surfaces at $\omega_x=-0.0015$ is colored in blue. The structural DMD modes in the above plot is the oblique view and the blow plot is the side view along span-wise direction.}
\end{figure*}

The effects of trailing edge and flexibility on the dynamic response and the flow characteristics of the pitching plate have been discussed in detail above. Based on the discussion, the wake structures are strongly affected by the deformation of the plate to further govern the propulsive performance. Thus, it is important to directly link the induced vortical structures with the structural deformation to understand the role of trailing edge shape and flexibility in the thrust generation. With the aid of the SP-DMD method, the $X$-vorticity and the displacements of the flexible plate are decomposed in a unified approach to extract the correlated DMD modes of the coupled system at the dominant frequency $f_p$. In Fig.~\ref{fig:DMD_mode3}, three types of plates with four representative $K_B$ values are selected to examine the correlated wake and structural modes. 

From the decomposed structural DMD modes shown in Fig.~\ref{fig:DMD_mode3} (a,e,i), we find that the flexible plate shows a chord-wise second mode at $K_B$=19.73. This chord-wise second mode is a natural selection of the coupling between the structural natural frequency and the applied pitching frequency. As $K_B$ increases to 51.8 to excite the resonance between the natural frequency of the first structural mode and the pitching frequency, the chord-wise second mode is suppressed and the chord-wise first mode is amplified. Meanwhile, more fluids in the transverse direction are affected by the pitching plate, which leads to wider transverse wake structures. The wake structures become more complex under the resonance condition. When the plate tends to be stiffer, the vibration amplitude of the chord-wise first mode becomes smaller, resulting in narrower wake structures in the transverse direction. 

The inclination between the wake structures and the centerline shows a growing trend with the increase of the trailing edge angle. Except for the resonance case, the wake structures start to split into two groups in the transverse direction to reduce the interaction of the vortices in the upper and lower branches when the trailing edge shape changes to convex. Meanwhile, the wake is slightly compressed in the span-wise direction due to the outward trailing edge shape.

\subsection{Unsteady momentum transfer and thrust generation} \label{unsteady_momentum}
The pitching flexible plate is interacted with the fluid to generate beneficial thrust by transferring momentum and energy to the fluid. To fully understand the thrust generation mechanism, the momentum creation and transport during pitching motion is analyzed via a control-volume formulation. We first apply this analysis to a nominal case to build a bridge between the spatial and temporal evolution of the vortical structures, the unsteady momentum transfer and the thrust generation. Then, the role of flexibility and trailing edge shape in the thrust-generating momentum is investigated by decomposing the total thrust into four terms with their physical interpretations.

\subsubsection{Decomposition of unsteady momentum}
% introduction of the control volume analysis based on the velocity-pressure method
To establish a quantitative connection between the flow dynamic behavior and the time-dependent force mechanisms, Noca et al. \cite{noca1997evaluation,noca1999comparison} derived a velocity/pressure equation to evaluate the time-dependent forces acting on a moving body immersed in an incompressible and viscous flow. The instantaneous force $\bm{F}(t)$ exerted by the unsteady fluid is calculated by integrating the momentum change within an arbitrary time-dependent control volume $V^f(t)$ and on the control surface $S^c(t)$ as well as the body surface $S^s(t)$
\begin{eqnarray}
&\bm{F}(t) = - \frac{\partial}{ \partial t}\int_{V^f(t)} \rho^f \bm{u}^f {\rm{d}}V - \oint_{S^c(t)} \bm{n} \cdot (\bm{u}^f - \bm{u}^{is}) \rho^f \bm{u}^f {\rm{d}} S \nonumber \\
&- \oint_{S^c(t)} p \bm{n} {\rm{d}} S + \oint_{S^c(t)} \boldsymbol{\tau} \bm{n} {\rm{d}} S \nonumber \\ 
&- \oint_{S^s(t)} \bm{n} \cdot (\bm{u}^f - \bm{u}^{is}) \rho^f \bm{u}^f {\rm{d}} S
\label{eq:momentum_equation_cva}
\end{eqnarray}
where $\bm{n}$ represents the unit vector normal to the integral surface. $\bm{u}^{is}$ denotes the velocity of the integral surface. $p$ and $\bm{I}$ are the pressure and unit tensor, respectively. $\boldsymbol{\tau}=\mu^f (\nabla \bm{u}^f + (\nabla \bm{u}^f)^T)$ is the viscous stress tensor and $\mu^f$ is the dynamic viscosity coefficient. 

In Eq.~(\ref{eq:momentum_equation_cva}), the first term on the right-hand side represents the rate of change of momentum within the control volume $V^f(t)$. The second term, the third term and the fourth term are the net momentum flux across the control surface, the instantaneous pressure force and the instantaneous viscous stresses acting on the control surface $S^c(t)$, respectively. The fifth term denotes the momentum flux across the body surface and it equals zero due to the no-through boundary condition of the body surface. In this study, a three-dimensional control volume with a size of $10c \times 10c \times 7.5c$ enclosed the pitching plate is constructed in a fixed frame to evaluate the time-dependent fluid loads. Thus, the velocity $\bm{u}^{is}$ of the control surface $S^c(t)$ is zero for a stationary control volume. The instantaneous thrust coefficient ${C_T}_w$ evaluated from the wake information in the control volume can be decomposed into four distinct terms written as
\begin{eqnarray}
& &{C_T}_w = -\frac{F_y}{\frac{1}{2} \rho^f U_{\infty}^2 S} = {C_T}_u + {C_T}_c + {C_T}_p + {C_T}_s \nonumber \\
& &{C_T}_u = \frac{1}{\frac{1}{2} \rho^f U_{\infty}^2 S} \frac{\partial}{\partial t}\int_{V^f(t)} (\rho^f \bm{u}^f) \cdot \bm{n}_y {\rm{d}}V \nonumber \\
& &{C_T}_c = \frac{1}{\frac{1}{2} \rho^f U_{\infty}^2 S} \oint_{S^c(t)}[ \rho^f \bm{u}^f (\bm{n} \cdot \bm{u}) ] \cdot \bm{n}_y {\rm{d}} S \nonumber \\
& &{C_T}_p = \frac{1}{\frac{1}{2} \rho^f U_{\infty}^2 S} \oint_{S^c(t)} (p \bm{n}) \cdot \bm{n}_y {\rm{d}} S \nonumber \\
& &{C_T}_s = -\frac{1}{\frac{1}{2} \rho^f U_{\infty}^2 S} \oint_{S^c(t)} (\boldsymbol{\tau} \bm{n}) \cdot \bm{n}_y {\rm{d}} S
\label{eq:thrust_decomposition}
\end{eqnarray}
where ${C_T}_u$, ${C_T}_c$, ${C_T}_p$ and ${C_T}_s$ are the unsteady term, the convective term, the pressure force term and the shear stress term, respectively. The flow variables obtained from the numerical simulation on each point of the body-fitted moving mesh are mapped to each point of the fixed reference mesh of the rectangular control volume shown in Fig.~\ref{fig:velocity_in_wake} to calculate each term in Eq.~(\ref{eq:thrust_decomposition}).

\begin{figure*}
	\subfloat[][]{\includegraphics[width=0.9\textwidth,height=0.4\textwidth]{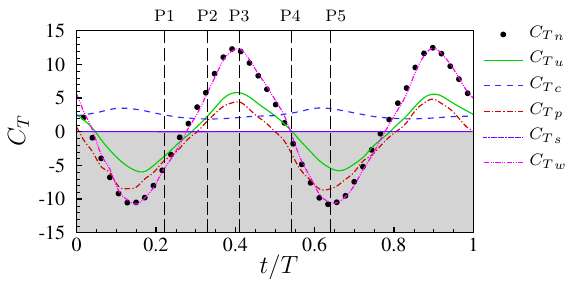}\label{fig:CT_decom_E1_05e6a}}
	\\
	\subfloat[][]{
	\includegraphics[width=0.24\textwidth,height=0.14\textwidth]{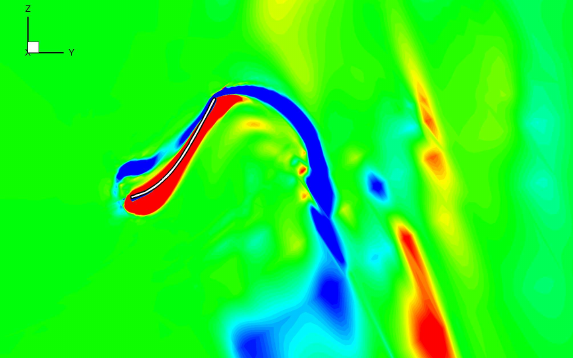}
	\
	\includegraphics[width=0.24\textwidth,height=0.14\textwidth]{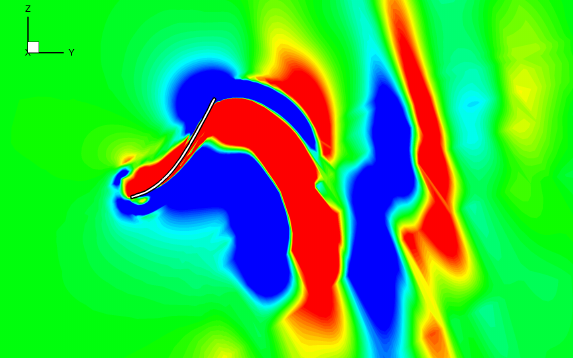}
	\includegraphics[width=0.24\textwidth,height=0.14\textwidth]{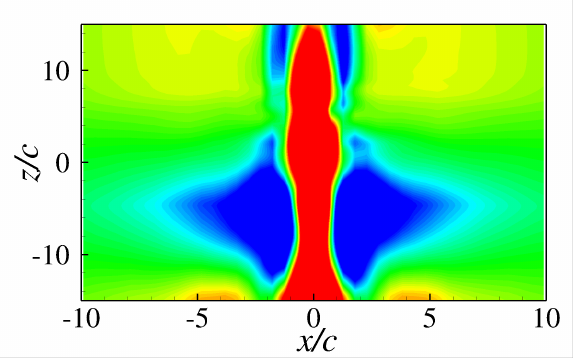}
	\includegraphics[width=0.24\textwidth,height=0.14\textwidth]{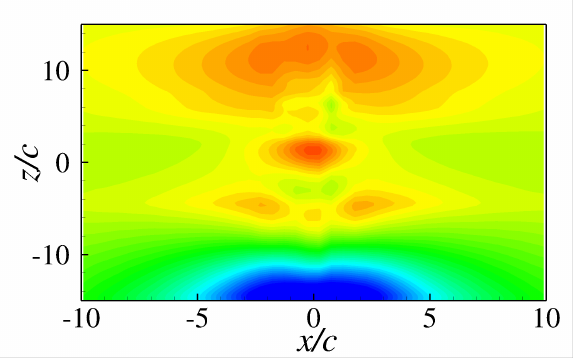}}
	\\
	\subfloat[][]{
	\includegraphics[width=0.24\textwidth,height=0.14\textwidth]{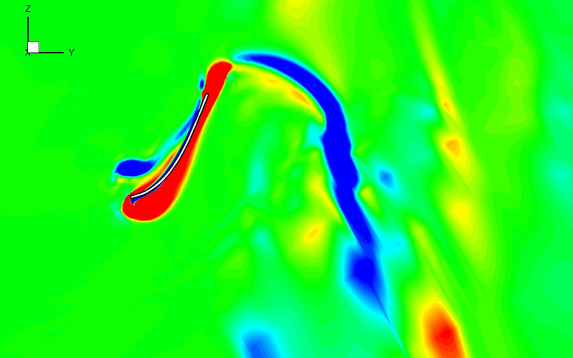}
	\
	\includegraphics[width=0.24\textwidth,height=0.14\textwidth]{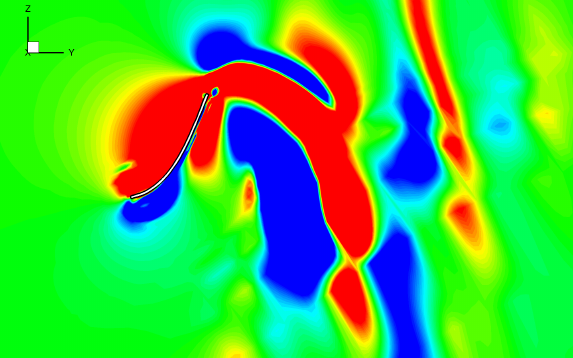}
	\includegraphics[width=0.24\textwidth,height=0.14\textwidth]{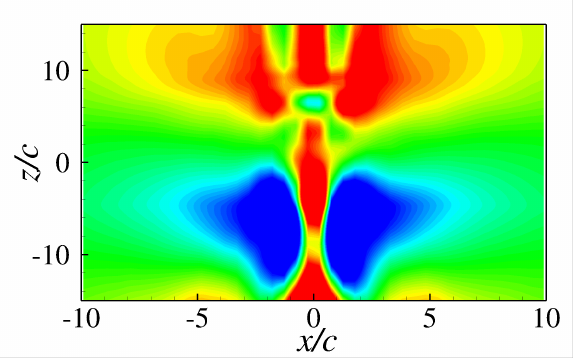}
	\includegraphics[width=0.24\textwidth,height=0.14\textwidth]{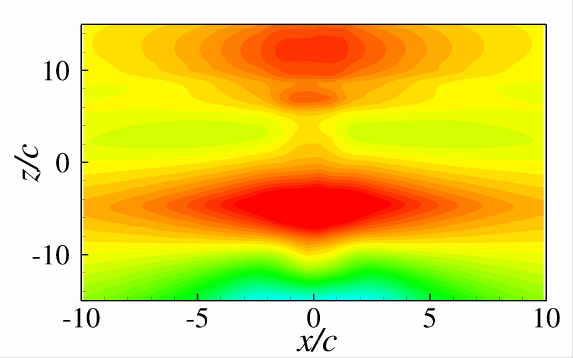}}
	\\
	\subfloat[][]{
	\includegraphics[width=0.24\textwidth,height=0.14\textwidth]{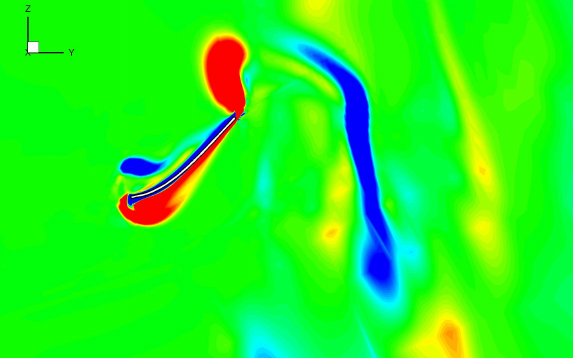}
	\
	\includegraphics[width=0.24\textwidth,height=0.14\textwidth]{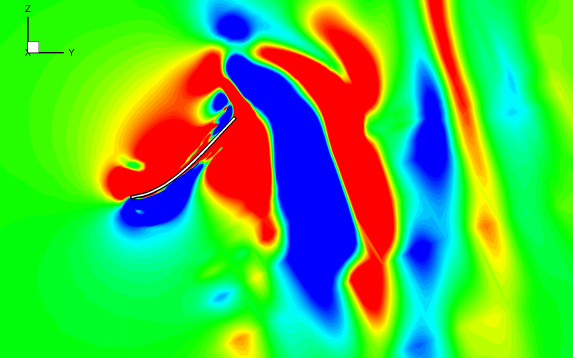}
	\includegraphics[width=0.24\textwidth,height=0.14\textwidth]{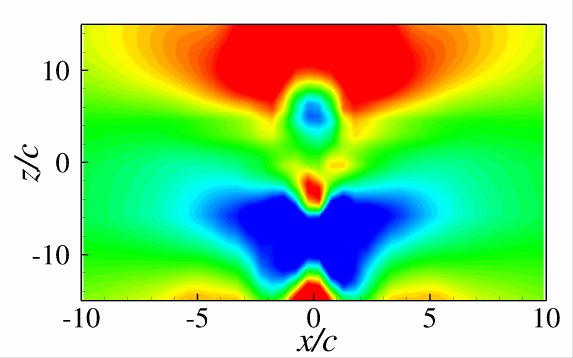}
	\includegraphics[width=0.24\textwidth,height=0.14\textwidth]{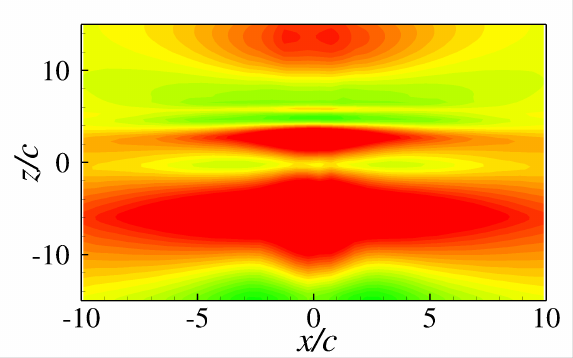}}
    \\
	\subfloat[][]{
	\includegraphics[width=0.24\textwidth,height=0.14\textwidth]{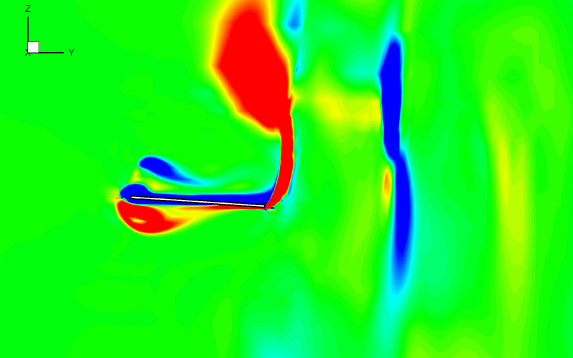}
	\
	\includegraphics[width=0.24\textwidth,height=0.14\textwidth]{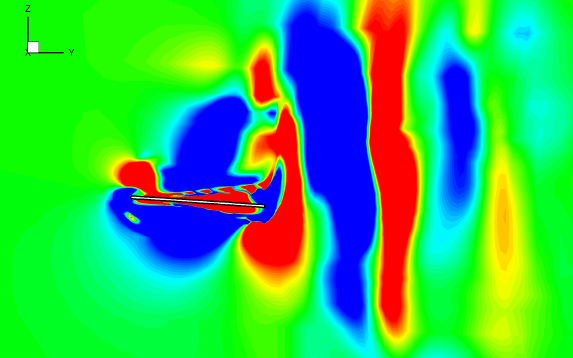}
	\includegraphics[width=0.24\textwidth,height=0.14\textwidth]{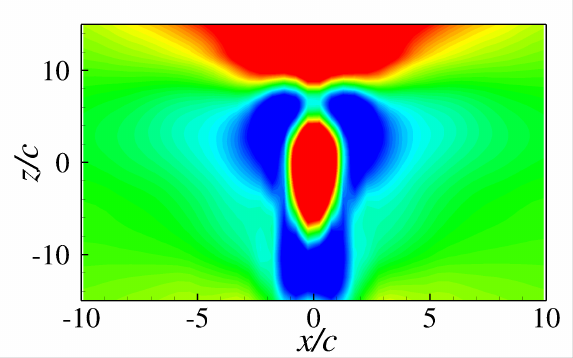}
	\includegraphics[width=0.24\textwidth,height=0.14\textwidth]{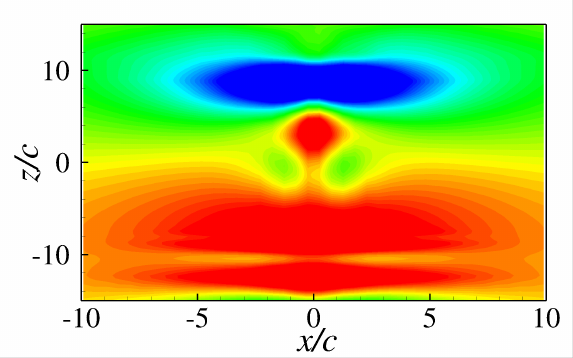}}
\end{figure*}

\begin{figure*}
	\addtocounter{subfigure}{5}
	\subfloat[][]{
	\includegraphics[width=0.24\textwidth,height=0.14\textwidth]{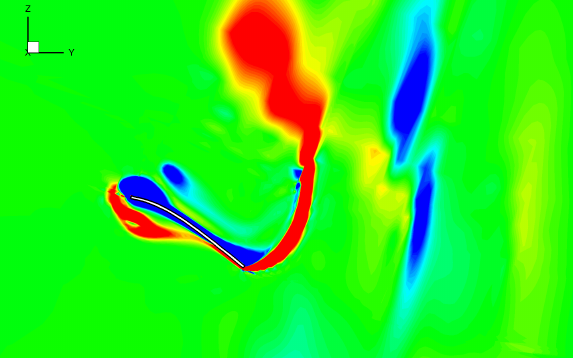}
	\
	\includegraphics[width=0.24\textwidth,height=0.14\textwidth]{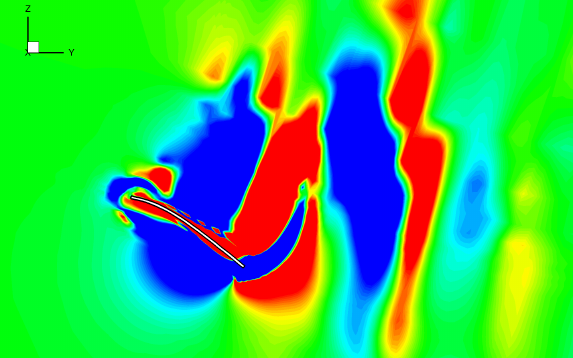}
	\includegraphics[width=0.24\textwidth,height=0.14\textwidth]{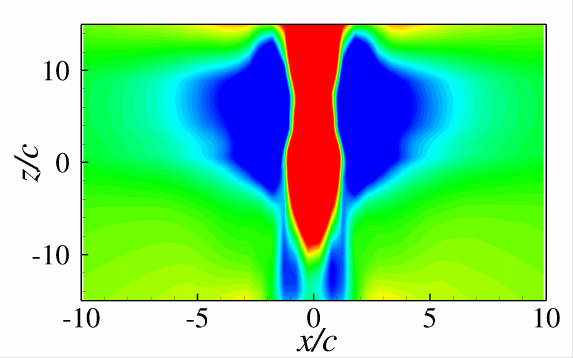}
	\includegraphics[width=0.24\textwidth,height=0.14\textwidth]{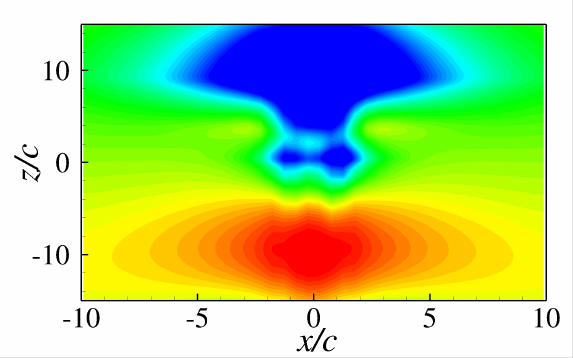}}
    \\
    \includegraphics[width=0.24\textwidth]{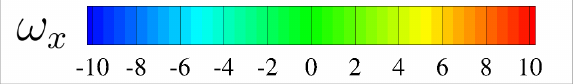}
    \quad
    \includegraphics[width=0.24\textwidth]{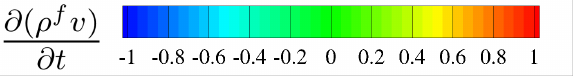}
    \quad
    \includegraphics[width=0.22\textwidth]{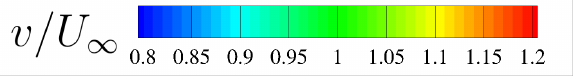}
    \
    \includegraphics[width=0.22\textwidth]{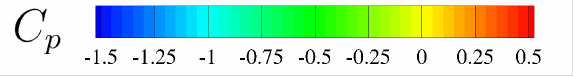}
	\caption{\label{fig:CT_decom_E1_05e6} (a) Thrust decomposition based on the momentum equation for flexible plate with trailing edge angle of $\Phi=135^\circ$ at $K_B=51.8$ as a nominal case. The flow features are plotted at non-dimensional time instants of (b) P1, (c) P2, (d) P3, (e) P4 and (f) P5. The instantaneous $X$-vorticity $\omega_x$ on the mid-span plane is presented in the first column. The instantaneous rate of change of momentum $\frac{\partial(\rho^f v)}{\partial t}$ on the mid-span plane is presented in the second column. The instantaneous non-dimensional streamwise velocity $v/U_{\infty}$ on the control surface at $y/c=2.5$ is presented in the third column. The instantaneous pressure coefficient $C_p$ on the control surface at $y/c=2.5$ is presented in the fourth column. $t/T=0$ corresponds to the pitching upward from the neutral position.}
\end{figure*}

The flexible convex plate immersed in the unsteady flow with the global maximum thrust under the resonance condition is considered as the nominal case to demonstrate and validate the momentum-based thrust evaluation formulation. Figure~\ref{fig:CT_decom_E1_05e6} \subref{fig:CT_decom_E1_05e6a} presents the decomposition of the four distinct thrust coefficient terms and the comparison of the total evaluated thrust coefficient ${C_T}_w$ and the total thrust coefficient ${C_T}_n$ calculated from Eq.~(\ref{eq:force_coefficient}) in the numerical simulation within one completed cycle. It can be seen that the evaluated thrust coefficient ${C_T}_w$ from the integration of the momentum equation within the control volume is consistent with the thrust coefficient ${C_T}_n$. In other words, the momentum-based control volume approach can accurately calculate the instantaneous thrust force based on the flow variables around and inside a control volume. Each decomposed thrust coefficient term has physical significances to correlate the instantaneous vortical structures and the resulting fluid loads. The unsteady term ${C_T}_u$ behaves an almost symmetrical feature with respect to the zero-thrust condition due to the symmetrical pitching motion. The variation of the unsteady term makes significant contributions to the instantaneous total thrust but leads to negligible contributions to the time-averaged net thrust. The convective term ${C_T}_c$ shows positive momentum flux across the control surface, which is related to the vortical structures induced by the pitching plate convecting downstream with high velocities. The pressure term ${C_T}_p$ presents comparable variations to the unsteady term and it plays a negative role in the mean thrust generation. The variation of the shear stress term ${C_T}_s$ is negligible compared to other terms due to the large control volume far from the moving plate. The unsteady term and the pressure term show a phase synchronized with the total thrust, while the convection term shows an opposite phase.

To shed light on the relationship between the instantaneous force and the evolution of the wake structures, the flow features within the control volume and on the control surface related to the decomposed thrust terms are plotted at five selected time instants (P1-P5) in Fig.~\ref{fig:CT_decom_E1_05e6} (b-f). It can be observed from the figures in the first column that the maximum total thrust is achieved at the time instant of P3 when the TEV detaches from the TE of the plate during downward movement. Meanwhile, the detachment and convection of the TEV induce the large time derivative of the momentum around the plate surface within the control volume shown in the second column, resulting in the optimal unsteady term. The low-velocity region presented in the third column expands on the control surface located at $y/c$=2.5. Thus, the convective term reduces to a lower value. In the fourth column, the greatest suction pressure difference is formed between the control surface at $y/c$=-5 and 2.5 to lead to the largest pressure term. As the plate continues to move downward from P3 to P5, the detached TEV convects further downstream and the vortex is gradually formed at the LE. Thus, the region with negative rates of change of momentum expands on the plate surface during the movement of the vortices from the LE to the TE, leading to the continuous reduction of the unsteady term. Conversely, the convective term grows up as the vortex rings containing high velocities reach the control surface at $y/c$=2.5. The suction effect between the front and back control surfaces is reversed when the low pressure region on the back control surface becomes larger. As a result, the pressure term is reduced and makes negative contributions to the instantaneous thrust generation.

\subsubsection{Effect of flexibility on unsteady momentum}
\begin{figure*}
	\subfloat[][]{\includegraphics[width=0.49\textwidth]{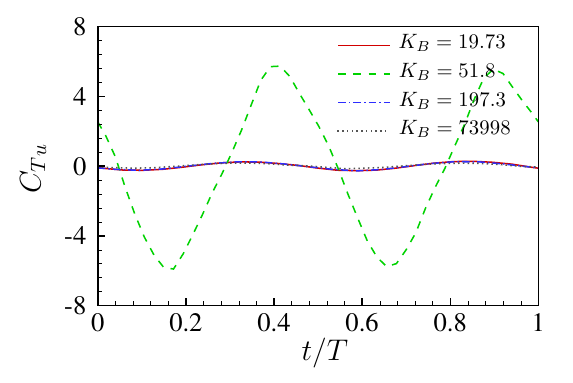}\label{fig:CT_KB_decoma}}
	\subfloat[][]{\includegraphics[width=0.49\textwidth]{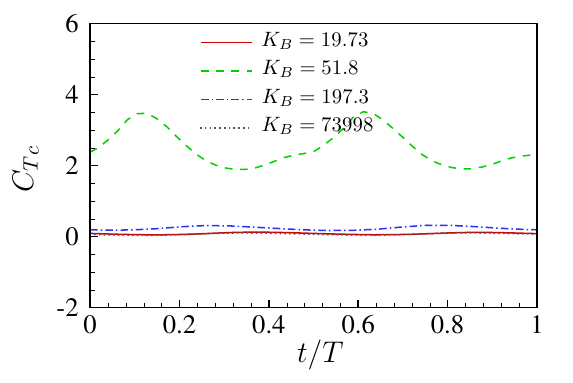}\label{fig:CT_KB_decomb}}
	\\
	\subfloat[][]{\includegraphics[width=0.49\textwidth]{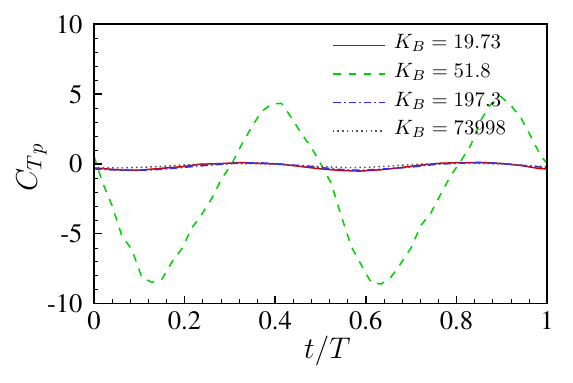}\label{fig:CT_KB_decomc}}
	\caption{\label{fig:CT_KB_decom} Comparison of (a) unsteady term, (b) convective term and (c) pressure term of pitching plate with trailing edge angle of $\Phi=135^\circ$ within one completed pitching cycle. $t/T=0$ corresponds to the pitching upward from the neutral position.}
\end{figure*}

It can be concluded from the control volume analysis for the nominal case that the unsteady term, the convective term and the pressure term are strongly related to the instantaneous thrust generation. The contribution of the shear stress term can be neglected. To further explore the effect of flexibility on the thrust generation, the instantaneous decomposed unsteady, convective and pressure terms evaluated by the momentum-based approach are compared for convex plates with four representative bending stiffness values in Fig.~\ref{fig:CT_KB_decom}. The convex plate with $K_B$=51.8 shows the largest amplitude of the unsteady term in Fig.~\ref{fig:CT_KB_decom} \subref{fig:CT_KB_decoma}. The large passive deformation excited under the resonance condition leads to the drastic change of the fluid acceleration (unsteady term) near the plate. Once the deformation amplitude reduces for flexible plates with low bending stiffness or high bending stiffness values, the variation of the unsteady term becomes weak. With respect to the convective term, the horseshoe-like vortical structures flowing across the control surface contain much higher velocities at $K_B$=51.8 observed in Fig.~\ref{fig:iso_vvel_rotate} and \ref{fig:wake_profile}, resulting in the largest momentum flux under the resonance condition. Compared to the highly rigid plate with $K_B$=73998, the flexible plate with moderate passive deformation is able to improve the momentum flux to enhance the thrust generation. However, the highly flexible case lacks the ability to impart large momentum to the wake. The pressure term is significantly enhanced under the resonance condition and shows negative contributions to the mean thrust. The difference between the pressure terms of the flexible plates is small under the off-resonance condition.

\subsubsection{Effect of trailing edge shape on unsteady momentum}
\begin{figure*}
	\subfloat[][]{\includegraphics[width=0.49\textwidth]{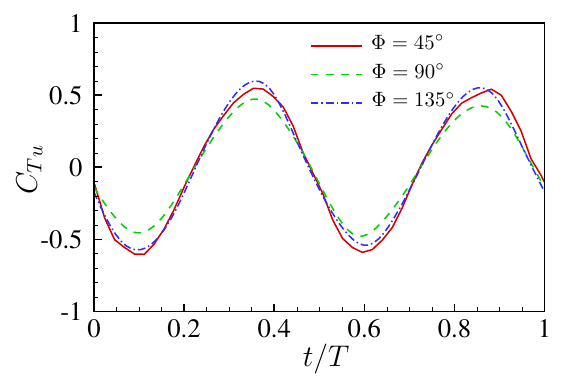}}
	\subfloat[][]{\includegraphics[width=0.49\textwidth]{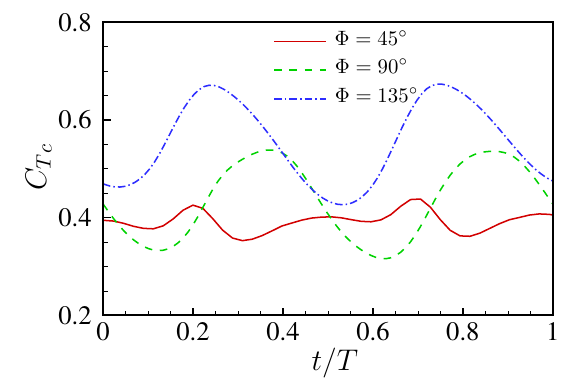}}
	\\
	\subfloat[][]{\includegraphics[width=0.49\textwidth]{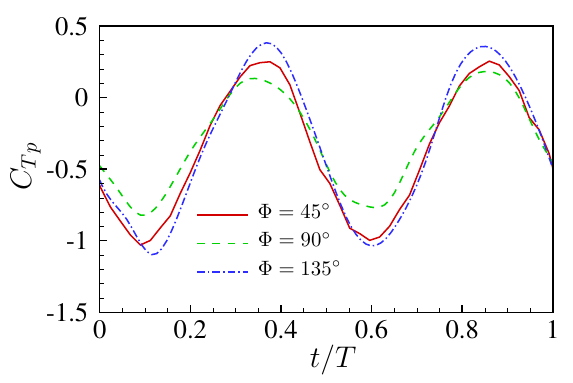}}
	\caption{\label{fig:CT_phi_decom} Comparison of (a) unsteady term, (b) convective term and (c) pressure term of pitching plate with $K_B=98.66$ within one completed pitching cycle. $t/T=0$ corresponds to the pitching upward from the neutral position.}
\end{figure*}

The comparison of the instantaneous decomposed three thrust terms for the flexible plate at a moderate bending stiffness of $K_B$=98.66 with different trailing edge shapes is shown in Fig.~\ref{fig:CT_phi_decom}. Compared to the concave plate, the absolute values of the minimum and maximum unsteady term of the rectangular plate become smaller, due to the lower acceleration caused by the shorter local chord. The stronger compression effect of the wake behind the convex plate leads to a larger local maximum and minimum values of the unsteady thrust term than those of the concave plate. The concave plate produces the overall smallest convective term while the convex plate is beneficial to enhance the momentum transfer. The surrounding flow can be strongly accelerated through the coupling effect near the mid-span location of the convex pitching plate, compared to the concave and rectangular plates. The pressure term of the rectangular plate has the smallest amplitude but the convex plate shows the largest amplitude. However, the pressure term of the rectangular plate has less contribution to the mean drag. The concave plate experiences the most mean drag penalty related to the pressure term.

\subsection{Drag-thrust transition}
Considering the complex vortical structures caused by the pitching plate, it is quite difficult to explore the mechanism of the drag-thrust transition by directly investigating the evolution of the wake structures. In addition to identifying the transition boundary, the variation of the mean thrust data as a function of trailing edge shape and flexibility is limited to provide further insight into the physical mechanisms related to the vortical structures. The momentum-based thrust evaluation approach breaks the barrier of the connection between the time-dependent or mean resulting forces and the vortical structures. With the aid of this effective method, the mechanism of the drag-thrust transition is quantitatively examined based on the variation of the mean decomposed thrust terms correlated to the time-averaged flow features.

\begin{figure*}
	\subfloat[][]{\includegraphics[width=0.49\textwidth]{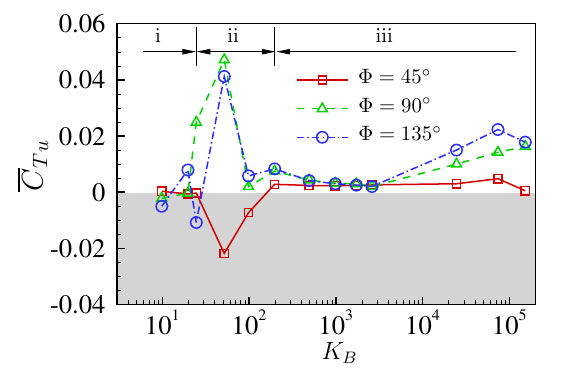}\label{fig:CT_phi_KB_decoma}}
	\subfloat[][]{\includegraphics[width=0.49\textwidth]{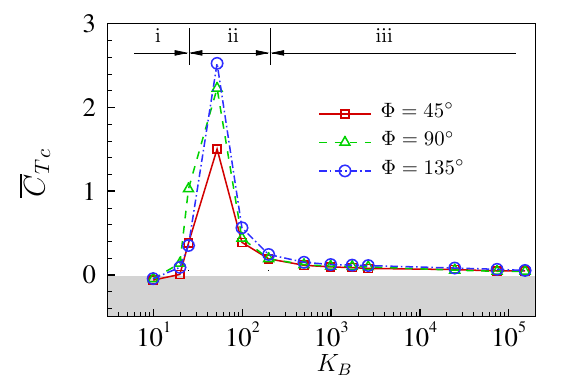}\label{fig:CT_phi_KB_decomb}}
	\\
	\subfloat[][]{\includegraphics[width=0.49\textwidth]{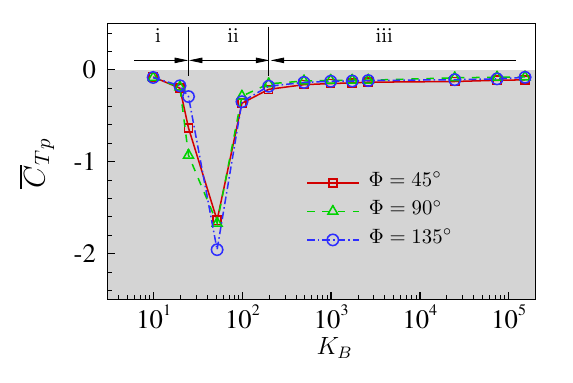}\label{fig:CT_phi_KB_decomc}}
	\caption{\label{fig:CT_phi_KB_decom} (a) Mean unsteady thrust coefficient ${\overline{C}_T}_u$, (b) mean convective thrust coefficient ${\overline{C}_T}_c$ and (c) mean pressure thrust coefficient ${\overline{C}_T}_p$ as a function of bending stiffness $K_B$ for pitching plates with varying trailing edge angles $\Phi$=$45^\circ$, $90^\circ$ and $135^\circ$ at $Re$ = 1000 and $St$ = 0.3.}
\end{figure*}

The variation of the mean unsteady term, the mean convective term and the mean pressure term as a function of trailing edge shape and flexibility is summarized in Fig.~\ref{fig:CT_phi_KB_decom}. The classification of the three distinctive regimes is added to the figures to help characterize the role of flexibility. It can be seen from Fig.~\ref{fig:CT_phi_KB_decom} \subref{fig:CT_phi_KB_decoma} that the contribution of the mean unsteady term to the total mean thrust is quite small, compared to the convective term and the pressure term. The DMD modes indicted by the $X$-vorticity shown in Fig.~\ref{fig:DMD_mode3} demonstrate that the variation of the vortical structures around the plate is almost symmetrical with respect to the $X$-$Y$ plane due to the sinusoidal prescribed motion. Thus, the time-averaged unsteady term is small. It is worth noting that the unsteady term reaches peak values under the resonance condition due to the symmetry breaking of the flow field.

In Fig.~\ref{fig:CT_phi_KB_decom} \subref{fig:CT_phi_KB_decomb}, the mean convective term of the flexible plate shows a peak near resonance, and then it reduces sharply and maintains an almost constant value finally when the plate becomes more rigid. By linking to the time-averaged velocity field in the wake shown in Fig.~\ref{fig:iso_vvel_rotate} and \ref{fig:wake_profile}, the surrounding flows are strongly accelerated by the large passive deformation under the resonance condition to induce the enhancement of the mean convective term. Since the passive deformation is suppressed for stiffer plates, the momentum flux convected downstream is reduced. With respect to the effect of trailing edge shape, it is observed that the rectangular plate can produce the largest mean convective thrust within the low bending stiffness regime. The concave and convex plates have similar contributions within this regime. As shown in Fig.~\ref{fig:plate_mode}, the excited chord-wise second mode reduces the effective areas related to the momentum transfer, which restricts the mean convective thrust generation. The convex plate has the largest contribution to the mean convective thrust at higher $K_B$ values. The concave plate is the most inefficient shape to produce mean momentum flux at moderate $K_B$ values. As the trailing edge angle increases, the improvement of the mean convective thrust is caused by the enhancement of the bifurcated jets containing high velocities observed in Fig.~\ref{fig:iso_vvel_rotate}. However, the effect of trailing edge shape on the mean convective thrust is small when the plate becomes stiff enough.

Similar trends of the pressure term as a function of flexibility are observed in Fig.~\ref{fig:CT_phi_KB_decom} \subref{fig:CT_phi_KB_decomc} for flexible plates with different trailing edge shapes. The contribution of the mean pressure term to the drag force becomes larger near resonance, and then reduces to almost constant values for stiffer plates. The rectangular plate, the convex plate and the concave plate show the largest contributions to the drag penalty within the low bending stiffness regime, the moderate bending stiffness regime and the high bending stiffness regime, respectively.

Among the three decomposed terms, the convective term makes positive contributions to the mean thrust generation and the pressure term leads to drag. The mean unsteady term is slightly larger than zero for most of the cases. The shear stress term is generally negligible. Consequently, the drag-thrust transition is mainly governed by the relative values of the convective term and the pressure term on the control surfaces. Increasing flexibility close to the resonance condition is able to largely improve the convective term. Within the optimal thrust region, the trailing edge shape has an opposite effect on the convective term and the pressure term. From Fig.~\ref{fig:thrust_efficiency}, two drag-thrust transition boundaries are observed as a function of flexibility. 
When varying from the low bending stiffness regime to the moderate bending stiffness regime near resonance, the transition from drag to thrust  is mainly caused by a significant increase in the momentum flux. The variation of the dominant structural mode from the chord-wise second mode to the first mode improves the effective areas and acceleration. As a result, this produces the vortical structures containing higher velocities. When the flexible plates approach to their rigid counterparts, the transition is triggered by the reduced passive deformation. With regard to the effect of trailing edge shape, the convex plate broadens the drag-thrust transition region as a function of flexibility by enhancing the momentum transfer to the wake. The concave plate has the narrowest transition region due to the negative unsteady thrust term within the moderate bending stiffness regime.

\subsection{Added mass effect on thrust generation}
\begin{figure*}
	\subfloat[][]{\includegraphics[width=0.7\textwidth]{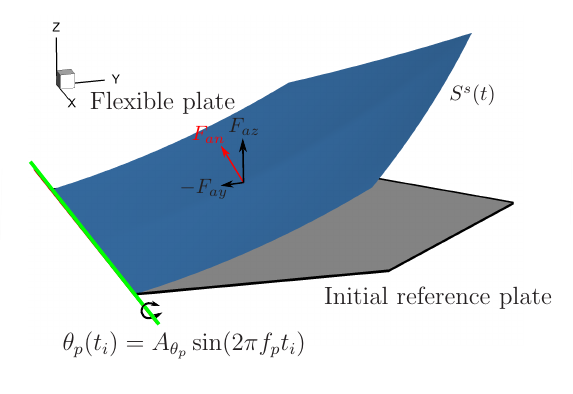}\label{fig:CT_added_mass_demoa}}
	\\
	\subfloat[][]{\includegraphics[width=0.9\textwidth]{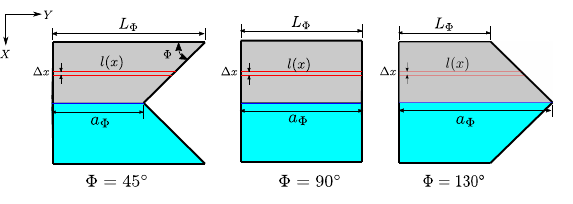}\label{fig:CT_added_mass_demob}}
	\caption{\label{fig:CT_added_mass_demo} Schematic of (a) force decomposition of the added mass force at time instant $t_i$ and (b) integral of the local added mass force $m_a (\bm{a}^s \cdot \bm{n}_c)$ at each Lagrangian point over the whole plate surface $S^s$ to calculate the normal component of the added mass thrust. $T=-F_{ay}$ indicates the thrust force and $F_{az}$ is the lateral force. $\theta_p(t_i)$ represents the instantaneous pitching angle at the leading edge at time instant $t_i$. $\Delta x$ denotes the local span length in the integral and $l(x)$ is the local chord length. $L_{\Phi}$ and $a_{\Phi}$ represent the chord length at the side and at the center, respectively.}
\end{figure*}

The added mass force plays an important role in thrust or lift generation for aquatic swimming and bird/insect flying \cite{bottom2016hydrodynamics}. To gain further insight into the effect of flexibility and trailing edge shape in the current study, the generated thrust force due to the added mass effect is estimated via an analytical formulation of a pitching flexible plate \cite{yadykin2003added}. Due to the added mass effect, the adjacent fluid can be accelerated by the flapping plate to generate a reaction force $\bm{F}_a$. The added mass force has a component in the thrust direction, thereby contributing to the thrust generation during flapping motion \cite{andro2009frequency,guvernyuk2020contribution}. Figure~\ref{fig:CT_added_mass_demo} \subref{fig:CT_added_mass_demoa} illustrates a schematic of the thrust generation contributed by the added mass force $\bm{F}_a$ at a selected time instant $t_i$ for the flexible plate with an instantaneous pitching angle of $\theta_p(t_i)$. The added mass force $\bm{F}_a$ can be decomposed into the normal component $F_{an}$ and the tangential component $F_{a \tau}$. In the current study, the tangential force $F_{a \tau}$ can be neglected for a relatively thin flexible plate. The normal force $F_{an}$ of the added mass force can be evaluated by \cite{yadykin2003added,guvernyuk2020contribution}
\begin{equation}
F_{an} =-\oint_{S^s(t)} m_a (\bm{a}^s \cdot \bm{n}_c) {\rm{d}} S 
%\\
%F_{at} =-\oint_{S^s(t)} m_a \bm{a}^s \cdot \bm{\tau}_c {\rm{d}} S
\label{eq:thrust_added}
\end{equation}
where $m_a=\frac{2 \rho^f l(x)}{m^s \pi}$ is the added mass coefficient of a surface plate per unit area, which is obtained from the analytical added mass tensor \cite{yadykin2003added}. Here $m^s$ denotes the chord-wise mode number and $l(x)$ is the local chord length at different span-wise locations. $\bm{a}^s$ is the acceleration of the plate at each Lagrangian point and $\bm{n}_c$ represents the unit vector normal to the chord of the moving plate surface $S^s(t)$. 
 In the current formulation, the variation of the chord-wise mode is observed and the influence of the mode shape on the added mass effect is considered.
 As illustrated in Fig.~\ref{fig:CT_added_mass_demo} \subref{fig:CT_added_mass_demoa}, the thrust force due to the added mass effect $T=-F_{ay}$ is calculated by projecting the normal component of the added mass force $F_{an}$ on the inversed freestream direction. Thus, the thrust coefficient due to the added mass effect ${C_T}_{a}$ can be written as
\begin{equation}
{C_T}_{a} = -\frac{F_{ay} }{\frac{1}{2} \rho^f U_{\infty}^2 S} = -\frac{F_{an} (\bm{n}_c \cdot \bm{n}_y) }{\frac{1}{2} \rho^f U_{\infty}^2 S}
\label{eq:thrust_added2}
\end{equation}
The thrust coefficient due to the added mass effect ${C_T}_{a}$ reflects a connection with the acceleration and geometry of the flexible plate. The residual thrust coefficient ${C_T}_{r}$ is calculated by subtracting ${C_T}_{a}$ from the total thrust coefficient ${C_T}$, which is defined as ${C_T}_{r} = {C_T} - {C_T}_{a} $. The residual thrust coefficient ${C_T}_{r}$ does not depend on the acceleration, but depends on the variation of the spatial vorticity and the fluid velocity in the fluid domain \cite{guvernyuk2020contribution}. 

\begin{figure*}
	\subfloat[][]{\includegraphics[width=0.49\textwidth]{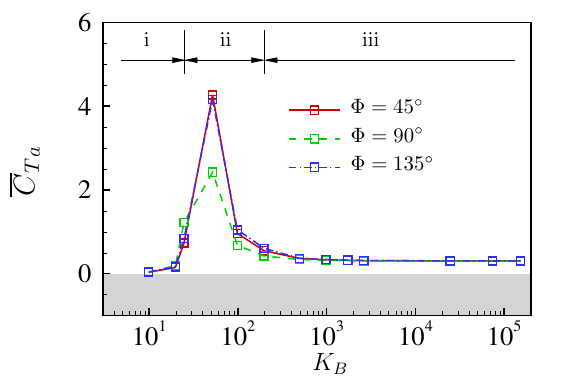}\label{fig:CT_phi_KB_ama}}
	\subfloat[][]{\includegraphics[width=0.49\textwidth]{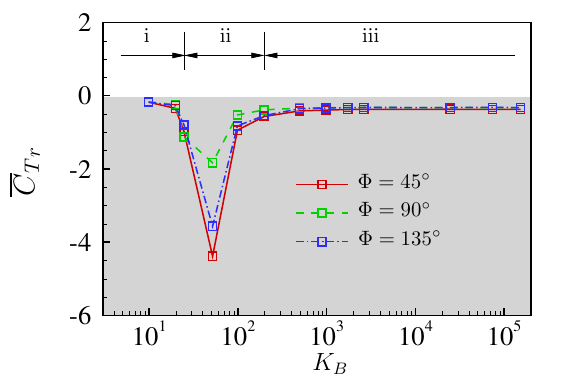}\label{fig:CT_phi_KB_amb}}
	\caption{\label{fig:CT_phi_KB_am} (a) Mean thrust coefficient due to added mass ${\overline{C}_T}_a$ and (b) mean residual thrust coefficient ${\overline{C}_T}_r$ as a function of bending stiffness $K_B$ for pitching plates with varying trailing edge angles $\Phi$=$45^\circ$, $90^\circ$ and $135^\circ$ at $Re$ = 1000 and $St$ = 0.3.}
\end{figure*}

With the aid of the analytical formulation of the added mass force, we evaluate the mean thrust coefficients contributed by the added mass effect and the residual term for pitching plates with varying flexibility and trailing edge angles, as shown in Fig.~\ref{fig:CT_phi_KB_am}. It can be seen that the mean thrust coefficient due to the added mass effect shows overall positive values in the studied parameter space. As illustrated in Fig.~\ref{fig:CT_added_mass_demo} \subref{fig:CT_added_mass_demob}, the geometry and acceleration distributions of the upper half part of the concave plate in gray color is the same as those of the lower half part of the convex plate in cyan color. Although the concave and convex plates in the current study show different trailing edge shapes, the integrated added mass thrust forces of the concave and convex plates using Eq.~(\ref{eq:thrust_added}) and (\ref{eq:thrust_added2}) are similar. The acceleration near the trailing edge of the rectangular plate is smaller than the non-flat plate due to the shorter local chord length, resulting in the overall lower thrust due to the added mass for the rectangular plate. Regardless of the trailing edge shape, the generated thrust due to the added mass force achieves peak values near the resonance and decreases when the plate becomes more rigid. We observe from Fig.~\ref{fig:structure_dynamic} that the plate exhibits significantly increased amplitude and acceleration under the resonance condition. Thus, the added mass effect becomes stronger when the resonance is established. It is worth noting that the variation of the trailing edge shape is limited to the three selected cases in the current study. The connection between the trailing edge shape and the added mass effect should be explored in detail over a wider range of trailing edge shapes in future studies.

The mean residual thrust coefficient related to the fluid velocity and the spatial vorticity distribution as a function of trailing edge shape and flexibility is shown in Fig.~\ref{fig:CT_phi_KB_am} \subref{fig:CT_phi_KB_amb}. The non-added mass effect has negative contributions to the mean residual thrust in the studied parameter space. We observe that the mean residual thrust coefficient of the concave plate is smaller than that of the convex plate within the studied bending stiffness range. As shown in Fig.~\ref{fig:iso_vvel_rotate} and \ref{fig:DMD_mode3}, the distributions of the fluid velocity and the spatial vorticity are strongly affected by the trailing edge shape. Thus, the mean residual thrust significantly depends on the trailing edge shape. The effect of the distribution of the flow features on the thrust generation is discussed in detail in Section \ref{unsteady_momentum} from the perspective of the unsteady momentum transfer. Although the mean added mass thrust between these two plates are similar, the difference of the mean residual thrust leads to the overall large mean total thrust of the convex plate than the concave plate. The mean residual thrust coefficient also shows peak values under the resonance condition. The reason can be attributed to the strong variation of the flow features near resonance.

\begin{figure*}
	\subfloat[][]{\includegraphics[width=0.49\textwidth]{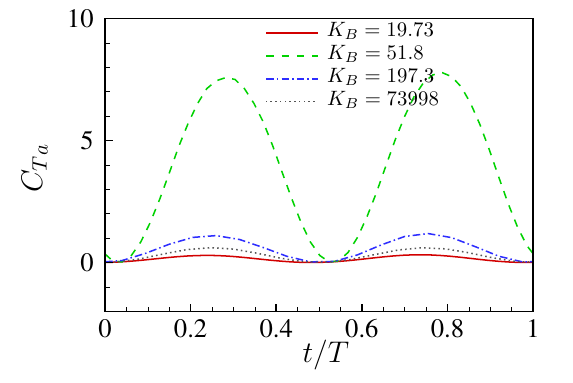}\label{fig:CT_KB_am_decoma}}
	\subfloat[][]{\includegraphics[width=0.49\textwidth]{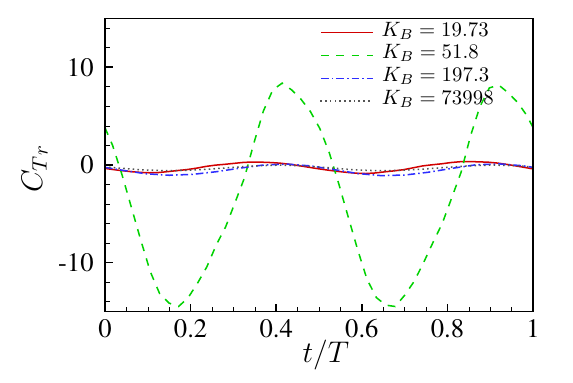}\label{fig:CT_KB_am_decomb}}
%	\\
%	\subfloat[][]{\includegraphics[width=0.49\textwidth]{./Figures/phi135/Re1000/CT_added_mass/CT_totalCT_KB_com_phi135.pdf}\label{fig:CT_KB_am_decomc}}
	\caption{\label{fig:CT_KB_am_decom} Comparison of (a) thrust coefficient due to added mass and (b) residual thrust coefficient of pitching plate with trailing edge angle of $\Phi=135^\circ$ within one completed pitching cycle. $t/T=0$ corresponds to the pitching upward from the neutral position.}
\end{figure*}

To characterize the added mass effect, Fig.~\ref{fig:CT_KB_am_decom} presents the comparison of the time-varying added mass thrust coefficient and the residual term as a function of bending stiffness for the convex plate. It can be seen from Fig.~\ref{fig:CT_KB_am_decom} \subref{fig:CT_KB_am_decoma} that the structural resonance can significantly amplify the added mass thrust due to the induced large acceleration. The contribution of the added mass to the thrust generation is suppressed when the plate becomes more rigid or softer under the off-resonance condition. The distributions of the fluid velocity and the spatial vorticity are significantly affected by interacting with the large-amplitude flapping plate near resonance. Thus, the residual thrust coefficient shows drastic changes under the resonance condition. 

\begin{figure*}
	\subfloat[][]{\includegraphics[width=0.49\textwidth]{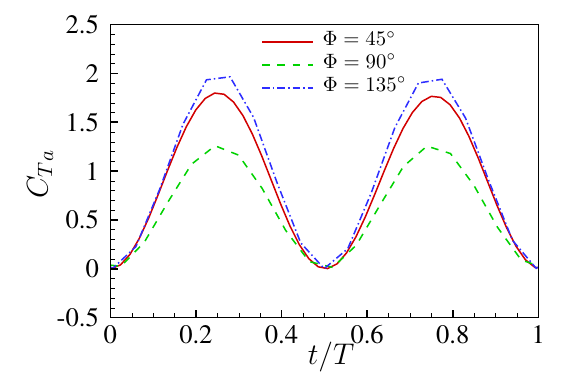}}
	\subfloat[][]{\includegraphics[width=0.49\textwidth]{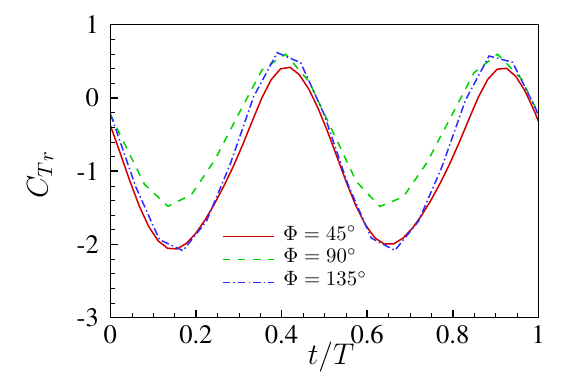}}
%	\\
%	\subfloat[][]{\includegraphics[width=0.49\textwidth]{./Figures/phi135/Re1000/CT_added_mass/CT_totalCT_phi_com_KB98.pdf}}
	\caption{\label{fig:CT_phi_am_decom} Comparison of (a) thrust coefficient due to added mass and (b) residual thrust coefficient of pitching plate with $K_B=98.66$ within one completed pitching cycle. $t/T=0$ corresponds to the pitching upward from the neutral position.}
\end{figure*}

Figure~\ref{fig:CT_phi_am_decom} presents the comparison of the time-varying thrust coefficient due to the added mass and residual terms as a function of trailing edge shape for the flexible plate with $K_B$=98.66. The added mass thrust coefficient of the convex plate has a slightly larger amplitude than that of the concave plate. It can be seen from Fig.~\ref{fig:structure_dynamic} that the convex plate exhibits a larger pitching amplitude at the trailing edge. Consequently, the difference in the added mass thrust between the convex and concave plates is caused by the larger acceleration of the convex plate. The added mass thrust of the rectangular plate is the smallest due to the lower acceleration and the short local chord lengths. The time-varying residual thrust coefficient of the rectangular plate presents the largest values among the three plates near $t/T=0.15$, while the concave plate shows the smallest values among all plates at $t/T=0.4$.

Through the systematic investigation of the role of flexibility, it can be concluded that the thrust generation is improved near $f^*=1$ corresponding to the resonance condition. Floryan \cite{floryan2018clarifying,floryan2019distributed} reported studies on the propulsive performance of a passively flexible structure with varying flexibility by employing a two-dimensional linear inviscid aerodynamic model coupled with the Euler-Bernoulli beam model. It was found that the structural resonance can lead to significant thrust gains and optimal efficiency. However, the studies were limited to low-frequency and small amplitude motions with no separation. In this simplified aerodynamic model \cite{floryan2018clarifying,floryan2019distributed}, the temporal and spatial evolution of the surrounding flow features and the vortex structure were not provided. Highly nonlinear coupled effects of the flapping motions, the instantaneous flow features and the time-dependent fluid loads were not captured. In the current study, we have employed our nonlinear body-fitted fluid-structure interaction solver for our high-fidelity numerical study on the three-dimensional pitching flexible plates with varying flexibility and trailing edge shapes. With the aid of the improved SP-DMD mode decomposition method, the momentum-based thrust evaluation approach and the analytical added mass model, the flapping characteristics and the unsteady fluid loads are quantified to understand the thrust-generating mechanism of a 3D flexible plate with varying flexibility and trailing edge shapes. Our findings show that the structural resonance can improve the thrust generation and promote the transition from drag to thrust by enhancing the unsteady momentum transfer and the added mass effect at large amplitude flapping motions. The convex trailing edge shape has more positive contributions to the thrust generation.

\section{Conclusions} \label{sec:section5}
We systematically explored the effect of flexibility and trailing edge shape on the propulsive performance of pitching flexible plates with a fixed actuated frequency. 
Based on the variation of the structural motions and the propulsive performance, three distinctive flapping motion regimes were classified as a function of flexibility, namely (i) low bending stiffness $K_B^{low}$, (ii) moderate bending stiffness $K_B^{moderate}$ near resonance, and (iii) high bending stiffness $K_B^{high}$. 
To examine the role of flexibility, the frequency ratio $f^*$ between the natural frequency of the flexible plate immersed in the unsteady flow and the pitching frequency was calculated. 
A combined SP-DMD method was employed to correlate the flow features and the structural motions. By examining the flapping dynamics and the flow features, we found that the maximum mean thrust was achieved about $f^* \approx 1$ corresponding to resonance. 
The large passive deformation governed by the flexibility effect can redistribute the pressure gradient to enhance the thrust generation. The optimal propulsive efficiency was observed for flexible plates with moderate passive deformation around $f^*$=1.54. 
Since maintaining the large passive deformation under the resonance condition required more input power, the efficiency was reduced compared to the system at a higher $K_B$ value. 
Based on the configurations undertaken in this work, the convex shape can help improve the mean thrust within $K_B^{moderate}$ and $K_B^{high}$ regimes, and achieve the optimal efficiency at low and high $K_B$ values. 
The rectangular shape has shown the largest mean thrust at low $K_B$ values, and was the most efficient propulsive system near resonance. 
The concave shape presented the poorest propulsive performance within the studied $K_B$ range. 
We employed a momentum-based thrust evaluation method to quantitatively examine the contributions of the evolution of the vortical structures to the produced thrust forces. Typically, the instantaneous maximum thrust was achieved when the vortex detached from the trailing edge. From the perspective of the unsteady momentum transfer, the reason was attributed to the large rate of change of the fluid momentum by the accelerating plate and the momentum convection process. 
The moderate flexibility near the resonance can greatly accelerate the surrounding fluid, thereby imparting more momentum to the fluid to enhance the thrust generation. 
The convex shape can generate vortical structures with higher velocities due to the longer local chord at the midspan location. 
Thus, the wake containing more momentum was convected downstream to improve the produced thrust. 
To shed light on the drag-thrust transition mechanism, we examined the variation of the decomposed thrust terms as functions of flexibility and trailing edge shape. By adjusting the flexibility value to make $f^*$ close to 1, the mean momentum convection was enhanced to promote the transition from drag to thrust. 
The increase of the trailing edge angle can further help the momentum convection process to broaden the range of flexibility that can generate positive mean thrust. Trough the investigation of the generated thrust force due to the added mass effect, we found that the moderate flexibility corresponding to resonance can enhance the added mass thrust. The non-flat trailing edge shape has more contributions to the added mass thrust, compared to the rectangular shape with the same area.

By examining the evolution of the temporal and spatial vortical structures and the time-dependent thrust, we found that the variation of the propulsive performance cannot be simply determined from the wake topologies when the flow features became complex and cannot be regarded as the typical reverse von Kármán vortex street. Some often cited thrust-generating explanations based on the vortex spacing and the typical wake patterns may not be suitable for the propulsion system with complex wake structures and varying Reynolds numbers. In this study, we employed a mode decomposition method called SP-DMD, a non-intrusive velocity/pressure momentum-based thrust evaluation approach and an analytical added mass model to reveal the thrust-generating mechanism for the coupled flapping system with complex wake structures. The unsteady momentum-based approach allows establishing a direct correlation between the flexible plate deformation, the temporal and spatial evolution of the vortical structures and the time-dependent fluid loads, while the added mass model estimates the reactive force due to the acceleration of the plate. The proposed nonlinear FSI study on the unsteady momentum transfer and the added mass effect provides a comprehensive understanding of the thrust-generating mechanism for efficient bio-inspired propulsion systems. This fundamental mechanism offered an effective way to design an optimal propulsion system with flexible wings. Some passive or active control methods can be used to transfer more momentum to the fluid and adjust the accelerations to enhance the thrust generation and the propulsive performance.

%\section*{DATA AVAILABILITY} \label{sec:section6}
%The data that support the findings of this study are available within this article.

\begin{acknowledgments}
The authors wish to acknowledge supports from the National University of Singapore
and the Ministry of Education, Singapore. The third author would like to acknowledge the support from the University of British Columbia and the Natural Sciences and Engineering Research Council of Canada (NSERC).
\end{acknowledgments}

\section*{APPENDIX A: Mesh convergence study and validation} \label{sec:section7}
In this appendix, we present a mesh convergence study and validation of our fluid-structure interaction solver.
%Before we proceed to investigate the role of the trailing edge shape and the flexibility on the propulsive performance of a pitching flexible foil, a mesh convergence study is conducted to ensure a sufficient mesh resolution. The generated thrust and propulsive efficiency for the pitching foil with a concave trailing edge shape obtained from the numerical simulation are compared against the experimental results to validate the numerical framework. The flow field details around a pitching foil with a convex trailing edge shape are also compared for validation purpose.
\subsection{Mesh convergence}
To perform the mesh convergence study and choose proper fluid and structure meshes with  sufficient mesh resolutions, we design three different meshes namely M1, M2 and M3 for the numerical simulation of the pitching plate problem. The unstructured finite element mesh is utilized to discretize the three-dimensional computational fluid domain into 268 030, 524 094 and 1 008 230 eight-node hexahedron elements for the three sets of meshes, respectively. A boundary layer mesh with a stretching ratio of 1.15 in the direction perpendicular to the plate surface is formed to ensure $y^+$ of the first layer less than 1.0. The pitching flat plate is modeled by the geometrically exact co-rotational shell elements with a prescribed rotation motion applied along the leading edge. A clamped boundary condition with fixed displacements is imposed along the leading edge, and only the relative rotation around the $X$-axis is allowed. Corresponding to the three sets of fluid meshes, we construct three sets of structure meshes for the flat plate, which consist of 40, 105 and 200 four-node quadrilaterals, respectively.

To keep consistent with the experimental condition, the plate is placed in the unsteady water medium with a freestream velocity of $U_{\infty} = 0.1$ m/s in the numerical simulation, which leads to a Reynolds number of $Re=10000$. The mass ratio is $m^*=0.03$ and the Young's modulus is set to $E=3.1 \times 10^9$ N/m$^2$. The pitching amplitude is $A_{\theta_p}=9^\circ$ and the pitching frequency is $f_p=1.1$ Hz for the prescribed pitching motion. Thus, the non-dimensional Strouhal number is $St=0.344$, which is relatively close to the Strouhal number $St$ corresponding to the pitching plate with the maximum propulsive efficiency measured in the water tunnel experiment \cite{VanBuren2017}. In the experiment, The input power of the pitching plate is defined as $P_{input}=\tau_X \dot{\theta}$, where $\tau_X$ is the torque along the span-wise direction around the leading edge and $\dot{\theta}$ denotes the angular velocity of the pitching motion. Thus, the Froude efficiency is calculated by $\eta_F=\frac{\overline{T} U_{\infty}}{\overline{P}_{input}}=\frac{\overline{C}_T}{\overline{C}_{power}}$. The non-dimensional time step is set to 0.01 for the numerical simulation, which allows sufficient resolution to capture the dynamics of the coupled system in each pitching cycle. The mesh characteristics and the performance results of the concave plate are summarized in Table~\ref{tab:mesh convergence}. The percentage differences for performance results of M1 and M2 are calculated with respect to those of M3. It can be concluded that the absolute errors for M2 are less than $2\%$. Hence, M2 is chosen for the further numerical validation study.

\begin{table*}
	\caption{\label{tab:mesh convergence}Mesh characteristics, mean thrust coefficient $\overline{C}_T$ and propulsive Froude efficiency $\eta_F$ for mesh convergence of a concave plate at $Re=10000$ with a Strouhal number of $St=0.344 $ and a trailing edge angle of $\Phi=45^\circ$. The percentage differences are calculated by using M3 results as the reference.}
	\begin{ruledtabular}
		\begin{tabular}{c c c c c c}
			Mesh & Structure elements & Fluid elements & $\overline{C}_T$ & ${C_L^{\prime}}^{rms}$ & $\eta_F$\\
			\hline
			M1 & 40 & 268 030  & 0.1167 (-1.93 \%) & 3.6312 (-0.77 \%)  & 0.0862 (-1.15 \%)  \\  
			M2 & 105 & 524 094 & 0.1176 (-1.17 \%) & 3.6266 (-0.90 \%)  & 0.0869 (-0.34 \%)  \\  
			M3 & 200 & 1 008 230 & 0.1190 & 3.6595 & 0.0872 \\  
		\end{tabular}
	\end{ruledtabular}
\end{table*}

\subsection{Validation}
\begin{figure*}
	\centering
	\subfloat[]{\includegraphics[width=0.49 \textwidth]{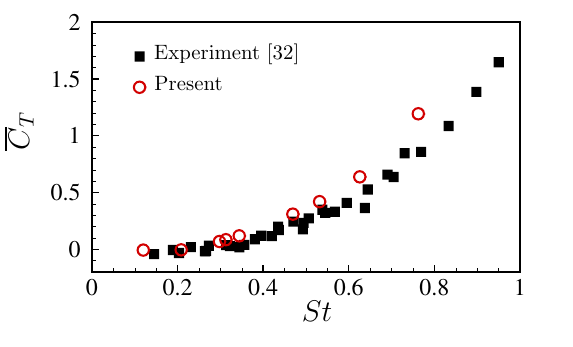}\label{thrust}}
	\subfloat[]{\includegraphics[width=0.49 \textwidth]{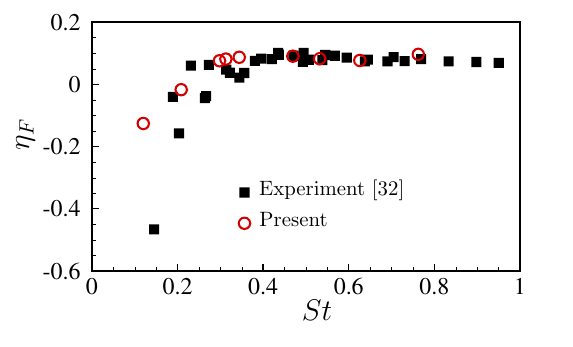}\label{efficiency}}
	\caption{\label{fig:comparison experiment}Flow past a pitching plate with a trailing edge angle of $\Phi=45^\circ$: comparison of the (a) mean thrust coefficient $\overline{C}_T$ and (b) propulsive Froude efficiency $\eta_F$ at different Strouhal numbers $St$ between the experimental data [\citenum{VanBuren2017}] and the numerical simulations at $Re=10000$.}
\end{figure*}

Based on the mesh convergence study, we select an optimal fluid and structure mesh combination M2 to further validate the developed numerical framework within a range of Strouhal numbers. The concave plate with a trailing edge angle of $\Phi=45^\circ$ is considered in the validation study. The simulation parameters for the coupled system are set the same as those in the mesh convergence study except for the Strouhal number. The pitching plate with a Strouhal number range of $St \in [0.12,0.76]$ is simulated by the developed fluid-structure interaction solver to calculate the propulsive performance, which is used to compare with the experimental results \cite{VanBuren2017} for validation purpose. Figure~\ref{fig:comparison experiment} depicts the mean thrust coefficient $\overline{C}_T$ and the propulsive Froude efficiency $\eta_F$ as a function of the Strouhal number $St$ for the numerical simulations and the water tunnel experiments \cite{VanBuren2017}. It can be seen that the overall trends of the simulated thrust coefficient and propulsive efficiency are well predicted, compared to the experimental results.

\section*{APPENDIX B: Sparsity-promoting dynamic mode decomposition of fluid-structure interaction} \label{sec:section8}
The DMD method provides an algorithm to identify the essential coherent wake structures in the complex fluid field and the structural modes by decomposing the numerical or experimental dataset into the non-orthogonal modes with a single temporal frequency for each mode. To select the most influential DMD modes from the whole modes without losing the quality of the approximation, sparsity-promoting dynamic mode decomposition (SP-DMD) proposed by Jovanovi{\'c} et al. \cite{jovanovic2014sparsity} seeks a sparsity structure and solves an optimization problem to achieve a desirable tradeoff between the approximation error and the number of influential modes by adjusting a user-defined regularization parameter $\gamma_{\rm{spdmd}}$ in the penalized term. We summarize the concise algorithm of the snapshot-based SP-DMD method in the following, and more details can be found in the literature \cite{jovanovic2014sparsity}.

Firstly, we collect the physical data $\boldsymbol{\chi}_i \in \mathbb{C}^{M \times 1}$ at the $M$ measurement spatial points with a equispaced time interval $\Delta t_{\rm{spdmd}}$ from the numerical simulation in a snapshot way for decomposition purpose. The collected data is sorted into two matrices shown as
\begin{eqnarray}
\boldsymbol{X_0} := [\boldsymbol{\chi}_0, \boldsymbol{\chi}_1, \boldsymbol{\chi}_2, \cdots, \boldsymbol{\chi}_{N-1}] \in \mathbb{C}^{M \times N},
\label{eq:spdmd1}
\\
\boldsymbol{X_1} := [\boldsymbol{\chi}_1, \boldsymbol{\chi}_2, \boldsymbol{\chi}_3, \cdots, \boldsymbol{\chi}_{N}] \in \mathbb{C}^{M \times N},
\label{eq:spdmd2}
\end{eqnarray}
where $N$ represents the total time sampling number. 

A linear relationship is assumed between these two time-sequential snapshots and a transformation matrix $\boldsymbol{A} \in \mathbb{C}^{M \times N}$ projects $\boldsymbol{X}_0$ to its next time step sequence $\boldsymbol{X}_1$, which is written as
\begin{equation}
\boldsymbol{X}_1 = \boldsymbol{A} \boldsymbol{X}_0
\label{eq:spdmd3}
\end{equation}

In the DMD algorithm, the transformation matrix $\boldsymbol{A}$ can be represented by an optimal similar matrix $\tilde{\boldsymbol{A}} \in \mathbb{C}^{r \times r}$ using a projection matrix $\boldsymbol{U}$.
\begin{equation}
%\tilde{\boldsymbol{A}} = \boldsymbol{U}^* \boldsymbol{A} \boldsymbol{U}
\boldsymbol{A} = \boldsymbol{U} \tilde{\boldsymbol{A}} \boldsymbol{U}^*
\label{eq:spdmd4}
\end{equation}
where $\boldsymbol{U} \in \mathbb{C}^{M \times r}$ is the POD modes of $\boldsymbol{X_0}$ calculated from a singular value decomposition (SVD) of $\boldsymbol{X_0} = \boldsymbol{U} \boldsymbol{\Sigma} \boldsymbol{V}^*$. In Eq.~(\ref{eq:spdmd4}), $\boldsymbol{U}^*$ is the complex-conjugate-transpose of $\boldsymbol{U}$. $\boldsymbol{\Sigma} \in \mathbb{C}^{r \times r}$ represents the singular values and $\boldsymbol{V}^*$ denotes the complex-conjugate-transpose of $\boldsymbol{V} \in \mathbb{C}^{r \times N}$ in SVD. Both $\boldsymbol{U}^* \boldsymbol{U} = \boldsymbol{I}$ and $\boldsymbol{V}^* \boldsymbol{V} = \boldsymbol{I}$ are satisfied.

To calculate the unknown matrix $\tilde{\boldsymbol{A}}$ for DMD, we construct an optimization problem shown below
\begin{equation}
\min(\tilde{\boldsymbol{A}}) : \lVert \boldsymbol{X_1} - \boldsymbol{U} \tilde{\boldsymbol{A}} \boldsymbol{\Sigma} \boldsymbol{V}^* \rVert^2_F
\end{equation}
where the optimal solution to this problem is $\tilde{\boldsymbol{A}}=\boldsymbol{U}^* \boldsymbol{X_1} \boldsymbol{V} \boldsymbol{\Sigma}^{-1}$. 

Once the matrix $\tilde{\boldsymbol{A}}$ is obtained, the DMD modes and their corresponding mode frequencies can be determined by solving the eigenvector $\boldsymbol{Y} \in \mathbb{C}^{r \times r}$ and eigenvalue $\boldsymbol{\Lambda} \in \mathbb{C}^{r \times r}$ of $\tilde{\boldsymbol{A}}$
\begin{equation}
\tilde{\boldsymbol{A}} \boldsymbol{Y} = \boldsymbol{Y} \boldsymbol{\Lambda}
\end{equation}

Thus, the DMD modes $\boldsymbol{\mathcal{V}} \in \mathbb{C}^{r \times r}$ are calculated by projecting the eigenvectors $\boldsymbol{Y}$ using the POD modes $\boldsymbol{U}$
\begin{equation}
\boldsymbol{\mathcal{V}} = [\boldsymbol{\upsilon}_1, \boldsymbol{\upsilon}_2, \boldsymbol{\upsilon}, \cdots, \boldsymbol{\upsilon}_r] = \boldsymbol{U} \boldsymbol{Y}
\end{equation}

The growth/decaying rate $\boldsymbol{\kappa}_{\rm{spdmd}}$ and the frequency $\boldsymbol{f}_{\rm{spdmd}}$ for the DMD modes are given below
\begin{eqnarray}
\boldsymbol{\kappa}_{\rm{spdmd}}= {\rm{real}}(\log (\boldsymbol{\Lambda}))
\\
\boldsymbol{f}_{\rm{spdmd}} = {\rm{imag}}(\log (\boldsymbol{\Lambda})/(2 \pi\Delta t_{\rm{spdmd}}))
\end{eqnarray}

Hence, the snapshot flow field can be reconstructed based on the DMD modes, which is given as
\begin{equation}
\boldsymbol{X}_0 \approx \boldsymbol{\mathcal{V}} \boldsymbol{D(\alpha)} \boldsymbol{V}_{\rm{and}}(\boldsymbol{\Lambda})
\end{equation}
where $\boldsymbol{D(\alpha)}$ represents the diagonal matrix of the DMD mode amplitude $\boldsymbol{\alpha}$. The standard Vandermonde matrix $\boldsymbol{V}_{\rm{and}}(\boldsymbol{\Lambda)}$ governs the temporal characteristics of the dynamic modes. 

The solution of the unknown mode amplitude $\boldsymbol{\alpha}$ can be achieved by solving an optimal optimization problem of minimizing the Frobenius norm of the difference between the collected snapshot matrix $\boldsymbol{X}_0$ and the reconstructed flow field based on the DMD modes, which is defined as
\begin{equation}
%{\mathop {\min} \limits_{\boldsymbol{\alpha}}} J (\boldsymbol{\alpha}) := \lvert \boldsymbol{X}_0 - \boldsymbol{D(\alpha)} \boldsymbol{V}_{\rm{and}}(\boldsymbol{\Lambda}) \rvert^2_F
\min({\boldsymbol{\alpha}}) : \lVert \boldsymbol{X}_0 - \boldsymbol{\mathcal{V}} \boldsymbol{D(\alpha)} \boldsymbol{V}_{\rm{and}}(\boldsymbol{\Lambda}) \rVert^2_F
\end{equation}

After the decomposition of the collected physical field based on the typical DMD method, the SP-DMD method proposed by Jovanovi{\'c} et al. \cite{jovanovic2014sparsity} is adopted to eliminate the undesirable modes with weak contributions to the physical data sequence and select the most dominant DMD modes by penalizing the $\ell_1$-norm of the non-zero amplitude $\boldsymbol{\alpha}$ of the DMD modes with a pre-defined parameter $\gamma_{\rm{spdmd}}$. Hence, the balance between the approximation error relative to the entire physical data sequence and the number of extracted DMD modes is achieved. The optimal values of the DMD amplitude can be determined from the following optimization problem
\begin{equation}
\min({\boldsymbol{\alpha}}) : \lVert \boldsymbol{X}_0 - \boldsymbol{\mathcal{V}} \boldsymbol{D(\alpha)} \boldsymbol{V}_{\rm{and}}(\boldsymbol{\Lambda}) \rVert^2_F + \gamma_{\rm{spdmd}} \sum_{i=1}^{r} |\alpha_i|
\end{equation}

The SP-DMD method allows us to flexibly identify the most influential DMD modes by adjusting the parameter $\gamma_{\rm{spdmd}}$ according to the specified requirement. Larger values of $\gamma_{\rm{spdmd}}$ selects fewer DMD modes, resulting in reduction of the approximation quality. To extract correlated structural motions and flow features from the fluid-flexible plate coupled system, a combined fluid-structure formulation based on mode decomposition technique is proposed by Goza et al. \cite{goza2018modal} to treat the fluid and structural fields together. In this study, we extend this combined formulation to the SP-DMD method to efficiently select the correlated DMD modes. The fluid variables of interest obtained from the fluid-structure interaction solver are mapped to a stationary reference mesh before the mode decomposition process. The projected fluid variables at $M^f$ spatial points are formed a sequential vector of $\boldsymbol{\chi}_i^f \in \mathbb{C}^{M^f \times 1}$. The structural displacement vector in the Lagrangian coordinate are collected at $M^s$ points on structural surface written as $\boldsymbol{\chi}_i^s \in \mathbb{C}^{M^s \times 1}$. The total vector for mode decomposition is expressed as $\boldsymbol{\chi}_i = \begin{bmatrix}
\boldsymbol{\chi}_i^f \\ \boldsymbol{\chi}_i^s \end{bmatrix} \in \mathbb{C}^{(M^f+M^s) \times 1}$. This total vector consisting of fluid variables and structural displacements is then written into the matrix forms shown in Eq.~(\ref{eq:spdmd1}) and (\ref{eq:spdmd2}) as the input for the mode decomposition. With the help of the combined SP-DMD method, the dominant wake structure behind the pitching plate and the structural modes can be identified from the complex coupled system together, so as to further understand the thrust generation mechanism in Section~\ref{sec:section4}.

\section*{DATA AVAILABILITY} \label{sec:section6}
The data that support the findings of this study are available within this article.

\nocite{*}
\bibliography{referenceBib}% Produces the bibliography via BibTeX.

\end{document}